\documentclass[preprint,review,3p,number]{elsarticle}

\usepackage{amsmath,amssymb,amsfonts,amsthm}
\usepackage{subfiles}
\usepackage{float}
\usepackage{booktabs}
\usepackage{enumitem}
\usepackage{pifont}

\usepackage[theorems,skins,breakable]{tcolorbox}
\usepackage[linesnumbered,noend]{algorithm2e}
\SetAlgoNoLine
\SetAlgoNoEnd
\DontPrintSemicolon
\SetKw{And}{and}
\SetKw{Or}{or}
\usepackage[colorlinks = true,linkcolor = blue,urlcolor = blue]{hyperref}

\newcommand{\mb}{\mathbf}
\newcommand{\mc}{\mathcal}
\newcommand{\ms}{\mathsf}

\newcommand{\lr}{\stackrel{\$}{\leftarrow}}

\newtheorem{thm}{Theorem}
\newtheorem{defn}[thm]{Definition}
\newtheorem{lem}[thm]{Lemma}
\newtheorem{prop}{Proposition}
\newproof{pf}{Proof}

\journal{Journal of Computer Standards and Interfaces}

\begin{document}

\begin{frontmatter}



\title{Traceable Signatures from Lattices}


\author[uow,csiro]{Nam Tran} 
\author[uow]{Khoa Nguyen}
\author[csiro]{Dongxi Liu}
\author[csiro,polish-aos]{Josef Pieprzyk}
\author[uow]{Willy Susilo}
\affiliation[uow]{organization={Institute of Cybersecurity and Cryptology, School of Computing and Information Technology, University of Wollongong},
            city={Wollongong},
            postcode={2500}, 
            state={NSW},
            country={Australia}}

\affiliation[csiro]{organization={CSIRO's Data61},
            addressline={Marsfield}, 
            city={Sydney},
            postcode={2122}, 
            state={NSW},
            country={Australia}}

\affiliation[polish-aos]{organization={Institute of Computer Science - Polish Academy of Sciences},
            country={Poland}}

\begin{abstract}
Traceable signatures (Kiayas \textit{et al.}, EUROCRYPT'04) is an anonymous digital signature system that extends the tracing power of opening authority in group signatures. There are many known constructions of traceable signatures but all are based on number-theoretical/pairing assumptions. For such reason, they may not be secure in the presence of quantum computers. This work revisits the notion of traceable signatures and presents a lattice-based construction provably secure in the quantum random oracle model (QROM).
\end{abstract}



\end{frontmatter}


\section{Introduction} \label{section:intro}
In today's world, balancing privacy with accountability has become a fundamental challenge for designing communication systems. Group signatures, introduced by Chaum and van Heyst \cite{CV91}, was an attempt at resolving this problem by allowing users to sign messages anonymously with a group manager who can trace signatures to their original signers. This dual capability makes group signatures particularly valuable in applications requiring privacy protection and accountability mechanisms. Since their introduction, group signatures have evolved drastically with constructions such as~\cite{ACJT00,BBS04,KY05,BW06,Groth07}, offering increasingly practical solutions. However, these schemes rely on number-theoretical assumptions (either RSA-based or Diffie-Hellman in groups equipped with bilinear pairings) and are vulnerable to attacks from quantum computers. This drawback motivated the search for post-quantum candidates, with Gordon \textit{et al.} \cite{GKV10} giving the first lattice-based group signature scheme in 2010. Since then, this has been an active research area that has developed many constructions with security and efficiency gradually approaching that of their classical counterparts.

Lattice-based constructions are the most dominant among proposals of post-quantum group signatures and remain the most advanced regarding security and efficiency. Although the initial proposal from Gordon \textit{et al.}~\cite{GKV10} only achieved weak anonymity with signature size linear order in the number of group members, subsequent works by Laguillaumie \textit{et al.}~\cite{LLLS13} and others~\cite{LLNW16,LNWX18,PLS18} made significant improvement. They simultaneously reduced the signature size to be logarithmic in group size and achieved anonymity in the strongest sense, where signers' secrets are exposed, and the adversary can open signatures of its choice. Current state-of-the-art constructions are from \cite{LNPS21,LNP22}, with a signature size of a few hundred kilobytes for a group size of roughly one million.

Alternatives other than lattices have also been explored, including isogeny-based \cite{BDKLP23}, hash-based \cite{BM18}, and code-based constructions \cite{ELLNW15,NTWZ19,BGM21}. The code-based scheme from \cite{NTWZ19} achieves full-anonymity and has a logarithmic signature size in the group size. 
Recently, \cite{OTX24} proposed a zero-knowledge protocol from the VOLE-in-the-Head framework, which substantially improved upon previous code-based protocols. As an application, they reduced signature size in the scheme from \cite{NTWZ19} to roughly half a megabyte for a group size of around a billion. 

Beyond basic user tracing, incorporating advanced functionalities in group signatures is a related topic that has raised much interest. The need for such features stems from specific privacy and security challenges that the existing means of group signatures may not easily deal with. For example, when the group manager expels some individuals from the group due to misbehavior, it should identify and revoke signatures produced from those. A simple method is to recollect and open every signature, but other users' anonymity would be affected. The notion of verifier local revocable (VLR) group signatures~\cite{BS04} deals with such challenge, by offering a membership-revoking mechanism so that signatures remain valid as long as signers are members of the group. There is also the scenario when the group manager is not fully trusted and needs to be responsible for action. Such is addressed by the notion of accountable tracing signatures~\cite{KohlweissM15}. Another interesting functionality is message-dependent opening~\cite{SEHKO12}, where the group manager can only open signatures on specific messages with the help of another third party (admitter). In the same spirit, many proposals of privacy-preserving signature systems generalize group signatures by enhancing the tracing power of authority while still protecting users' privacy to a certain extent \cite{KGK14,LNPY21,NGSY22}. This research direction also results in a long line of work extending post-quantum group signatures by incorporating sophisticated and expressive tracing features~\cite{LMN16,CNW16,LNRW18,LNWX19,LNPY21,NGSY22}.

This work focuses on traceable signatures, proposed by Kiayias, Tsiounis, and Yung \cite{KTY04}. In essence, traceable signatures are group signatures but offer more tracing power to the group manager and users. In addition to standard tracing capability, group manager also has a \textit{user-specific} tracing power that can be widely distributed among the system. Roughly speaking, the group manager possesses information called \textit{tracing trapdoor} unique to each user. When delegated to sub-openers, the tracing trapdoor helps recognize signatures from the corresponding user while not affecting the privacy of others. In this way, user-specific tracing can be performed concurrently and without recollecting all of the previously issued signatures. In addition, traceable signatures also offer a \textit{self-tracing} mechanism, allowing a user to claim authorship of previously issued signatures without revealing any sensitive information that may compromise its anonymity in the past or the future. 

Traceable signatures may find application in anonymous systems where simple tracing is not enough to hold users accountable. For example, in financial transaction systems, there could be situations where a user conducts illegal trading activities. In such cases, the system administrator and other users should be able to invalidate the user's previous and future transactions. A user-specific tracing method is best for this scenario, which is offered by the means of traceable signatures. Furthermore, a self-tracing capability can allow users to argue that they conducted past transactions faithfully, preventing trading assets from any ownership dispute while maintaining privacy for their other activities.\\

\begin{table}[H]
\hspace*{-1cm}
\centering
\setlength{\tabcolsep}{2pt}
\begin{tabular}{|c|c|c|}
  \hline
  \rule{0pt}{2.8ex}
  \textbf{Functionality} & \textbf{Construction} & 
    \begin{tabular}{c}
    \textbf{Post-quantum Assumption} \\
    \textbf{and Security Model}
    \end{tabular}
 \\
  \hline
  \rule{0pt}{2.8ex}
Verifier local-revocation & \cite{BS04,LNRW18,IEHT18,ZLHZJ19} &  \begin{tabular}{c}
    Lattice-based + ROM \cite{LNRW18,ZLHZJ19}
     \end{tabular}  \\
  \hline
  \rule{0pt}{2.8ex}
Accountable tracing & \cite{KohlweissM15,LNWX19} & Lattice-based + ROM \cite{LNWX19} \\
  \hline
  \rule{0pt}{2.8ex}
Message-dependent opening
   & \cite{SEHKO12,LMN16,CCMC23} & Lattice-based + ROM \cite{LMN16,CCMC23} 
    \\
    \hline
  \rule{0pt}{2.8ex}
\begin{tabular}{c}
    User-specific tracing \\
    and self-tracing
\end{tabular}
   & \cite{KTY04,LY09,BP12} & \textcolor{red}{\ding{55}}
    \\
    \hline
\end{tabular}
\caption{An overview of some advanced tracing functionalities in group signatures with the corresponding proposals. For constructions from post-quantum assumptions, we list the assumption type and security model in the third column. We remark that for group signatures with user-specific tracing and self-tracing (i.e., traceable signatures), all known constructions are from number-theoretic assumptions. To the best of our knowledge, only \cite{PA20} proposed a lattice-based traceable signature scheme. However, we identify a critical flaw which will be explained later.} \label{tab:group-sign-function}
\end{table}

\noindent
\textbf{Formalization of Traceable Signatures}. Kiayias \textit{et al.} \cite{KTY04} defined the syntax and security requirement of traceable signatures, which we refer to as the KTY model. Informally, a traceable signature scheme in a collection of efficient algorithms ($\ms{Setup}$, $\ms{Join}$, $\ms{Sign}$, $\ms{Verify}$, $\ms{Open}$, $\ms{Reveal}$, $\ms{Trace}$, $\ms{Claim}$, $\ms{ClaimVerify}$). While defining a dynamic group signature scheme, these algorithms extend its functionality by formalizing tracing and claiming mechanisms. The group manager executes $\ms{Reveal}$ to obtain a user-specific tracing trapdoor. Algorithm $\ms{Trace}$ allows authorized parties to identify signatures from a specific user using their trapdoor. Algorithm $\ms{Claim}$ enables users to create claims of authorship for their signatures, using their signing keys as secret input. Such claims can be publicly verified by $\ms{ClaimVerify}$.

The KTY security model consists of \textit{traceability}, \textit{non-frameability} and \textit{anonymity}. These are similar to the security notions of group signatures but with subtleties due to the presence of tracing and claiming mechanisms. For example, non-frameability requires that a dishonest group manager or opening authority cannot create a valid signature that is traced or opened to an honest user. Moreover, users cannot illegally claim authorship of signatures they did not create, even when in coalition with the group manager and/or the opening authority. On the other hand, anonymity states that signatures reveal nothing about signers as long as users' secrets and tracing trapdoors are not exposed. This notion is similar to \textit{selfless-anonymity} in dynamic group signatures, put forward by \cite{CG04}.

There exist extensions of the KTY security model, which generalize the above security notions in a way that resembles those of group signatures. Libert and Yung \cite{LY09} considered a notion of anonymity similar to \textit{CCA-anonymity}, when the adversary can observe claims and open signatures of its choices. The work of Blazy and Pointcheval \cite{BP12} generalized the KTY model by separating the roles of opening authority and group manager similar to dynamic group signatures \cite{BSZ05}. In their model (the BP model), the opening authority can recover tracing trapdoors from users' public information, and users can deny authorship of signatures. \\

\noindent
{\textbf{Previous Constructions.}} Existing traceable signature schemes have predominantly relied on classical cryptographic assumptions. Kiayas \textit{et al.} \cite{KTY04} proposes the first traceable signature schemes based on pairings and RSA assumptions. Subsequently, constructions such as \cite{CPY06,DChow09,LY09,BP12} all were based on variants of Diffie-Hellman-like assumptions
and/or required pairings. Surprisingly, to our knowledge, there are no known constructions of traceable signatures from post-quantum assumptions, except for \cite{PA20}. However, there is a flaw that we will discuss shortly after.

The high-level approach to designing tracing/claiming mechanisms is including a \textit{tracing tag} when generating a signature. Typically, this tag is computed by some pseudorandom function (PRF), evaluated on a seed (or PRF's key) known only by the user and group manager. Constructions such as \cite{CPY06,DChow09,LY09,BP12} implicitly employed modified variants of Dodis-Yampolskiy pairing-based verifiable random function~\cite{DY05}, which is a special case of PRF. In these constructions, the seed (or PRF's key) will serve as the user-specific tracing trapdoor. Once released, anyone can ``test" the tag by either re-evaluating the PRF, or verifying specific equalities the tag must satisfy.

To claim ownership, the user generates another \textit{self-tracing} tag from signing secret and includes it in the signature. Signers create a claim by running a non-interactive zero-knowledge (NIZK) proof showing well-formedness of the tag. Such proof can be obtained from Fiat-Shamir heuristics~\cite{FS86C} or by using specific proof systems such as Groth-Sahai proof~\cite{GS08}.

We stress that \cite{PA20} proposed a lattice-based traceable signature scheme and proved security in the random oracle model (ROM). However, in the construction in \cite{PA20}, the tracing tag is computed \textit{deterministically} from the user's secret and remains the same in every user's signature. As such, this does not guarantee anonymity. \\

\noindent
\textbf{Our Contributions.} This paper proposes a traceable signature scheme from lattice assumptions and is provably secure in the quantum random oracle model (QROM). Hence, our scheme is the first traceable signature with complete post-quantum security. 

Our construction implements the generic framework of dynamic group signatures \cite{BSZ05}, building upon lattice-based cryptographic primitives including signature scheme by Jeudy \textit{et al.} \cite{JRS23} and GPV identity-based encryption (IBE) scheme \cite{GPV08}. The non-trivial task is realizing user-specific tracing and self-tracing mechanisms via lattice-based tools. To this end, we adapt the revocation technique of the VLR group signature scheme from \cite{LNRW18}.

In our scheme, we use the ZK knowledge argument by  Yang \textit{et al.}  \cite{YAZ19} to prove the well-formedness of users' signatures. The relations defining signatures are handled with suitable transformations to make them compatible with the ZK framework. To make the argument a non-interactive argument of knowledge (\textsf{NIZKAoK}) secure in QROM, we employ generalized Unruh transformation~\cite{Unruh15,FLW19}. Security of the scheme is proven under the formulation of the BP model \cite{BP12}, which is more expressive and general than that of the KTY model \cite{KTY04}. We stress that the proof is not direct and certain details should be treated carefully using non-blackbox techniques. For example, the anonymity proof uses a technique for simulating GPV-IBE ciphertext in QROM \cite{KYY18}.  \\

\noindent
\textbf{Overview of Our Techniques.} We sketch the high-level idea of tracing and claiming in our construction. For user-specific tracing, we turn to a technique employed by the VLR lattice-based group signature scheme in \cite{LNRW18}. Recall that a VLR group signature scheme comes with a membership-revoking functionality. Roughly speaking, the group manager sets up a secret list of \textit{revocation tokens} corresponding to users. A user is revoked from the group when the group manager publishes its token, and signatures are valid as long as the signers' tokens are not revealed. In a sense, this implies user-specific tracing, with revocation tokens playing the same role as tracing trapdoors. 

In the VLR group signature scheme of \cite{LNRW18}, signer includes in signature a tag $\mb{t}$ in the form of an LWE sample, i.e. $\mb{t} = \mb{M} \cdot \mb{x} + \mb{e} \bmod q$, where $\mb{e} \in \mathbb{Z}^m$ is a randomized error term and $\mb{x} \in \mathbb{Z}_q^n$ is the revocation token. The matrix $\mb{M} \in \mathbb{Z}_q^{m \times n}$ is also included in the signature and freshly chosen by hashing a short randomness. Based on the LWE assumption, the tag $\mb{t}$ hides the token $\mb{x}$, guaranteeing the anonymity of users that are not revoked. 

Looking differently, $\mb{t}$ is a ``tag" computed by an LWE-based randomized PRF with ``key" $\mb{x}$ and ``input" $(\mb{M}, \mb{e})$. From this observation, the revocation technique of \cite{LNRW18} can be transformed into a user-specific mechanism, mimicking the PRF-based design in previous traceable schemes. Particularly, in our scheme $\mb{t}$ acts as the tracing tag, and $\mb{x}$ acts as the user's tracing key. For the self-tracing mechanism, we bind tracing key $\mb{x}$ to the user's secret via a suitable one-way function. Signer can claim ownership of a signature by generating an NIZK proof proving knowledge of tracing key $\mb{x}$ generating $\mb{t}$, and knowledge of the secret that is bound to $\mb{x}$. For this purpose, we instantiate the one-way function with the family based on the SIS problem \cite{Ajtai96}. In this way, $\mb{t}$ also acts as the self-tracing tag.

Therefore, with the above modification to the revocation technique of \cite{LNRW18}, we realize user-specific tracing and self-tracing functionalities using lattice-based cryptographic tools. To construct a full-fledge traceable scheme, we follow the high-level design of dynamic group signatures in \cite{BSZ05}. The building blocks consist of an ordinary digital signature scheme for the group manager to certify members when joining the group, a CCA-secure public key encryption scheme for users to encrypt their identities, and a simulation-sound NIZK proof system for proving relations when generating signatures. For efficiency, we instantiate these components with the lattice-based signature scheme in \cite{JRS23} and the GPV-IBE scheme \cite{GPV08}, converted to a CCA-secure encryption with the help of CHK transformation \cite{CHK04}.

Regarding the NIZK layer, we employ the ZK argument system in \cite{YAZ19} for handling various lattice-based relations. In addition, its inverse polynomial soundness error is an advantage compared to Stern-like ZK framework~\cite{Stern96} employed by previous VLR schemes \cite{LNRW18,ZLHZJ19}. In particular, one needs $\mc{O}(\lambda/\log \ms{poly}(\lambda))$ parallel repetitions in order to boost the soundness error of the employed ZK argument system - improving a factor $\log \ms{poly}(\lambda)$ to that of Stern ZK protocol.  Our design choices give a nearly constant signature size if the group size is a priori bounded. \\

\noindent
\textbf{Other Related Work and Further Discussion}. In \cite{DChow09}, a different model of traceable signatures was proposed. In this model, tracing agents can compute a set of identifying tokens from trapdoors, allowing them to recognize signatures from specific users. Abe \textit{et al.} \cite{ACHO11} provided a generic construction of traceable signatures via a primitive called \textit{double-trapdoor anonymous tag system}. 

Although VLR group signatures and traceable signatures bear certain similarities between membership revocation and user-specific tracing, there are a few differences separating the two. First, VLR group signatures can only support static groups, while traceable signatures support dynamic groups. Regarding security, VLR group signatures can achieve the standard security notion of \textit{full-anonymity}~\cite{IEHT18}. In contrast, the result from \cite{PS19} potentially suggested that selfless-anonymity is the strongest security traceable signatures can obtain.

Additionally, the syntax of VLR group signatures \cite{BS04} does not require an opening algorithm. However, this can be done with the help of revocation tokens as follows: the group manager checks for which revocation token such that the signature is not valid when verified; if such token is found, the signature opens to the respective user. In traceable signatures, this is equivalent to opening signatures by trying all available tracing trapdoors instead of using the master opening key. In a sense, traceable signatures imply a form of VLR group signatures for dynamic groups. 

From a certain point of view, the result of this paper provides a ``conversion'' of the VLR signature scheme of \cite{LNRW18} to a traceable scheme. Nevertheless, our techniques are non-trivial and depend significantly on the mathematical structure of revocation tags and tokens. Generally speaking, it is unclear whether VLR group signatures can be transformed into traceable signatures. Even the generic bluerprint of \cite{IEHT18} lacks a proper mechanism linking signers to its signatures to enable ownership claiming. Thus, this may suggest that such a conversion should be non-blackbox and carefully crafted from the underlying design components. 


\noindent
\textbf{Organization}. In Section~\ref{section:prelim}, we recall some results from lattice-based cryptography and the definitions of traceable signatures. In Section~\ref{section:lattice-ts}, we present our lattice-based traceable signature scheme and then give a detailed analysis of its efficiency and security in Section~\ref{section:analysis}.  \\

\noindent
\textbf{Acknowledgement}. Nam Tran is supported by CSIRO's Data61 PhD Scholarship program. Willy Susilo is supported by the Australian Research Council Laureate Fellowship (FL230100033). 


\section{Preliminaries}\label{section:prelim}
\subsection{Notations}
Let $[n]$ be the set of positive integers up to $n$, i.e. $[n] = \{1, \ldots, n\}$. We work with column vectors and denote them by bold, lower-case letters. Matrices are denoted by bold, upper-case letters. The concatenation of two vectors $\mb{v}_1, \mb{v}_2$ is a column vector and denoted as $(\mb{v}_1, \mb{v}_2)$. 

For a vector $\mb{x} \in \mathbb{R}^n$, the $\ell_2$-norm and $\ell_\infty$-norm is defined as
\begin{align*}
    \left\| \mb{x} \right\|_2 &:= \sqrt{\sum \limits_{i=1}^{n} \vert \mb{x}[i] \vert^2},\\
    \left\| \mb{x} \right\|_\infty &:= \max\{\left| \mb{x}[1] \right|, \ldots, \left| \mb{x}[n] \right| \}.
\end{align*}
For $\mb{B} = (\mb{b}_1 \vert \ldots \vert \mb{b}_m) \in \mathbb{R}^{n \times m}$, we let $\left\|\mb{B}\right\|_2 = \max \limits_{i \in [m]}\left\|\mb{b}_i\right\|_2$ and $\left\|\mb{B}\right\|_\infty = \max \limits_{i \in [m]}\left\|\mb{b}_i\right\|_\infty$.

We write $x \leftarrow D$ when $x$ is sampled from a probability distribution $D$ over some space $S$. In the case $D$ is uniform over some set $S$, we simply write $x \lr S$. For a positive integer $m$, the $m$-th dimensional distribution $D^m$ over the space $S^m$ is defined as the distribution of $\mb{s} = (s_1, \ldots, s_m) \in S^m$, where the components $s_i$'s are independently, identically distributed following $D$.

\subsection{Lattices}
\subsection{Hardness Assumptions}
We recall the SIS and the LWE problems, which serve as the hardness assumptions for proving security.
\begin{defn} \label{def:search-sis} 
Let $\mb{A} \lr \mathbb{Z}_q^{n \times m}$ and $i \in \{2, \infty\}$, parameters $n, m, q > 0$ and norm bound $0 < B < q$. The SIS problem, denoted by $\ms{SIS}_{n,m,q,B}^{(i)}$, asks to find $\mb{z} \in \mathbb{Z}_q^m$ such that $\mb{A} \cdot \mb{z} = \mb{0}$ over $\mathbb{Z}_q$ and $0 < \|\mb{z}\|_i \leq B$. 
\end{defn}

\begin{defn} \label{def:lwe} Given $n, q > 0$ and an error distribution $\chi$ over $\mathbb{Z}$, the LWE distribution with uniform secret $\mb{s} \lr \mathbb{Z}_q^n$ is defined as the distribution $\mc{A}_{\mb{s}, \chi}$ that works as follows:
\begin{itemize}
    \item Sample $\mb{a} \lr \mathbb{Z}_q^{n}$;
    \item Sample $\mb{e} \leftarrow \chi$;
    \item Compute $\mb{b} = \mb{a}^\top \cdot \mb{s} + \mb{e} \bmod q$;
    \item Output $(\mb{a}, \mb{b}) \in \mathbb{Z}_q^{n} \times \mathbb{Z}_q$.
\end{itemize}
The LWE problem with parameters $n, q > 0$ and an error distribution $\chi$ over $\mathbb{Z}$, denoted by $\ms{LWE}_{n,q,\chi}$, asks a computationally-bounded adversary $\mathcal{A}$ to distinguish between $m = \ms{poly}(n)$ samples of $\mc{A}_{\mb{s}, \chi}$ and $m$ samples drawed from uniform distribution over $\mathbb{Z}_q^{n} \times \mathbb{Z}_q$.
\end{defn}
We also work with a variant of LWE \cite{KYY21}, in which the adversary has access to a quantum random oracle. 
\begin{defn}[\cite{KYY21}]
Let $n, q, \chi, A_{\mathbf{s}, \chi}$ be defined as in Definition~\ref{def:lwe} and let $\mc{H}: \{0, 1\}^a \rightarrow \{0, 1\}^b$ be a function chosen uniformly at random. The $\textsf{LWE}_{n, q, \chi, \mc{H}}$ problem asks to distinguish~$m = \ms{poly}(n)$ samples chosen according to $\mathcal{A}_{\mathbf{s},\chi}$ (for $\mathbf{s} \xleftarrow{\$} \mathbb{Z}_q^n$) and $m$ samples chosen according to the uniform distribution over $\mathbb{Z}_q^n \times \mathbb{Z}_q$, given oracle access to $\mc{H}$.
\end{defn}
It is proven in \cite{KYY21} that a quantum adversary against $\textsf{LWE}_{n, q,\chi, \sf{QRO}_{a,b}}$ can be converted to a quantum adversary against $\textsf{LWE}_{n, q,\chi}$ with an exact advantage and a loss of the time for reduction depending on the number of queries to the quantum random oracle.
\subsubsection{Statistical Lemmas}
\begin{lem}[\protect{\cite[Lemma 4]{LLNW14}}] \label{lem:norm-bound}
Let $\beta = \ms{poly}(n)$, $q \geq (4\beta + 1)^2$ and $m \geq 3n$. Then, over the randomness of $\mathbf{B} \stackrel{\$}{\leftarrow} \mathbb{Z}_q^{m \times n}$, we have
\[
\Pr[\hspace{2pt} \exists\mathbf{s} \in \mathbb{Z}_q^n\backslash\{\mb{0}\} : \|\mb{B} \cdot \mb{s} \bmod q \|_{\infty}  \leq 2\beta \hspace{2pt}] \leq q^{-n}.
\]
\end{lem}
\begin{lem}[Leftover Hash Lemma, adapted from \protect{\cite[Lemma~1]{GKPV10}}] \label{lem:lhl}
Let $q \geq 2$ be a prime and $\mc{D}$ be a distribution over $\mathbb{Z}_q^{m}$ with min-entropy $k$. For any $\varepsilon > 0$ and $k \geq n\log q + 2\log(1/\varepsilon) + \mathcal{O}(1)$, the statistical distance between the two distributions
\begin{align*}
    \left\{\left(\mb{A},  \mb{A} \cdot \mb{y} \bmod q\right) : \mathbf{A} \lr \mathbb{Z}_q^{n \times m}; \mb{y} \leftarrow \mathcal{D} \right\}, \\
    \left\{\left(\mb{A}, \mb{u} \right) : \mb{A} \lr \mathbb{Z}_q^{n \times m}; \mb{u} \lr \mathbb{Z}_q^n \right\}    
\end{align*}
is at most $\varepsilon$.
\end{lem}
\subsection{Discrete Gaussian and Trapdoor Sampling}
For a full-rank lattice $\mc{L} \subset \mathbb{Z}^n$, a parameter $\sigma > 0$ and a vector $\mb{c} \in \mathbb{R}^m$, the discrete Gaussian distribution over $\mc{L}$ with width $\sigma$ and center $\mb{c}$, denoted as $\mc{D}_{\mc{L}, \sigma, \mb{c}}$ is defined such that for all $\mb{x} \in \mc{L}$ the value $\mc{D}_{\mc{L}, \sigma, \mb{c}}(\mb{x})$ is proportional to $\rho_{\sigma, \mb{c}}(\mb{x})$ where $\rho_{\sigma, \mb{c}}$ is the Gaussian function:  \[\rho_{\sigma, \mb{c}}(\mb{x}) = \exp\left(-\dfrac{\pi \left\|\mb{x} - \mb{c} \right\|^2}{\sigma^2} \right), \forall x \in \mathbb{R}^n. \]
When the center $\mb{c}$ is zero, the discrete Gaussian is denoted by $\mc{D}_{\mc{L}, \sigma}$. When $\mc{L} = \mathbb{Z}^n$, we write $\mc{D}_{\sigma}^n$ to denote the discrete Gaussian $\mc{D}_{\mathbb{Z}^n, \sigma}$. We stress that $\mc{D}_\sigma^n$ is identical to the distribution $(\mc{D}_{\mathbb{Z}, \sigma})^n$.

\begin{thm}[\protect{\cite{Ban93,Regev05}}] \label{thm:discrete-gaussian}
For a full-rank lattice $\mc{L} \subset \mathbb{Z}^n$ and $\sigma > 0$, the following hold
\begin{enumerate}
    \item $\Pr_{\mathbf{v} \leftarrow \mc{D}_{\mc{L}, \sigma}} [\left\|\mb{v}\right\|_2 > \sigma \cdot \sqrt{n} ] \leq 2^{-2n}$;
    \item A set of $\mc{O}(n^2)$ vectors independently sampled from $\mc{D}_{\mc{L}, \sigma}$ contains $n$ linearly independent vectors except with probability at most $\ms{negl}(n)$.
\end{enumerate}
\end{thm}

We also recall the notion of \textit{gadget matrix} in \cite{MP12}, that is a matrix $\mb{G}_n \in \mathbb{Z}_q^{n \times m}$ of the form  \[\left(\begin{array}{c|c}
    \mb{I}_n \otimes \begin{pmatrix}
        1 & & 2 & & \ldots & & 2^{\lceil \log q \rceil-1}
    \end{pmatrix} & \mb{0}^{m - n \lceil \log q \rceil}
\end{array} \right) \in \mathbb{Z}_q^{n \times m}. \]
The row vector $\mb{g} := \begin{pmatrix}
        1 & & 2 & & \ldots & & 2^{\lceil \log q \rceil-1}
    \end{pmatrix}$ is called \textit{gadget vector}. We remark that for any $\mb{u} \in \mathbb{Z}_q^{n}$, there exists a unique vector $\mb{v} \in \{0, 1\}^m$ such that $\mb{G}_n \cdot \mb{v} = \mb{u} \bmod{q}$. We call $\mb{v}$ the binary decomposition of $\mb{u}$ along $\mb{G}_n$ and denote as $\ms{bin}(\mb{u})$.

For a matrix $\mb{A} \in \mathbb{Z}_q^{n \times m}$, a $\mb{G}$-trapdoor of $\mb{A}$ with an invertible tag matrix $\mb{H} \in \mathbb{Z}_q^n$ is a short matrix $\mb{R} \in \mathbb{Z}^{(m-n\lceil \log q \rceil) \times n\lceil \log q \rceil}$ such that $\mb{A} \cdot \begin{pmatrix}
    \mb{R} \\
    \mb{I}_{n\lfloor \log q \rfloor}
\end{pmatrix} = \mb{H} \cdot \mb{G}$. 

We also make use of the $\mb{G}$-trapdoor sampling algorithm from \cite{MP12}.
\begin{prop}[\protect{\cite{MP12}}] \label{prop:sampled}
There exists an algorithm $\ms{SampleD}$ that takes as input a matrix $\mb{R} \in \mathbb{Z}^{m \times n\lceil\log q\rceil}$, a matrix $\mb{A} \in \mathbb{Z}_q^{n \times m}$, an invertible matrix $\mb{H} \in \mathbb{Z}_q^{n \times n}$, a syndrome $\mb{u} \in \mathbb{Z}_q^n$ and a standard deviation $\sigma \geq \omega(\sqrt{\log n}) \cdot\sqrt{1 + \|\mb{R}\|_2^2}$, outputs $\mb{v} \in \mathbb{Z}_q^{m + n\lceil \log q \rceil}$. The distribution of $\mb{v}$ is at a statistical distance at most $\ms{negl}(n)$  to $\mc{D}_{\sigma}^{m+n\lceil\log q\rceil}$ conditioned on $\left(\mb{A}|\mb{H}\mb{G} - \mb{A}\mb{R}\right)\mb{v} = \mb{u} \bmod q$.
\end{prop}

\subsection{Traceable Signatures} \label{section:traceable-sign}
The syntax of \textit{traceable signatures} (TS) follows the BP model \cite{BP12}. A TS scheme $\mc{TS}$ is a tuple of polynomial-time algorithms $(\ms{Setup}$, $\ms{KeyGen}$, $\ms{Join}$, $\ms{Sign}$, $\ms{Verify}$, $\ms{Open}$, $\ms{Reveal}$, $\ms{Trace}$, $\ms{Claim}$, $\ms{ClaimVerify})$ such that:

\begin{description}
\item [$\ms{Setup}(1^\lambda, N) \rightarrow \ms{pp}$.]
It is a probabilistic algorithm that takes a security parameter $\lambda$ and a group size $N$ as input; and outputs public parameters $\ms{pp}$ which are available to all users. Algorithms defining $\mc{TS}$ may implicitly take $\ms{pp}$ as input.

\item [$\ms{KeyGen}(1^\lambda) \rightarrow (\ms{gpk}, \ms{gsk}, \ms{osk}, \ms{reg})$.]
It is a probabilistic algorithm that takes as input security parameter $\lambda$ and generates a group public key $\ms{gpk}$, a secret key $\ms{gsk}$ of a group manager $\mc{GM}$ and a secret key $\ms{osk}$ of an opening authority $\mc{OA}$. It also initialized a registry table $\ms{reg}$ recording public information of users.

\smallskip
\item  [$\ms{Join}\left\langle \ms{GM}(\ms{gsk}), \mc{U} \right\rangle(\ms{gpk})$.] 
This is an interactive protocol between $\mc{GM}$ (holding $\ms{gsk}$) and a user $\mc{U}$ who wants to join the group. Both parties rely on $\ms{gpk}$ as the common information. At the end of $\ms{Join}$:
\smallskip
\begin{itemize}
\item user $\mc{U}$ obtains $(\ms{id}, \ms{usk}_{\ms{id}}, \ms{cert}_{\ms{id}})$, 
where $\ms{id}$ is a unique, public user identifier, $\ms{usk}_{\ms{id}}$ is secret known only by the user and $\ms{cert}_{\ms{id}}$ is a membership certificate. \smallskip
\item $\mc{GM}$ records the transcript $\ms{transcript}_\ms{id}$ of $\ms{Join}$ in $\ms{reg}$. We remark that $\ms{transcript}_\ms{id}$ is uniquely determined by $\ms{id}$.
\end{itemize}

\smallskip

\item [$\ms{Sign}(\ms{gpk},(\ms{usk}_{\ms{id}}, \ms{cert}_\ms{id}), M)\rightarrow \Sigma$.] 
This is a probabilistic algorithm that takes as input group public key
$\ms{gpk}$, user's secret $\ms{usk}_{\ms{id}}$, user's certificate $\ms{cert}_{\ms{id}}$ and a message $M$. The algorithm returns a signature $\Sigma$.

\smallskip
\item [$\ms{Verify}(\ms{gpk},  M, \Sigma)\rightarrow 1/0$.] 
This deterministic algorithm, on input group public key $\ms{gpk}$, a signature $\Sigma$ and a message $M$,  outputs either 1/0 indicating $\Sigma$ is valid/invalid.

\smallskip

\item [$\ms{Open}(\ms{gpk},\ms{osk}, M, \Sigma)\rightarrow \ms{id}/\bot$.] 
This algorithm takes public parameters $\ms{pp}$, the secret key $\ms{osk}$ of $\mc{OA}$, a \textit{valid} message-signature pair $(M,\Sigma)$, outputs a user identifier $\ms{id}$ or a failure symbol $\bot$.

\smallskip

\item [$\ms{Reveal}(\ms{gpk}, \ms{osk}, \ms{id}) \rightarrow \ms{trace}_\ms{id}$.] This algorithm is invoked by $\mc{OA}$, and returns a tracing trapdoor $\ms{trace}_\ms{id}$ corresponding to user $\ms{id}$.

\smallskip

\item [$\ms{Trace}(\ms{gpk}, \ms{trace}_\ms{id}, \Sigma) \rightarrow 1/0$. ] This deterministic algorithm takes as input $\ms{gpk}$, a trapdoor $\ms{trace}_\ms{id}$, a signature $\Sigma$ and outputs 1/0.

\smallskip

\item [$\ms{Claim}(\ms{gpk}, (\ms{usk}_{\ms{id}}, \ms{cert}_\ms{id}), (M, \Sigma)) \rightarrow \chi$. ] This algorithm takes as input $\ms{gpk}$, a valid message-signature pair $(M, \Sigma)$ generated from $(\ms{usk}_{\ms{id}}, \ms{cert}_\ms{id})$, and outputs a claim $\chi$.

\smallskip

\item [$\ms{ClaimVerify}(\ms{gpk}, (M, \Sigma), \chi) \rightarrow 1/0$. ] Given group public key $\ms{gpk}$, a valid message-signature pair $(M, \Sigma)$ and a claim $\chi$, this deterministic algorithm outputs 1 (accept) or 0 (reject).
\end{description}

\begin{defn}[Correctness]
The scheme $\mathcal{TS}$ is \textbf{correct} if for any message $M$, for any user $\ms{id}$ and any signature $\Sigma \leftarrow \ms{Sign}(\ms{gpk},(\ms{usk}_{\ms{id}}, \ms{cert}_\ms{id}), M)$, the following four conditions are satisfied except with a negligible probability in security parameter $\lambda$:

\begin{enumerate}
    \item \textbf{Sign-correctness:} $\ms{Verify}(\ms{gpk}, M, \Sigma) = 1$;
    
    \item \textbf{Open-correctness:} $\ms{Open}(\ms{gpk},\ms{osk}, M, \Sigma) = \ms{id};$
    
    \item \textbf{Trace-correctness:} $\ms{Trace}(\ms{gpk}, \ms{trace}_\ms{id}, \Sigma) = 1$ where $\ms{trace}_\ms{id} \leftarrow \ms{Reveal}(\ms{gpk}, \ms{osk}, \ms{id})$; and for any $\Sigma^\prime$ not output from $\ms{Sign}$ on input $(\ms{usk}_{\ms{id}}, \ms{cert}_\ms{id})$, we have that $\ms{Trace}(\ms{gpk}, \ms{trace}_\ms{id}, \Sigma^\prime) = 0$;
    
    \item \textbf{Claim-correctness:} let $\chi \leftarrow \ms{Claim}(\ms{gpk}, (\ms{usk}_{\ms{id}}, \ms{cert}_\ms{id}),(M, \Sigma) )$; then $\ms{ClaimVerify}(\ms{gpk}, (M, \Sigma), \chi) = 1$.
\end{enumerate}
\end{defn}
For security, a TS scheme should be \textit{traceable}, \textit{non-frameable} and \textit{anonymous}. \ref{appendix:security-requirement} provides the full description of these security notions.

\section{A Traceable Signature Scheme from Lattices}\label{section:lattice-ts}
\subsection{Technical Overview} \label{section:lattice-ts-technical}
At a high level, our lattice-based TS scheme is similar to the generic construction of dynamic group signatures in \cite{BSZ05}. The building blocks include a digital signature scheme, a CCA-secure public key encryption scheme, and a simulation-sound NIZK proof system. The group manager uses the digital signature scheme to certify membership. The NIZK proof system is invoked by users when they have to prove certain relations in zero knowledge while generating signatures. For these two components, we employ the lattice-based signature scheme by Jeudy \textit{et al.}~\cite{JRS23} and the ZK argument system by  Yang \textit{et al.}~\cite{YAZ19}. For the encryption layer, we apply the CHK transformation \cite{CHK04} to GPV-IBE scheme \cite{GPV08} to obtain a CCA-secure encryption.

As explained in Section~\ref{section:intro}, our construction adapts the revocation technique in \cite{LNRW18} for tracing/claiming mechanisms. Recall that the user's tracing trapdoor is a vector $\mb{x} \in \mathbb{Z}_q^n$, and the tracing tag will be an LWE sample with $\mb{x}$ as secret. In constructions following the KTY model \cite{KTY04}, the tracing trapdoor is sent directly to the group manager. However, in the BP model~\cite{BP12}, an adversary can passively observe the $\ms{Join}$ protocol and thus may use the obtained tracing trapdoor to distinguish between users' signatures. To hide the trapdoors, the idea of~\cite{BP12} is to use public key encryption, and to let the group manager certify the encryption of tracing trapdoors instead. However, rather than public key encryption, we can employ a trapdoor one-way function with sufficient pseudorandomness. An LWE-based function suffices: observe that if $\mb{x} \in \mathbb{Z}_q^n$ is a tracing key, then a user can ``encrypt" $\mb{x}$ as $\mb{B} \cdot \mb{x} + \mb{e} \mod q$ where $\mb{e}$ is an error term. Note that by the hardness of LWE, the adversary observing $\ms{Join}$ learns nothing about $\mb{x}$ from $\mb{B} \cdot \mb{x} + \mb{e} \mod q $. Additionally, $\mb{x}$ can be recovered if we know many short vectors $\mb{s} \in \mathbb{Z}^{m_\mb{B}}$ such that $\mb{B}^\top \cdot \mb{s} = 0 \bmod q$. From this observation, we let $\mb{B}$ serve as the public key to the LWE-based one-way function and also as the master public key in GPV-IBE scheme as well. This design choice reduces the size of the opening authority's secret.

When signing a message $M$, the signer encrypts its unique identifier issued by the group manager and computes the tracing tag as $\mb{t} = \mb{M} \cdot \mb{x} + \mb{e}_\mb{t} \bmod q$. The matrix $\mb{M}$ is freshly sampled per signature by hashing a short randomness $\rho$. The user then generates an \textsf{NIZKAoK} proving that: 1) the tracing trapdoor is bound to the user's secret, 2) an ``encryption" of the tracing trapdoor is certified, 3) the ``encryption" of the tracing trapdoor is correct, 4) the encryption of its identifier is correct and 5) the tag is correct. For condition 1), we let the tracing trapdoor $\mb{x}$  be the output of an SIS-based one-way function evaluated on the user's secret. We remark that proving condition 1 is crucial to our construction. Otherwise, malicious signers may defeat traceability and non-framebility of the system.

For the NIZK layer, the ZK argument by Yang \textit{et al.} can be adapted to handle the signature-defining relations above, using techniques presented in \cite{LNSW13}. To make the argument non-interactive in QROM, we apply Unruh transformation \cite{Unruh15,FLW19}. 

\smallskip
\noindent \textbf{Supporting ZK argument.} The TS scheme uses two prime moduli $q$ and $q^\prime$. The modulus $q$, typically large, is used for the ZK argument of \cite{YAZ19} and the lattice-based signature scheme of \cite{JRS23}. In contrast, the modulus $q^\prime$ is much smaller than $q$ and used for the other lattice-based components. 

In the construction, we apply ZK argument by Yang \textit{et al.} \cite{YAZ19} to prove the knowledge of a tuple ($\ms{id}$, $\mb{z}$, $\mb{x}$, $\mb{e}$, $\mathsf{cert}_{\ms{id}} = (\mb{y}, \mb{v}_1, \mb{v}_2)$, $\mb{r}_\mb{c}$, $\mb{e}_\mb{t}$) such that:
\begin{enumerate}[label=(\roman*)] 
    \item $(\mb{z}, \mb{x})$ is a preimage-image pair of an SIS-based one-way function; \label{item-1}
    
    \item $\ms{id}$ and $(\mb{v}_1, \mb{v}_2)$ form a tag-signature pair on $\ms{bin}(\mb{y})$ under the signature scheme of \cite{JRS23}; here $\ms{bin}(\mb{y})$ is the binary decomposition of $\mb{y}$ along a gadget matrix; \label{item-2}

    \item $\mb{y}$ is an LWE-sample generated from secret $\mb{x}$ with error $\mb{e}$; \label{item-3}
    
    \item A GPV-IBE ciphertext $\mb{c}$ is an encryption of $\ms{bin}(\ms{id})$, with randomnesses $\mb{r}_\mb{c}$; \label{item-4}
    
    \item An LWE sample $\mb{t}$ is generated from the secret $\mb{x}$ and error $\mb{e}_\mb{t}$. \label{item-5}
\end{enumerate}
Proving statement \ref{item-2} is equivalently to proving knowledge of message-signature pair in the scheme from \cite{JRS23} and thus can be handled by the ZK framework from~\cite{YAZ19}. We briefly explain how to prove in ZK the statements \ref{item-1}, \ref{item-3}, \ref{item-4} and \ref{item-5}. In our construction, statements \ref{item-1}, \ref{item-3}, \ref{item-4}, \ref{item-5} consist of equation modulo $q^\prime$, while the ZK system works with a different modulo $q$. Therefore, we need to appropriately transform these statements into some sets of equations modulo $q$. Moreover, we observe that statements \ref{item-1}, \ref{item-3}, \ref{item-4}, \ref{item-5} are special cases of the following relation
\begin{align*}
\mathcal{R}_{m, n, q^\prime, \beta} &= \{((\mb{A}, \mb{y}), \mb{x}) \in \mathbb{Z}_{q^\prime}^{m \times n} \times \mathbb{Z}_{q^\prime}^m \times \mathbb{Z}_{q^\prime}^n : \mb{y} = \mb{A} \cdot \mb{x} \bmod{{q^\prime}}; \left\|\mb{x}\right\|_\infty \leq \beta \},    
\end{align*}  
which, by following the binary decomposition technique of \cite{LNSW13}, can be transformed to the following linear relation with binary constraint over the witness \[ 
\mc{R}_\ms{bin} = \{((\overline{\mb{A}^\prime}, \mb{y}^\prime), \overline{\mb{x}^\prime}) \in \mathbb{Z}_q^{m^\prime \times n^\prime} \times \mathbb{Z}_q^{n^\prime} \times \{0, 1\}^{m^\prime} : \mb{y} = \overline{\mb{A}^\prime} \cdot \overline{\mb{x}^\prime} \bmod{q}\}.
\] 
The above relation can be proved in ZK using the ZK argument of \cite{YAZ19}. We devote \ref{appendix:proof-statements} for the detailed transformations applied for each of the statements \ref{item-1}, \ref{item-3}, \ref{item-4}, \ref{item-5} to a case of $\mc{R}_\ms{bin}$. 

\subsection{Description of the Scheme} \label{section:lattice-ts-description} 
Let $\lambda$ be a security parameter and let $n = \Theta(\lambda)$ be a lattice dimension. Let $N = 2^\ell - 1 \in \ms{poly}(\lambda)$ be the maximum group size, which also defines the identity space $[N]$. The public parameter $\ms{pp}$ output from $\ms{Setup}$ consists of parameters and distributions defined from $\lambda, n$ and $N$. They are set so that all algorithms run in polynomial time and are correct with overwhelming probability. Furthermore, the parameter regimes defining the LWE and the SIS problems, on which security of the scheme is based on, are secure against known attacks. The parameters and distributions are as follows:
\begin{itemize}
    \item Prime moduli $q, q^\prime \in \mathsf{poly}(\lambda)$ such that $q > N$;
    \item Dimension $m_\mb{F} = n \lceil \log q^\prime \rceil + \Theta(\lambda)$;
    \item Dimension $m_\mb{B} = 2n \lceil \log q^\prime \rceil + \Theta(\lambda)$;
    \item Dimensions $m_1, m_2$ such that $m_1 \log 3 = n \log q + \Theta(\lambda)$ and $m_2 = n \lceil \log q \rceil$;
    \item Dimension $m_\mb{M} = 3n$;
    \item Gaussian widths $\sigma_\ms{sign}, 
    \sigma_\mathsf{com} > 0$; these parameters are the preimage sampling widths and commitment randomness sampling widths in the signing algorithm of the signature scheme of \cite{JRS23}. In addition, let $\sigma_\mathsf{verif} = \sqrt{\sigma_\mathsf{com}^2 + \sigma_\mathsf{sign}^2}$;
    \item $\ell_\infty$-norm bounds $\beta_1 = \sigma_\ms{verif} \log m_1 $ and $\beta_2 = \sigma_\ms{sign} \log m_2$;
    \item  Parameters $\alpha_\ms{GPV}$ and $\sigma_\ms{GPV} > \sqrt{n + \log m_\mb{B}}$. The former defines a discrete Gaussian distribution $\mc{D}_{\alpha_\ms{GPV} \cdot q^\prime}$ for sampling GPV-IBE encryption randomness, the latter defines a discrete Gaussian distribution $\mc{D}_{\sigma_\ms{GPV}}$ for sampling decryption key. In our scheme, GPV-IBE encrypts messages in $\{0, 1, \ldots, N\}$, therefore we require that $\alpha_\ms{GPV} \cdot q^\prime \cdot \sqrt{n} \cdot \sigma_\ms{GPV} \cdot \sqrt{m_\mb{B} + 1} < q^\prime/(4(N+1))$ for decryption to be correct. We also let $B_\ms{GPV} = \alpha_\ms{GPV} \cdot q^\prime \cdot \sqrt{m_\mb{B}+1}$;
    \item A distribution ${D}_\ms{LWE}$ over $\mathbb{Z}$ such that elements sampled from ${D}_\ms{LWE}$ has absolute values upper bounded by $B_\ms{LWE}$ with overwhelming probability in $\lambda$. We require $(4B_\ms{LWE} + 1)^2 < q^\prime$ and $\sigma_\ms{GPV} \cdot \sqrt{m_\mb{B}} \cdot B_\ms{LWE} < q/2$;
    \item Parameters defining the public key to the BDLOP commitment scheme \cite{BDLOP18}, which serves as the common reference string $\ms{crs}$ to the ZK argument of \cite{YAZ19};
    \item A repetition parameter $\kappa = \Theta(\lambda/\log \lambda)$, here $p \in \ms{poly}(\lambda)$ is a small integer determining the soundness error $2/(2p+1)$ of the ZK argument of~\cite{YAZ19};
\end{itemize}
The public parameter $\ms{pp}$ is given as \[
\ms{pp} = \left(n, q, q^\prime, m_\mb{F}, m_\mb{B}, m_1, m_2, m_\mb{M}, \sigma_\ms{sign}, \sigma_\ms{com}, \alpha_\ms{GPV}, \sigma_\ms{GPV}, D_\ms{LWE}, \ms{crs}, \kappa \right).
\]
To sign a message, signer needs to generate an \textsf{NIZKAoK} for the following relation:
\begin{defn}\label{def:ts-relation-sign}
Define 
\begin{eqnarray*}
\mc{R}_\ms{Sign} = 
\Big\{\Big(\big(
\mb{A}, \mb{A}^\prime, \mb{D}, \mb{u}, \mb{F}, \mb{B}, \mb{v}, \mb{M}, \mb{c}, \mb{t}
\big),\big(\ms{id}, \mb{z}, \mb{x}, \mb{e}, \mathsf{cert}_{\ms{id}} = (\mb{y}, \mb{v}_1, \mb{v}_2), \mb{r}, \mb{e}_\mb{c}, \mb{e}_\mb{t}\big)\Big)\Big\}
\end{eqnarray*}
as a relation, where
\begin{enumerate}
    \item[(i)] $\mb{A} \in \mathbb{Z}_q^{n \times m_1}, \mb{A}^\prime \in \mathbb{Z}_q^{n \times m_2}, \mb{D} \in \mathbb{Z}_q^{n \times m_\mb{B} \lceil \log q^\prime \rceil}, \mb{B} \in \mathbb{Z}_{q^\prime}^{n \times m_\mb{B}}, \mb{v} \in \mathbb{Z}_{q^\prime}^{n}, \mb{F} \in \mathbb{Z}_{q^\prime}^{n \times m_\mb{F}}$, $\mb{M} \in \mathbb{Z}_{q^\prime}^{m_\mb{M} \times n}, $ $\mb{u} \in  \mathbb{Z}_q^n$, $\mb{c} \in \mathbb{Z}_{q^\prime}^{m_\mb{B} + \ell}$, $\mb{t} \in \mathbb{Z}_{q^\prime}^{m_\mb{M}}$;
    \item[(ii)] $\ms{id} \in \{1, \ldots, N\}$, $\mb{z} \in \{0,1\}^{m_\mb{F}}$, $\mb{x} \in \mathbb{Z}_{q^\prime}^n$, $\mb{e} \in [-B_\ms{LWE}, B_\ms{LWE}]^{m_\mb{B}}$, $\mb{y} \in \mathbb{Z}_{q^\prime}^{m_\mb{B}}$, $\mb{v}_1 \in [-\beta_1, \beta_1]^{m_1}$, $\mb{v}_2 \in [-\beta_2, \beta_2]^{m_2}$, $\mb{r} \in \mathbb{Z}_{q^\prime}^n$, $\mb{e}_\mb{c} \in [-B_\ms{GPV}, B_\ms{GPV}]^{m_\mb{B} + 1}$;
    $\mb{e}_\mb{t} \in [-B_\ms{LWE}, B_\ms{LWE}]^{m_\mb{M}}$;
\item[(iii)] $\mb{x} = \mb{F} \cdot \mb{z} \bmod {q^\prime}$;
\item[(iv)] $\big[\mb{A} \mid \ms{id} \cdot \mb{G}_n + \mb{A}^\prime\big]\cdot (\mb{v}_1, \mb{v}_2)^\top = \mb{u} + \mb{D} \cdot \ms{bin}(\mb{y}) \bmod q$, where $\mb{G}_n \in \mathbb{Z}_q^{n \times m_2}$ is a gadget matrix; 
\item[(v)] $\mb{B}^\top \cdot  \mb{x} + \mb{e} = \mb{y} \bmod q^\prime$;
\item[(vi)] $\begin{pmatrix}
        \mb{B}^{\top} \\
        \mb{v}^{\top}
    \end{pmatrix} \cdot  \mb{r} + \mb{e}_\mb{c} +  \begin{pmatrix}
        \mb{0}^{n} \\
        \lceil q^\prime/2(N+1) \rfloor \cdot \ms{id}
    \end{pmatrix}  = \mb{c} \bmod q^\prime$;
\item[(vii)] $\mb{t} = \mb{M} \cdot \mb{x} + \mb{e}_\mb{t} \bmod q^\prime$.
\end{enumerate}
\end{defn}
By the techniques discussed in~\ref{appendix:proof-statements}, $\mc{R}_{\ms{Sign}}$ can be transformed into a case of the following relation 
\[
\left\{((\mb{A}, \mb{y}, \mathcal{S}), \mb{x}): \mb{A} \cdot \mb{x} = \mb{y} \bmod{q} \land \forall(h, i, j) \in \mathcal{S}: \mb{x}[h] = \mb{x}[i] \cdot \mb{x}[j] \bmod{q} \right\}, 
\]
where $\mb{A} \in \mathbb{Z}_q^{m^\prime + n^\prime}, \mb{y} \in \mathbb{Z}_q^{m^\prime}$ and $\mc{S}$ is a set of 3-tuples of size $s$ each of which consists of 3 integers in $[1, n^\prime]$. By applying the Unruh transformation to the ZK argument~\cite{YAZ19}, we obtain an \textsf{NIZKAoK} for $\mc{R}_\ms{Sign}$.

In addition, as the signer claims a signature by generating an \textsf{NIZKAoK}, the relation signer needs to prove is defined as follows:
\begin{defn}\label{def:ts-relation-claim}
Define 
\begin{eqnarray*}
\mc{R}_\ms{Claim} = 
\Big\{
\Big(\big(\mb{F}, \mb{M}, \mb{t}\big), \big(\mb{z}, \mb{x}, \mb{e}_\mb{t}\big)\Big)
\Big\}
\end{eqnarray*}
as a relation, where
\begin{enumerate}
    \item[(i)] $\mb{F} \in \mathbb{Z}_{q^\prime}^{n \times m_\mb{F}}$, $\mb{M} \in \mathbb{Z}_{q^\prime}^{m_\mb{M} \times n}$, $\mb{t} \in \mathbb{Z}_{q^\prime}^{m_\mb{M}}$;
    \item[(ii)] $\mb{z} \in \{0,1\}^{m_\mb{F}}$, $\mb{x} \in \mathbb{Z}_{q^\prime}^n$, $\mb{e} \in [-B_\ms{LWE}, B_\ms{LWE}]^{m_\mb{M}}$;
\item[(iii)] $\mb{x} = \mb{F} \cdot \mb{z} \bmod {q^\prime}$;
\item[(iv)] $\mb{t} = \mb{M} \cdot \mb{x} + \mb{e}_\mb{t} \bmod q^\prime$.
\end{enumerate}
\end{defn}
Similarly, we obtain an \textsf{NIZKAoK} for $\mc{R}_\ms{Claim}$ by applying Unruh transformation to the ZK argument~\cite{YAZ19}.

Now we describe the remaining algorithms in our lattice-based TS scheme:
\begin{itemize}
\item $\ms{KeyGen}(1^\lambda) \rightarrow (\mathsf{gpk}, \ms{gsk}, \ms{osk}, \ms{reg})$: this algorithm performs the following steps.
\begin{enumerate}
\item Generate verification key $$(\mb{A}, \mb{A}^\prime, \mb{D}, \mb{u}) \in \mathbb{Z}_q^{n \times m_1} \times \mathbb{Z}_q^{n \times m_2} \times \mathbb{Z}_q^{n \times  m_\mb{B} \lceil \log q^\prime \rceil} \times \mathbb{Z}_q^n$$ and signing key $\mb{R_A}$ for the signature scheme of \cite{JRS23} as follows:
\begin{itemize}
    \item Sample $\mb{A} \lr \mathbb{Z}_q^{n \times m_1}$;
    \item Sample $\mb{R_A} \lr \{-1, 0, 1\}^{m_1 \times m_2}$;
    \item Compute $\mb{A}^\prime = -\mb{A} \cdot \mb{R}_\mb{A} \bmod q$;
    \item Sample $\mb{D} \lr \mathbb{Z}_q^{n \times m_\mb{B} \lceil \log q^\prime \rceil}$.
\end{itemize}
\smallskip
\item Sample matrices $\overline{\mb{B}} \lr \mathbb{Z}_{q^\prime}^{n \times (m_\mb{B} - n\lceil \log q^\prime \rceil)}$ and $\mb{S_B} \lr \{0, 1\}^{(m_\mb{B} - n\lceil \log q^\prime \rceil) \times n\lceil \log q^\prime \rceil}$. Restart if $\overline{\mb{B}}$ is not full-rank. Define $\mb{B} = \left(
    \overline{\mb{B}} \vert \mb{G}_n - \overline{\mb{B}} \cdot \mb{S}_\mb{B} 
\right) \in \mathbb{Z}_{q^\prime}^{n \times m_\mb{B}}$;
\smallskip
\item Sample $\mb{F} \lr \mathbb{Z}_{q^\prime}^{n \times m_\mb{F}}$  
that defines a SIS-based one-way function. 
\smallskip
\item Select a strongly secure one-time signature scheme $\mc{OTS} = (\ms{KeyGen}, \ms{Sign}, \ms{Verif})$. Let $l = \ms{poly}(\lambda)$ be the bit length of the verification key output by $\mc{OTS}.\ms{KeyGen}$.
\smallskip
\item Select hash functions $\mc{H}_\ms{GPV}: \{0, 1\}^l \rightarrow \mathbb{Z}_{q^\prime}^n$, $\mc{H}_\ms{LWE}: \{0, 1\}^\lambda \rightarrow \mathbb{Z}_{q^\prime}^{m_\mb{M} \times n}$, $\mc{H}_\ms{Sign}^{(1)}: \{0,1\}^* \rightarrow \{1, 2, 3, 4\}^\kappa$, $\mc{H}_\ms{Sign}^{(2)}: D_\ms{Sign} \rightarrow D_\ms{Sign}$, $\mc{H}_\ms{Claim}^{(1)}: \{0,1\}^* \rightarrow \{1, 2, 3, 4\}^\kappa, \mc{H}_\ms{Claim}^{(2)}: D_\ms{Claim} \rightarrow D_\ms{Claim}$ that serve as our random oracles. Here $D_\ms{Sign}$ and $D_\ms{Claim}$ denotes the set of all prover's responses in the underlying ZK system for $\mc{R}_\ms{Sign}$ and $\mc{R}_\ms{Claim}$ respectively. Looking forward, $\mc{H}_\ms{Sign}^{(1)}, \mc{H}_\ms{Sign}^{(2)}, \mc{H}_\ms{Claim}^{(1)}$ and $\mc{H}_\ms{Claim}^{(2)}$ are used in Unruh transformation to convert the ZK arguments into non-interactive versions.
\end{enumerate}
 Output $\ms{gsk} = \mb{R_A}$, $\ms{osk} = \mb{S}_\mb{B}$, $\ms{reg} = \epsilon$ and 
    \[\hspace{-0.5cm}\ms{gpk} = (\mb{A}, \mb{A}^\prime, \mb{D}, \mb{u}, \mb{B}, \mb{F}, \mc{OTS}, \mc{H}_\ms{GPV}, \mc{H}_\ms{LWE}, \mc{H}_\ms{Sign}^{(1)}, \mc{H}_\ms{Sign}^{(2)}, \mc{H}_\ms{Claim}^{(1)}, \mc{H}_\ms{Claim}^{(2)});\]

\smallskip

\item $\ms{Join}\left\langle \mc{GM}(\ms{gsk}), \mc{U} \right\rangle(\ms{gpk})$: group manager $\mc{GM}$ initializes a counter $\ms{st}$ at 0. A prospective user $\mc{U}$, who wants to join the group, interacts with $\mc{GM}$ following these steps.
\smallskip

\begin{enumerate}
    \item The user selects $\mb{z} \xleftarrow{\$} \{0,1\}^{m_\mb{F}}$, computes $\mb{x} = \mb{F} \cdot \mb{z} \bmod q^\prime \in \mathbb{Z}_{q^\prime}^n$, samples $\mb{e} \leftarrow D_\ms{LWE}^m$ and computes $\mb{y} = \mb{B}^\top \cdot  \mb{x}  +  \mb{e}  \bmod q^\prime \in \mathbb{Z}_{q^\prime}^{m_\mb{B}}$, then signs $\ms{bin}(\mb{y}) \in \{0, 1\}^{m_\mb{B} \lceil \log q^\prime \rceil}$ using an ordinary signature. The message $\ms{bin}(\mb{y})$ and the signature $\sigma$ are sent to $\mc{GM}$;
    
    \smallskip
    \item $\mc{GM}$ checks the registry table whether $\ms{bin}(\mb{y})$ is not previously ceritified registered and $\sigma$ is valid. If both checks are successful, $\mc{GM}$ increases $\ms{st}$ by 1 and assigns the identifer $\ms{id} = \ms{st}$. It then signs $\ms{bin}(\mb{y}) \in \{0, 1\}^{m_\mb{B} \lceil \log q^\prime \rceil}$ under the signature scheme of~\cite{JRS23}, using $\ms{gsk}=\mb{R}_\mb{A}$ as signing key and $\ms{id}$ as tag. The signature has the form $(\ms{id}, \mb{v}_1, \mb{v}_2) \in \mathbb{Z}_q\backslash\{0\} \times [-\beta_1, \beta_1]^{m_1} \times [-\beta_2, \beta_2]^{m_2}$, and satisfies 
    \begin{eqnarray} \label{eq:cert}
    \big[\mb{A} \mid \ms{id} \cdot \mb{G}_n + \mb{A}^\prime\big]\cdot \left(\mb{v}_1, \mb{v}_2\right) = \mb{u} + \mb{D} \cdot \ms{bin}(\mb{y}) \bmod q.
    \end{eqnarray}
    $\mc{GM}$ then sends $(\ms{id}, \mb{v}_1, \mb{v}_2)$ to user $\mc{U}$;    
    \smallskip
    
    \item User $\mc{U}$ verifies $(\ms{id}, \mb{v}_1, \mb{v}_2)$ by checking \eqref{eq:cert}. It sets $\ms{usk}_\ms{id} = \mb{z} \in \{0, 1\}^{m_\mb{F}}$ and \[\ms{cert}_\ms{id} = (\mb{y}, \mb{v}_1, \mb{v}_2) \in \mathbb{Z}_{q^\prime}^{m_\mb{B}} \times [-\beta_1, \beta_1]^{m_1} \times [-\beta_2, \beta_2]^{m_2};\]
    \item $\mc{GM}$ sets $\ms{transcript}_{\ms{id}} = (\ms{id},\sigma, \ms{cert}_{\ms{id}})$ and updates the registry table $\ms{reg}: = \ms{reg} \left\| \ms{transcript}_{\ms{id}} \right.$.
\end{enumerate}

\item $\ms{Sign}(\ms{gpk},\ms{usk}_{\ms{id}},\ms{cert}_\ms{id}, M)$: user $\ms{id}$ executes this algorithm on input $\ms{gpk}$, secret $\ms{usk}_{\ms{id}} = \mb{z} \in \{0, 1\}^{m_\mb{F}}$, certificate $\ms{cert}_\ms{id} = (\mb{y}, \mb{v}_1, \mb{v}_2) \in \mathbb{Z}_{q^\prime}^{m_\mb{B}} \times [-\beta_1, \beta_1]^{m_1} \times [-\beta_2, \beta_2]^{m_2}$ and a message $M \in \{0, 1\}^*$. It performs these steps:
\begin{enumerate}
    \item Compute $\mb{x} = \mb{F} \cdot \mb{z} \bmod q^\prime$;
    \smallskip
    \item Compute $\mb{e} = \mb{y} - \mb{B}^\top \cdot \mb{x} \bmod q^\prime$. Note that $\left\|\mb{e}\right\|_\infty \leq B_\ms{LWE}$;
    \item Generate a key pair $(vk, sk) \leftarrow \mc{OTS}.\ms{KeyGen}(1^\lambda)$. Then encrypt $\ms{id}$ with respect to ``identity" $vk$ by computing $\mb{v} = \mc{H}_\ms{GPV}(vk) \in \mathbb{Z}_q^n$; sampling $\mb{r} \lr \mathbb{Z}_q^{n}$ and $\mb{e}_\mb{c} \leftarrow \mc{D}_{\alpha_\ms{GPV} \cdot q^\prime}^{m_\mb{B} + 1}$. Let the final ciphertext be
    \smallskip
    \[
    \mb{c}  = \begin{pmatrix}
        \mb{B}^{\top} \\
        \mb{v}^{\top}
    \end{pmatrix} \cdot  \mb{r} + \mb{e}_\mb{c} +  \begin{pmatrix}
        \mb{0}^{m_\mb{B}} \\
        \lceil q^\prime/2(N+1) \rfloor \cdot \ms{id}
    \end{pmatrix} \bmod q^\prime;
    \]
    \smallskip
    \item Sample randomness $\rho \lr \{0, 1\}^\lambda$ and compute $\mb{M} = \mc{H}_\ms{LWE}(\rho) \in \mathbb{Z}_{q^\prime}^{m_\mb{M} \times n}$, sample $\mb{e}_\mb{t} \leftarrow D_\ms{LWE}^{m_\mb{M}}$ and set $\mb{t} = \mb{M} \cdot \mb{x} + \mb{e}_\mb{t} \bmod q^\prime$;
    \smallskip
    \item Using the witness $\big(\ms{id}, \mb{z}, \mb{x}, \mb{e}, \mathsf{cert}_{\ms{id}} = (\mb{y}, \mb{v}_1, \mb{v}_2), \mb{r}, \mb{e}_\mb{c}, \mb{e}_\mb{t}\big)$, 
    generate a ``signature of knowledge" (an \textsf{NIZKAoK}) $\pi$ for the relation $\mc{R}_{\ms{Sign}}$ (Definition~\ref{def:ts-relation-sign}), on the message $M$. This is done by applying the Unruh transformation for generalized $\Sigma$-protocol \cite{FLW19} using hash functions $\mc{H}_\ms{Sign}^{(1)}$ and $\mc{H}_\ms{Sign}^{(2)}$. Specifically, when applying the transformation, it computes 
    \[\mc{H}_\ms{Sign}^{(1)}\left(\ms{crs}, M, \rho, x, (\ms{com}_i)_{[\kappa]}, (\ms{ch}_{i, j})_{[\kappa] \times [4]}, (h_{i, j})_{[\kappa] \times [4]}  \right), \]
    where $x$ is the statement of $\mc{R}_\ms{Sign}$, $\ms{com}_i$; and $\ms{ch}_{i, j}, h_{i, j} = \mc{H}_\ms{Sign}^{(2)}(\ms{rsp_{i, j}})$ denote the commitment, challenge, hash of response generated by the prover. Let $\pi$ be the final proof.
    \smallskip
    \item Compute a one-time signature $sig \leftarrow \mc{OTS}.\ms{Sign}(sk, (\rho, \mb{c}, \mb{t}, \pi))$;
\end{enumerate}
The algorithm outputs the signature 
\begin{eqnarray}\label{eq:ts-sig}
\Sigma = (\rho, \mb{c}, \mb{t}, \pi, vk, sig). 
\end{eqnarray}

\item  $\ms{Verify}(\ms{gpk}, {M}, \Sigma)$: this algorithm parses the input signature $\Sigma$ as in \eqref{eq:ts-sig}. The algorithm returns 1 if and only if $\mc{OTS}.\ms{Verif}(vk, (\rho, \mb{c}, \mb{t}, \pi), sig) = 1$ and $\pi$ is a valid  \textsf{NIZKAoK} for $\mc{R}_\ms{Sign}$ as in Definition~\ref{def:ts-relation-sign}.

\item $\ms{Open}(\ms{gpk},\ms{osk} = \mb{S}_\mb{B}, M, \Sigma)$:  return $\bot$ if $\Sigma$ does not verify with $M$. Otherwise, the algorithm proceeds as follows: 
\begin{enumerate}
    \item Compute $\mb{v} = \mc{H}_\ms{GPV}(vk) \in \mathbb{Z}_{q^\prime}^{n}$.
        Next, using $\mb{S}_\mb{B}$, sample $\mb{e}_{vk} \leftarrow \mc{D}_{\sigma_\ms{GPV}}^{m_\mb{B}}$ such that $\mb{B}\cdot \mb{e}_{vk} = \mb{v} \bmod {q^\prime}$. This is done by the algorithm of Proposition~\ref{prop:sampled};
    
    \item Decrypt $\mb{c}$ by computing $\lceil\left(-\mb{e}_{vk}^\top \vert 1\right) \cdot \mb{c}_1 \bmod q^\prime)/ (q^\prime/2(N+1)) \rfloor$. Let $\ms{id} \in \{0, \ldots, N\}$ be the result, output $\ms{id}$. 
\end{enumerate}

\item $\ms{Reveal}(\ms{gpk}, \ms{osk}, \ms{id})$: on input an identifier $\ms{id} \in \{1, \ldots, N\}$ and $\ms{osk} = \mb{S}_\mb{B}$, the algorithm checks the public registry table for an entry starting with $\ms{id}$. If such an entry does not exist, output $\bot$. Otherwise, it retrieves the component $\mb{y} \in \mathbb{Z}_{q^\prime}^{m_\mb{B}}$ in $\ms{cert}_\ms{id}$ and inverts the LWE sample $\mb{y}$ as follows:
\begin{enumerate}
    \item Using the algorithm of Proposition~\ref{prop:sampled}, sample a set of $\mathbb{R}$-linearly independent vectors $\mb{s}_1, \ldots, \mb{s}_{m_\mb{B}}$ from $\mc{D}_{\sigma_\ms{GPV}}^{m_\mb{B}}$ such that $\mb{B} \cdot \mb{s}_i = \mb{0} \bmod q$ for $i = 1, \ldots, m_\mb{B}$. Let $\mb{S} = \left(\mb{s}_1 \vert \ldots \vert \mb{s}_{m_\mb{B}}\right) \in \mathbb{Z}^{m_\mb{B} \times m_\mb{B}}$. Note that by Theorem~\ref{thm:discrete-gaussian}, the algorithm of Proposition~\ref{prop:sampled} should be invoked $\mc{O}(m_\mb{B}^2)$ times so that the $\mb{s}_i$'s can be found;
    \item Compute $\mb{S}^\top \cdot \mb{y} \bmod q^\prime$. Note that if $\mb{y} = \mb{B}^\top \cdot \mb{x} + \mb{e} \bmod q^\prime$, then $\mb{S}^\top \cdot \mb{y} \bmod q^\prime = \mb{S}^\top \cdot \mb{e} \bmod q^\prime = \mb{S}^\top \cdot \mb{e}$. This is because $\left\|\mb{S}^\top \cdot \mb{e}\right\|_2 \leq \left\|\mb{S}\right\|_2 \cdot \left\|\mb{e}\right\|_\infty \leq \sigma_\ms{GPV} \cdot \sqrt{m_\mb{B}} \cdot B_\ms{LWE} < q/2$;
    \item Using linear algebra, solve the linear equation $\mb{S}^\top \cdot \mb{e} = \mb{S}^\top \cdot \mb{y} \bmod q^\prime$ over $\mathbb{Z}$. Output $\bot$ if $\mb{e}$ does not exist or $\left\|\mb{e} \right\|_\infty > B_\ms{LWE}$;
    \item Using linear algebra, solve the linear equation $\mb{B}^\top \cdot \mb{x} = \mb{y} - \mb{e} \bmod q^\prime$ over $\mathbb{Z}_{q^\prime}$. Note that as $\mb{B}^\top$ is full-rank, there exists at most one solution $\mb{x}$;
    \item Output $\ms{trace}_\ms{id} = \mb{x}$. If no such $\mb{x}$ exists then output $\bot$.
\end{enumerate}

\item $\ms{Trace}(\ms{gpk}, \ms{trace}_\ms{id}, \Sigma)$: on input $\ms{gpk}$, trapdoor $\ms{trace}_\ms{id} =\mb{x} \in \mathbb{Z}_{q^\prime}^{n}$ and the signature $\Sigma$, the algorithm parses $\Sigma$ as in \eqref{eq:ts-sig} and computes $\mb{M} = \mc{H}_\ms{LWE}(\rho)$. Returns 1 iff $\left\|(\mb{t} - \mb{M} \cdot \mb{x} \bmod q^\prime)\right\|_\infty \leq B_\ms{LWE}$.

\item $\ms{Claim}(\ms{gpk}, \ms{usk}_\ms{id}, \ms{cert}_\ms{id}, (M, \Sigma))$: The algorithm takes as input public key $\ms{gpk}$, secret $\ms{usk}_\ms{id} = \mb{z}$, certificate $\ms{cert}_\ms{id} = (\mb{y}, \mb{v}_1, \mb{v}_2)$ and a message-signature pair $(M, \Sigma)$. It parses $\Sigma$ as in (\ref{eq:ts-sig}), computes $\mb{M} = \mc{H}_\ms{LWE}(\rho)$, $\mb{x} = \mb{F} \cdot \mb{z} \bmod q^\prime$ and $\mb{e}_\mb{t} = \mb{t} - \mb{M} \cdot \mb{x} \bmod q^\prime$. Returns $\bot$ if $\left\|\mb{e}_\mb{t} \right\|_\infty > B_\ms{LWE}$. Else, it generates an \textsf{NIZKAoK} $\chi$ for the relation $\mc{R}_\ms{Claim}$ defined in Definition~\ref{def:ts-relation-claim}, using witness $(\mb{x}, \mb{z}, \mb{e}_\mb{t})$. Similar to $\ms{Sign}$, the \textsf{NIZKAoK} is obtained via the Unruh transformation with random oracles $\mc{H}_\ms{Claim}^{(1)}$ and $\mc{H}_\ms{Claim}^{(2)}$. When applying the transformation, it computes 
\begin{eqnarray*}
\mc{H}_\ms{Claim}^{(1)}\left(\ms{crs}, M, \rho, x, (\ms{com}_i)_{[\kappa]}, (\ms{ch}_{i, j})_{[\kappa] \times [4]}, (h_{i, j})_{[\kappa] \times [4]}  \right),  
\end{eqnarray*}
where $x$ is the statement of $\mc{R}_\ms{Claim}$; and $\ms{com}_i, \ms{ch}_{i, j}, h_{i, j} = \mc{H}_\ms{Claim}^{(2)}(\ms{rsp}_{i, j})$ are commitment, challenge, hash of response generated by the prover. Let $\chi$ be the final proof. The algorithm outputs $\chi$.

\item $\ms{ClaimVerify}(\ms{gpk}, (M, \Sigma), \chi)$: the algorithm takes as input $\ms{gpk}$, a valid message-signature pair $(M, \Sigma)$ and a claim $\chi$. It parses $\Sigma$ as in \eqref{eq:ts-sig}, then from $\ms{gpk}$ and $\Sigma$ it reconstructs the statement $x \in \mc{R}_\ms{Claim}$. It outputs 1 if the claim $\chi$ is an valid \textsf{NIZKAoK} for $x$, and 0 otherwise.
\end{itemize}
 
\section{Analysis of the Scheme} \label{section:analysis}
\subsection{Correctness and Efficiency} \label{section:correctness-efficiency}
\subsubsection{Correctness} The sign/open/claim-correctness of the TS scheme in Section~\ref{section:lattice-ts-description} follows from the correctness of its building blocks: the lattice-based signature scheme of~\cite{JRS23}, GPV-IBE scheme~\cite{GPV08}, the ZK argument system of~\cite{YAZ19} and the NIZK from generalized Unruh transformation~\cite{FLW19}. Tracing correctness is implied by the following lemma.
\begin{lem}\label{lem:ts-trace-correct}
Let $\ms{trace}_\ms{id} = \mb{x}, \ms{trace}_{\ms{id}^\prime} = \mb{x}^\prime \in \mathbb{Z}_q^n$ be the output of $\ms{Reveal}$ with respect to different users $\ms{id}, \ms{id}^\prime \in \{1, \ldots, N\}$. Assume that $\Sigma$ is a signature created by $\ms{id}$, then $\ms{Trace}(\ms{gpk}, \ms{trace}_{\ms{id}^\prime}, \Sigma) = 1$ with negligible probability.
\end{lem}
\begin{pf}
First note that $\ms{Reveal}$ correctly recovers the tracing trapdoors $\mb{x}$ and $\mb{x}^\prime$ with overwhelming probability. In addition, $\mb{x}$ and $\mb{x}^\prime$ are independent and distributed statistically close to uniform (by Lemma~\ref{lem:lhl}). Therfore we have that $\mb{x} \neq \mb{x}^\prime$ except with probability at most $2^{-\lambda} + {q^\prime}^{-n} = \ms{negl}(\lambda)$. 

Parse the signature $\Sigma$ as in \eqref{eq:ts-sig}. The condition $\ms{Trace}(\ms{gpk}, \ms{trace}_{\ms{id}^\prime}, \Sigma) = 1$ is equivalent to $\left\|\mb{t} - \mb{M} \cdot \mb{x}^\prime \right\| \leq B_\ms{LWE}$. It follows that \[ 
 \left\|\mb{M} \cdot (\mb{x} - \mb{x}^\prime) \right\|_\infty \leq \left\|\mb{t} - \mb{M} \cdot \mb{x} \right\|_\infty + \left\|\mb{t} - \mb{M} \cdot \mb{x}^\prime \right\|_\infty \leq 2B_\ms{LWE}.
 \]  
Since $\mb{M}$ is uniformly distributed, the above only happens with probability at most ${q^\prime}^{-n} = \ms{negl}(\lambda)$, by Lemma~\ref{lem:norm-bound}. \qed
\end{pf}
 
\subsubsection{Efficiency} We analyze the asymptotic efficiency of the proposed scheme regarding key and signature sizes. The size of $\ms{gpk}$ is dominated by the size of verification key of the signature scheme from~\cite{JRS23}. The user secret key $\ms{usk}_{\ms{id}}$ is a binary vector of length $m_\mb{F} = {\mc{O}}(n \log q^\prime) = \mc{O}(\lambda \log \lambda)$. The group manager and opening authority secrets are matrices with bit sizes of order $\mc{O}((n\log q)^2) = \mc{O}(\lambda^2\log^2\lambda)$ and $\mc{O}((n \log q^\prime)^2) = \mc{O}(\lambda^2 \log^2\lambda)$ respectively. 

To estimate the signature size, we observe that in a signature $\Sigma$ of the form in \eqref{eq:ts-sig}, the most dominant parts are the verification key $vk$ and the signature $sig$ in the one-time signature scheme $\mc{OTS}$. In turn, the size of $vk$ and the size of $sig$ depend on the bit-length $l_\mc{OTS}$ of the message that is signed under $\mc{OTS}$. Note that we employ $\mc{OTS}$ to sign the randomness $\rho$, the ciphertext $\mb{c}$ encrypting $\ms{id}$, the tag $\mb{t}$ and the NIZK proof $\pi$ altogether. Among these four components, the bit-size of $\pi$ dominates and is $4 \kappa$ times the size of prover's response in the ZK argument~\cite{YAZ19} for $\mc{R}_\ms{Sign}$. As the size of prover's response is $\mc{O}(\lambda \log^3\lambda)$ (see~\ref{appendix:asymptotic-size}) and $\kappa = \Theta(\lambda/\log\lambda)$, it follows that the size of $\pi$ is $\mc{O}(\lambda^2 \log^2\lambda)$. By instantiating $\mc{OTS}$ with the lattice-based one-time signature in \cite{Mohassel10}, we obtain one-time signature size of order $\mc{O}(\lambda \log \lambda)$ and one-time verification key size of order $\mc{O}(\lambda^2 \log^2 \lambda + \lambda \log \lambda \cdot l_\mc{OTS})$. Therefore, we obtain a traceable signature scheme with signature size of order $\mc{O}(\lambda^3 \log^3 \lambda)$.

For the size of a claim, we note that the size of prover's response in the ZK argument~\cite{YAZ19} for $\mc{R}_\ms{Claim}$ is of order $\mc{O}(\lambda \log^2 \lambda)$ (see~\ref{appendix:asymptotic-size}). Since the size of the claim is $4\kappa$ times the size of prover's response in the ZK argument~\cite{YAZ19} for $\mc{R}_\ms{Claim}$, it follows that the bit-size of a claim in our traceable scheme is of order $\mc{O}(\lambda^2 \log \lambda)$.

\subsection{Security}
The security of our TS scheme is stated in the following theorem
\begin{thm}\label{thm:lattices-TS}
Suppose that $\ms{SIS}$ and $\ms{LWE}$ assumptions hold. Then the lattice-based TS scheme presented in Section~\ref{section:lattice-ts-description} is CCA-anonymous, traceable and non-frameable in QROM.
\end{thm}

The proof of Theorem~\ref{thm:lattices-TS} relies on a result in \cite{Zhandry2015}, which states that polynomial and truly random functions are perfectly indistinguishable against PPT quantum adversaries. 

\begin{prop}[\protect{\cite{Zhandry2015,FLW19}}]\label{prop:qrom-simulation}
A uniformly random polynomial function of the degree at least $2q - 1$ is perfectly indistinguishable from a random function for any PPT quantum algorithm performing at most $q$ queries.
\end{prop}

Proposition~\ref{prop:qrom-simulation} is particularly useful for proofs in QROM. It provides a way to perfectly simulate a quantum random oracle, which is crucial for the zero-knowledge and simulation-sound online-extractable of NIZK from generalized Unruh transformation \cite{FLW19}.

We prove Theorem~\ref{thm:lattices-TS} by proving a sequence of lemmas. The first is as follows.
\begin{lem}[Anonymity]\label{lem:ts-lattice-anon}
Let $N \in \ms{poly}(\lambda)$ be the maximum group size. Let $q_\ms{Sign}, q_\ms{Open}, q_{\mc{H}_\ms{GPV}} \in \ms{poly}(\lambda)$ be the maximum numbers of adversary's queries to $\mc{O}_\ms{Sign}$, $\mc{O}_\ms{Open}$ and $\mc{H}_\ms{GPV}$ respectively. Suppose that 
\begin{itemize}
    \item[(i)] The $\mathsf{LWE}_{n, q^\prime, {D}_\ms{LWE}}$ assumption holds;
    \item[(ii)] The $\mathsf{LWE}_{n, q^\prime, \mc{D}_{\alpha \cdot q^\prime}}$ assumption holds, for $\alpha$ satisfying $\alpha_\ms{GPV}/2\alpha > \sqrt{m_\mb{B} \cdot \sigma_\ms{GPV}^2 + 1}$;
    \item[(iii)] $\mc{OTS}$ is a strongly secure one-time signature scheme;
    \item[(iv)] The employed NIZK argument is zero-knowledge.
\end{itemize} 
Then the TS scheme presented in Section~\ref{section:lattice-ts-description} is CCA-anonymous against any PPT adversary in the QROM.
\end{lem}
\begin{pf}
We consider a sequence of games. The first game is the experiment $\ms{Exp}_{\mc{TS}, \mc{A}}^\ms{anon}$ defining CCA-anonymity (see Figure~\ref{fig:exp-anon}) executed between a PPT adversary $\mc{A}$ and a challenger. The last game is the experiment $\ms{Exp}_{\mc{TS}, \mc{A}}^\ms{anon}$, modified so that the advantage of $\mc{A}$ is negligible. In addition, we let $W_i$ be the probability that $\ms{Exp}_{\mc{TS}, \mc{A}}^\ms{anon}$ returns 1, and the advantage of $\mc{A}$ in \textbf{Game}-$i$ is $\ms{Adv}_i = \left| \Pr[W_i] - 1/2 \right|$. We argue that all the games are indistinguishable.

\smallskip
\noindent
{\bf Game}-0: This is the original $\ms{Exp}_{\mc{TS}, \mc{A}}^\ms{anon}$. In this game, the challenger generates $(\ms{gpk}, \ms{gsk}, \ms{osk}, \ms{reg} = \epsilon)$ by running $\ms{KeyGen}$ and gives $(\ms{gpk},\ms{gsk})$ to the adversary $\mathcal{A}$. With the oracles provided in the experiment, $\mathcal{A}$ outputs a tuple $(M^\star, \ms{id}_0^\star, \ms{id}_1^\star)$. The challenger checks $\ms{id}_0^\star, \ms{id}_1^\star$ as in line 3 in Figure~\ref{fig:exp-anon}, then chooses a uniformly random bit $b$ and outputs the challenge signature $\Sigma^\star$ by running $\ms{Sign}(\ms{gpk}, \ms{usk}_{\ms{id}_b^\star}, \ms{cert}_{\ms{id}_b^\star})$. Note that, $\Sigma^\star$ has the form  \[\Sigma^\star = (\rho^\star, \mb{c}^\star, \mb{t}^\star, \pi^\star, vk^\star, sig^\star)\] as in \eqref{eq:ts-sig}. Adversary $\mc{A}$ is still allowed oracle accesses except for trivial reveal/claim/opening queries. Finally, $\mc{A}$ outputs a bit $b^\prime$ and wins the game if $b^\prime = b$. Obviously, $\ms{Adv}_0= \ms{Adv}_{\mc{TS},\mc{A}}^\ms{anon}$. We note that at the start of \textbf{Game}-0, the random oracles are chosen as truly random functions. 

\smallskip
\noindent
{\bf Game}-1: In this game, the challenger chooses uniformly at random an unordered pair of identifiers $(\ms{id}_0^\prime, \ms{id}_1^\prime)$ at the beginning of the experiment. This is a guess for the targeted identifiers $(\ms{id}_0^\star, \ms{id}_1^\star)$ that will be chosen by $\mc{A}$. The challenger then interacts with $\mc{A}$ faithfully. When $\mc{A}$ submits its challenge identifiers, challenger checks if the guess is correct. If not, the challenger aborts. Else, it continues the experiment.

Obviously, {\bf Game}-1 is identical to {\bf Game}-0 with the addition that the challenger guesses the targeted identifiers. The guess is correct with a probability at least $2/(N(N-1))$, since this is independent with $\mc{A}$'s view. It follows that $\ms{Adv}_1 \geq 2/(N(N-1)) \cdot \ms{Adv}_0$, or $\ms{Adv}_0 \leq N(N-1)/2 \cdot \ms{Adv}_1$.

\smallskip
\noindent
{\bf Game}-2: In this game, we change how the challenge signature is generated. Recall that the challenge signature has the form $\Sigma^\star = (\rho^\star, \mb{c}^\star, \mb{t}^\star, \pi^\star, vk^\star, sig^\star)$. Now in {\bf Game}-2, the one-time signature key pair $(vk^\star, sk^\star)$ is generated in the start of the game. During the game, if $\mathcal{A}$ requests for opening of valid signatures of the form $\Sigma=(\rho, \mb{c}, \mb{t}, \pi, vk, sig)$, where ${vk}= {vk}^\star$ then the challenger outputs a random bit and aborts. 

It follows that {\bf Game}-2 and {\bf Game}-1 are indistinguishable. Note that before the challenge phase, ${vk}^\star$ is independent of $\mc{A}$'s view, and thus, the probability that ${vk}^\star$ shows up in $\mathcal{A}$'s requests is negligible. Furthermore, after seeing $\Sigma^\star$, if $\mc{A}$ comes up with a valid signature $\Sigma=(\rho, \mb{c}, \mb{t}, \pi, vk, sig)$ such that ${vk}= vk^\star$, then this implies a forgery against $\mc{OTS}$. Therefore, the probability that the challenger aborts in this experiment is negligible and in particular  \[
\left|\ms{Adv}_2 - \ms{Adv}_1 \right| \leq \ms{Adv}_\mc{A}^{\mc{OTS}} + \ms{negl}(\lambda),
\]
where $\ms{Adv}_\mc{A}^{\mc{OTS}}$ is the advantage of $\mc{A}$ against strong one-time unforgeability of $\mc{OTS}$. In the subsequent games, we can assume that $\mathcal{A}$ does not request for opening of valid signatures that include ${vk}^\star$.

\smallskip
\noindent
{\bf Game}-3: In this game, we change how $\mc{O}_\ms{pJoin}$ functions when an identifier $\ms{id} \in \{\ms{id}_0^\prime, \ms{id}_1^\prime\}$ is issued. In the private execution of $\ms{Join}$ when $\ms{id}$ is issued, the vector $\mb{y}$ is chosen uniformly random in $\mathbb{Z}_{q^\prime}^{m_\mb{B}}$, instead of being in the form $\mb{B}^\top \cdot \mb{x} + \mb{e} \bmod{q^\prime}$. In addition, challenger chooses tracing trapdoor $\mb{x}$ uniformly random in $\mathbb{Z}_q^n$, instead of faithfully computing $\mb{x} = \mb{F} \cdot \mb{z} \bmod{q^\prime}$ where $\mb{z} = \ms{usk}_\ms{id}$. By Lemma~\ref{lem:lhl}, the change in the distribution of $\mb{x}$ is noticeable with an advantage up to $\mc{O}(2^{-\lambda})$. Additionally, the change in the distribution of $\mb{y}$ is noticeable up to a negligible advantage, assuming $\ms{LWE}_{n, q^\prime, D_\ms{LWE}}$ is hard. This gives 
 \[
\left|\ms{Adv}_3 - \ms{Adv}_2 \right| \leq  \ms{Adv}_\mc{A}^{\ms{LWE}_{n, q^\prime, D_\ms{LWE}}} + \mc{O}(2^{-\lambda}),
\]
where $ \ms{Adv}_\mc{A}^{\ms{LWE}_{n, q^\prime, D_\ms{LWE}}}$ is the advantage of $\mc{A}$ against  $\ms{LWE}_{n, q^\prime, D_\ms{LWE}}$ . 

\smallskip
\noindent
{\bf Game}-4: This game is identical to \textbf{Game}-3, but we change the way the \textsf{NIZKAoK} $\pi$ in signatures of $\ms{id}_0^\prime$ and $\ms{id}_1^\prime$ is created. At the start of the game, the challenger chooses descriptions for $\mc{H}_\ms{Sign}^{(1)}$, $\mc{H}_\ms{Sign}^{(2)}$ as in Proposition~\ref{prop:qrom-simulation}. When $\mc{A}$ requests a signature on behalf of $\ms{id}_1^\prime$ or $\ms{id}_0^\prime$ (including the challenge signature), the \textsf{NIZKAoK} $\pi$ is simulated by the NIZK simulator, using the description of $\mc{H}_\ms{Sign}^{(1)}$ and $\mc{H}_\ms{Sign}^{(2)}$. The view of $\mc{A}$ can only changed by a negligible quantity compared to \textbf{Game}-3, otherwise $\mc{A}$ is a distinguisher against the zero-knowledge property of the NIZK argument. In particular, we have that \[
\left|\ms{Adv}_4 - \ms{Adv}_3 \right| \leq \ms{Adv}_{\mc{A}}^{\ms{zk},\mc{R}_\ms{Sign}},
\]
where $\ms{Adv}_\mc{A}^{\ms{zk}, \mc{R}_\ms{Sign}}$ denotes the advantage of $\mc{A}$ against zero-knowledge property of the NIZK argument system for relation $\mc{R}_\ms{Sign}$.

\smallskip
\noindent
{\bf Game}-5: This game is identical to \textbf{Game}-4, but we change the way $\mc{O}_\ms{Claim}$ responses to queries. At the start of the game, the challenger choosing descriptions for $\mc{H}_\ms{Claim}^{(1)}$, $\mc{H}_\ms{Claim}^{(2)}$ as in Proposition~\ref{prop:qrom-simulation}. Whenever the adversary queries $\mc{O}_\ms{Claim}$ on $(\ms{id}, M, \Sigma)$ such that $\ms{id} \in \{\ms{id}_1^\prime, \ms{id}_0^\prime\}$, the claim $\chi$ on the tag $\mb{t}$ in $\Sigma$ is simulated by the NIZK simulator, using the description of $\mc{H}_\ms{Claim}^{(1)}$ and $\mc{H}_\ms{Claim}^{(2)}$. The view of $\mc{A}$ can only changed up to a negligible quantity compared to \textbf{Game}-4, otherwise $\mc{A}$ is a distinguisher against the zero-knowledge property of the NIZK argument. In particular, we have that \[
\left|\ms{Adv}_5 - \ms{Adv}_4 \right| \leq \ms{Adv}_{\mc{A}}^{\ms{zk},\mc{R}_\ms{Claim}},
\]
where $\ms{Adv}_\mc{A}^{\ms{zk}, \mc{R}_\ms{Claim}}$ denotes the advantage of $\mc{A}$ against zero-knowledge property of the NIZK argument system for relation $\mc{R}_\ms{Claim}$.

\smallskip
\noindent
{\bf Game}-6: In this game, we change the way the tags $\mb{t}$ are computed in the signatures of user $\ms{id}_0^\prime$ and $\ms{id}_1^\prime$. Instead of computing $\mb{t}$ faithfully, when the adversary $\mc{A}$ requests a signature on behalf of $\ms{id}_0^\prime$ or $\ms{id}_1^\prime$ (including the challenge signature), the challenger simply samples $\mb{t} \lr \mathbb{Z}_{q^\prime}^{m_\mb{M}}$ as the tag.

We analyze the change of $\mc{A}$'s view. In \textbf{Game}-5, the tags included in the signatures from either $\ms{id}_0^\prime$ or $\ms{id}_1^\prime$ are in the form $\mb{M} \cdot \mb{x} + \mb{e}$, for $\mb{e} \leftarrow {D}_\ms{LWE}^{m_\mb{M}}$, $\mb{x} \in \mathbb{Z}_q^n$ that is uniformly distributed  and $\mb{M}$ computed by random oracle $\mc{H}_\ms{LWE}$. As $\mc{A}$ can obtain at most $q_\ms{Sign}$ signatures from either $\ms{id}_0^\prime$ or $\ms{id}_1^\prime$, any non-negligible change in $\mc{A}$'s view immediately implies a distinguisher for $\ms{LWE}_{n, q^\prime, D_\ms{LWE}, \mc{H}_\ms{LWE}}$. In particular, \[\left| \ms{Adv}_6 - \ms{Adv}_5 \right| \leq  q_\ms{Sign} \cdot \ms{Adv}_\mc{A}^{\ms{LWE}_{n, q^\prime, D_\ms{LWE}, \mc{H}_\ms{LWE}}} =  q_\ms{Sign} \cdot \ms{Adv}_\mc{A}^{\ms{LWE}_{n, q^\prime, D_\ms{LWE}}}, \]
where $\ms{Adv}_\mc{A}^{\ms{LWE}_{n, q^\prime, D_\ms{LWE}}}$ denotes the advantage of $\mc{A}$ against $\ms{LWE}_{n, q^\prime, D_\ms{LWE}}$.

\smallskip
\noindent
{\bf Game}-7: In this game, we modify the generation of matrix $\mb{B} \in \mathbb{Z}_{q^\prime}^{n \times m_\mb{B}}$ and change the way $\mc{O}_\ms{Reveal}$ responses to queries. Instead of generating $\mb{B}$ along with a $\mb{G}$-trapdoor $\mb{S}_\mb{B}$, we now sample $\mb{B} \lr \mathbb{Z}_{q^\prime}^{n \times m_\mb{B}}$. Note that, by Lemma~\ref{lem:lhl}, the matrix $\mb{B}$ in {\bf Game}-6 has a distribution statistically close to uniform. For answering queries to $\mc{O}_\ms{Reveal}$ on input $\ms{id} \not \in \{\ms{id}^\prime_0, \ms{id}^\prime_1\}$ (without $\mb{S}_\mb{B}$, which is not available in this game), the challenger simply retrieves $\ms{usk}_\ms{id} = \mb{z} \in \{0, 1\}^{m_\mb{B}}$ and returns $\mb{x} = \mb{F} \cdot \mb{z} \bmod q^\prime$. We remark this change to $\mc{O}_\ms{Reveal}$ is syntactical, since $\mc{O}_\ms{Reveal}$ only accepts queries on honest identities and thus challenger knows corresponding secret $\ms{usk}_\ms{id}$. In addition, with overwhelming probability, $\mc{O}_\ms{Reveal}$ returns the same $\mb{x}$ if challenger uses $\mb{S}_\mb{B}$ to invert the LWE sample $\mb{y}$. Therefore, \[
\left|\ms{Adv}_7 - \ms{Adv}_6 \right| \leq \ms{negl}(\lambda).
\]

\smallskip
\noindent
{\bf Game}-8: In this game, we program the random oracle $\mc{H}_\ms{GPV}$ as follows: at the start of the game, the challenger choose a random oracle $\mc{H}_\ms{GPV}^\prime$ that maps input in $\{0, 1\}^l$ to output in the randomness space of distribution $\mc{D}_{\sigma_\ms{GPV}}^{m_\mb{B}}$; the random oracle $\mc{H}_\ms{GPV}$, when queried on input $vk \in \{0, 1\}^l$, returns $\mb{v} = \mb{B} \cdot \mb{e} \bmod q^\prime$, where $\mb{e}$ is sampled from  $\mc{D}_{\sigma_\ms{GPV}}^{m_\mb{B}}$ using randomness $\mc{H}_\ms{GPV}^\prime(vk)$. The challenger also records $(vk, \mb{v}, \mb{e})$ to answer queries to $\mc{O}_\ms{Open}$ that involves $vk$. Following the proof of \cite[Theorem 2]{KYY18}, we have that 
\[
\left|\ms{Adv}_8 - \ms{Adv}_7 \right| \leq Q^2 \cdot 2^{-\Omega(n)},
\]
assuming $\mc{H}_\ms{GPV}$ is accessed at most $Q$ times. Observe that challenger invokes $\mc{H}_\ms{GPV}$ whenever $\mc{A}$ queries $\mc{O}_\ms{Sign}, \mc{O}_\ms{Open}$ and $\mc{H}_\ms{GPV}$ directly. Therefore $Q \leq q_{\ms{Sign}} + q_\ms{Open} + q_{\mc{H}_\ms{GPV}}$ and we have \[
\left|\ms{Adv}_8 - \ms{Adv}_7 \right| \leq (q_{\ms{Sign}} + q_\ms{Open} + q_{\mc{H}_\ms{GPV}})^2 \cdot 2^{-\Omega(m_\mb{B})}.
\]

\smallskip
\noindent
{\bf Game}-9: In this game, we change how the ciphertext $\mb{c}^\star$ in the challenge signature $\Sigma^\star$ is generated. Note that in {\bf Game}-8, we have \[
    \mb{c}^\star  = \begin{pmatrix}
        \mb{B}^{\top} \\
        {\mb{v}^\star}^{\top}
    \end{pmatrix} \cdot  \mb{r}^\star + \mb{e}_\mb{c}^\star +  \begin{pmatrix}
        \mb{0}^{n} \\
        \lceil q^\prime/2(N+1) \rfloor \cdot \ms{id}_b^\star
    \end{pmatrix} \bmod q^\prime = \begin{pmatrix}
        \mb{c}_1^\star \\
        c_1^\star 
    \end{pmatrix} \in \mathbb{Z}_{q^\prime}^{m_\mb{B}} \times \mathbb{Z}_{q^\prime};
    \]
wher $\mb{v}^\star = \mc{H}_\ms{GPV}(vk^\star)$, $\mb{r}^\star \lr \mathbb{Z}_{q^\prime}^n$ and $\mb{e}_\mb{c}^\star \leftarrow \mc{D}_{\alpha_\ms{GPV} \cdot q^\prime}^{m_\mb{B}+1}$. Now the challenger generates $\mb{c}^\star$ as follows:
\begin{enumerate}
    \item Sample $\mb{r}^\prime \lr \mathbb{Z}_{q^\prime}^n$;
    \item Sample $\mb{e}^\prime \leftarrow \mc{D}_{\alpha \cdot q^\prime}^{m_\mb{B}}$;
    \item Apply the algorithm \textsf{ReRand} in \cite[Lemma 7]{KYY18} on input $\mb{V} = \left(\mb{I}_{n} \vert \mb{v}^\star\right)$, $\mb{b} = \mb{B}^\top \cdot \mb{r}^\prime \bmod q^\prime$, $\mb{z} = \mb{e}^\prime$, $r = \alpha \cdot q^\prime$ and $\sigma = \alpha_\ms{GPV}/2\alpha$ to generate $\mb{c}^\prime = (\mb{c}_1^\prime, c_1^\prime) \in \mathbb{Z}_{q^\prime}^{m_\mb{B}} \times \mathbb{Z}_{q^\prime}$. This algorithm ensures that $\mb{c}^\prime$ has the form \[\mb{V}^\top \cdot \mb{b} + \mb{e} = \begin{pmatrix}
        \mb{B}^\top \\
        {\mb{v}^\star}^\top
    \end{pmatrix} \cdot \mb{r}^\prime + \mb{e}, \] where $\mb{e} \in \mathbb{Z}^{m_\mb{B}+1}$ is distributed within a statistical distance at most $2^{-\Omega(n)}$ to $\mc{D}_{\alpha_\ms{GPV} \cdot q^\prime}^{m_\mb{B}}$;
    \item Output $\mb{c}^\star = \mb{c}^\prime + (\mb{0},  \lceil q^\prime/2(N+1) \rfloor \cdot \ms{id}_b^\star) \bmod q^\prime$.
\end{enumerate}
It follows that
 \[
\left|\ms{Adv}_9 - \ms{Adv}_8 \right| \leq 2^{-\Omega(n)}.
\]

\smallskip
\noindent
{\bf Game}-10: In this game, we change how the ciphertext $\mb{c}^\star$ in the challenge signature $\Sigma^\star$ is generated. The process is identical to {\bf Game}-9 except that the vector $\mb{b} = \mb{B}^\top \cdot \mb{r}^\prime \bmod q^\prime \in \mathbb{Z}_{q^\prime}^{m_\mb{B}}$ is now replaced by a uniformly random $\mb{b} \lr \mathbb{Z}_{q^\prime}^{m_\mb{B}}$. It follows that the vector $\mb{c}^\prime = (\mb{c}_1^\prime, c_1^\prime) \in \mathbb{Z}_{q^\prime}^{m_\mb{B}} \times \mathbb{Z}_{q^\prime}$ output by \textsf{ReRand} has the form \[
\mb{V}^\top \cdot \mb{b} + \mb{e} = \begin{pmatrix}
    \mb{I}_n \\
    {\mb{v}^\star}^\top
\end{pmatrix} \cdot \mb{b} + \mb{e} \bmod q^\prime,
\]
where $\mb{e} \in \mathbb{Z}^{m_\mb{B}+1}$ is distributed within a statistical distance at most $2^{-\Omega(n)}$ to $\mc{D}_{\alpha_\ms{GPV} \cdot q^\prime}^{m_\mb{B}}$. By a similar argument in the proof of \cite[Theorem 2]{KYY18}, we have that {\bf Game}-10 and {\bf Game}-9 are indistinguishable assuming $\ms{LWE}_{n, q^\prime, \mc{D}_{\alpha \cdot q^\prime}, \mc{H}_\ms{GPV}^\prime}$ is hard. Thus,
 \[
\left|\ms{Adv}_{10} - \ms{Adv}_9 \right| \leq \ms{Adv}_\mc{A}^{\ms{LWE}_{n, q^\prime, \mc{D}_{\alpha \cdot q^\prime}}},
\]
where $\ms{Adv}_\mc{A}^{\ms{LWE}_{n, q^\prime, \mc{D}_{\alpha \cdot q^\prime}}}$ denotes the advantage of $\mc{A}$ against $\ms{LWE}_{n, q^\prime, \mc{D}_{\alpha \cdot q^\prime}}$.

\smallskip
\noindent
{\bf Game}-11: In this game, we change how the ciphertext $\mb{c}^\star$ in the challenge signature $\Sigma^\star$ is generated as follows: sample $\mb{b} \lr \mathbb{Z}_{q^\prime}^{m_\mb{B}}$, sample $\mb{e} \leftarrow \mc{D}_{\alpha_\ms{GPV} \cdot q^\prime}^{m_\mb{B}}$, compute \[\mb{c}^\prime = \begin{pmatrix}
    \mb{I}_n \\
    {\mb{v}^\star}^\top
\end{pmatrix} \cdot \mb{b} + \mb{e} \bmod q^\prime \in \mathbb{Z}_{q^\prime}^{m_\mb{B}+1},\] and output $\mb{c}^\star = \mb{c}^\prime + (\mb{0},  \lceil q^\prime/2(N+1) \rfloor \cdot \ms{id}_b^\star) \bmod q^\prime$. By a similar argument in the proof of \cite[Theorem 2]{KYY18},  {\bf Game}-11 and {\bf Game}-10 are distinguishable up to an advantage of order $2^{-\Omega(n)}$ and in addition, $\mb{c}^\star$ has a distribution statistically close to uniform over $\mathbb{Z}_{q^\prime}^{m_\mb{B} + 1}$. \\

Finally, $\mc{A}$ can only wins {\bf Game}-11 with probability negligibly close to 1/2. To see this, note that to $\mc{A}$'s view, the certificates of $\ms{id}_0^\star$ and $\ms{id}_1^\star$ reveal nothing about the respective tracing trapdoors. Any claim $\chi$ on behalf of $\ms{id}_0^\star$ or $\ms{id}_1^\star$ is simulated. In the signatures of $\ms{id}_0^\star$ or $\ms{id}_1^\star$, the proof $\pi$ is simulated and the tag $\mb{t}$ is uniformly distributed. As such, oracle outputs involving $\ms{id}_0^\star$ or $\ms{id}_1^\star$ reveals nothing about their tracing keys and their secrets. Furthermore, every component of the challenge signature is generated independently with challenger's random bit $b$ and the ciphertext $\mb{c}^\star$ statistically hides $\ms{id}_b^\star$. It follows that the traceable scheme in Section~\ref{section:lattice-ts-description} is CCA-anonymous. \qed
\end{pf}

\begin{lem}[Traceability]\label{lem:ts-lattice-trace}
The TS scheme in Section~\ref{section:lattice-ts-description} is traceable in QROM, assuming that:
(i) the signature scheme of~\cite{JRS23} is unforgeable 
under chosen-message attacks; 
(ii) the NIZK argument system is simulation-sound online-extractable; 
and (iii) the $\ms{SIS}_{n, m_\mb{F}+1, q, 1}^\infty$ problem is hard.
\end{lem}
\begin{pf}
Consider the experiment $\ms{Exp}_{\mc{TS},\mc{A}}^{\ms{trace}}(\lambda)$ presented in Figure~\ref{fig:exp-trace}. A challenger $\mc{C}$ sets up the public key and provides the adversary $\mc{A}$ with input and oracle access (by interacting with $\mc{A}$, acting as group manager). In addition, it programs the random oracles $\mc{H}_\ms{Sign}^{(1)}$, $\mc{H}_\ms{Sign}^{(2)}$ using Proposition~\ref{prop:qrom-simulation}. Since the simulation can be done perfectly, the view of any PPT quantum adversary $\mc{A}$ remains unchanged.

Whenever $\mc{A}$ outputs a valid signature $(M^\star, \Sigma^\star)$, the challenger $\mc{C}$ parses $\Sigma^\star= (vk^\star, \rho^\star, \mb{c}^\star, \mb{t}^\star, \pi^\star, sig^
\star)$ as in \eqref{eq:ts-sig}. Then it reconstructs the statement of relation $\mc{R}_\ms{Sign}$ in Definition~\ref{def:ts-relation-sign} and invokes the NIZK extractor on the \textsf{NIZKAoK} $\pi^\star$ with the oracle description of $\mc{H}_\ms{Sign}^{(1)}$, $\mc{H}_\ms{Sign}^{(2)}$. Let $\zeta = \big(\ms{id}^\star, \mb{z}^\star, \mb{x}^\star, \mb{e}^\star, \mathsf{cert}_{\ms{id}}^\star = (\mb{y}^\star, \mb{v}_1^\star, \mb{v}_2^\star), \mb{r}^\star, \mb{e}^\star_\mb{c}, \mb{e}^\star_\mb{t}\big)\Big)$ be the output of NIZK extractor. As the NIZK system is online-extractable in QROM, $\zeta$ is a valid witness of $\mc{R}_{\ms{Sign}}$ with overwhelming probability. In particular:
\begin{itemize}
\item $\mb{x}^\star = \mb{F} \cdot \mb{z}^\star \bmod {q^\prime}$;
\item $\big(\mb{A} \mid \ms{id}^\star \cdot \mb{G}_n + \mb{A}^\prime\big)\cdot (\mb{v}_1^\star, \mb{v}_2^\star) = \mb{u} + \mb{D} \cdot \ms{bin}(\mb{y}^\star) \bmod q$; 
\item $\mb{y}^\star = \mb{B}^\top \cdot \mb{x}^\star + \mb{e}^\star \bmod q^\prime$;
\item $\mb{c}^\star = \begin{pmatrix}
        \mb{B}^\top \\
        \mb{v}^\top
    \end{pmatrix} \cdot \mb{r}^\star + \mb{e}_\mb{c}^\star + \begin{pmatrix}
        \mb{0}^{m_\mb{B}} \\
        \lceil q/(2(N+1)) \rfloor \cdot \ms{id}^\star
    \end{pmatrix} \bmod q^\prime$;
\item$\mb{t}^\star = \mb{M}^\star \cdot \mb{x}^\star + \mb{e}_\mb{t}^\star \bmod q^\prime$;
\end{itemize}
where $\mb{M}^\star = \mc{H}_\ms{PRF}(\rho^\star)$.

Note that with overwhelming probability, $(\ms{id}^\star, \mb{v}_1^\star, \mb{v}_2^\star, \mb{y}^\star)$ belongs to some user which joined the group via either $\mc{O}_{\ms{pJoin}}$ or $\mc{O}_\ms{aJoin}$. If this is not the case, $\mc{C}$ immediately obtains a forgery of JRS signature scheme. Furthermore, the equalities \[\mb{y}^\star = \mb{B}^\top \cdot \mb{x}^\star + \mb{e}^\star \bmod q^\prime\]
and \[
\mb{t}^\star = \mb{M}^\star \cdot \mb{x}^\star + \mb{e}_\mb{t}^\star \bmod q^\prime
\]
imply that the output of $\ms{Reveal}$ on $\ms{id}^\star$ is $\mb{x}^\star$, and $\Sigma^\star$ is traced to $\ms{id}^\star$ when running $\ms{Trace}$ on tracing key $\mb{x}^\star$. In addition, the equality \[ \mb{c}^\star = \begin{pmatrix}
        \mb{B}^\top \\
        \mb{v}^\top
    \end{pmatrix} \cdot \mb{r}^\star + \mb{e}_\mb{c}^\star + \begin{pmatrix}
        \mb{0}^{m_\mb{B}} \\
        \lceil q/(2(N+1)) \rfloor \cdot \ms{id}^\star
    \end{pmatrix} \bmod q^\prime \] 
implies that $\Sigma^\star$ is opened to $\ms{id}^\star$. Either way, $\mc{A}$ wins if and only if $\ms{id}^\star \in \ms{Honest}$, i.e. $\ms{id}^\star$ was issued in an execution of $\mc{O}_\ms{pJoin}$ and was not queried to $\mc{O}_\ms{Corrupt}$. We now show that in this case $\mc{C}$ can solve $\ms{SIS}_{n, m+1, q, 1}^\infty$, using $\mc{A}$ as subroutine. The details are as follows. \\

\noindent \textbf{Setup:} $\mc{C}$ receives an SIS instance $\mb{A}_\ms{SIS} \lr \mathbb{Z}_{q^\prime}^{n \times (m_\mb{F}+1)}$. It parses $\mb{A}_\ms{SIS} = \left(\mb{F}^\prime \mid -\mb{x}^\prime \right) \in \mathbb{Z}_{q^\prime}^{n \times m_\mb{F}} \times \mathbb{Z}_{q^\prime}^n$. Next, $\mc{C}$ sets $\mb{F} = \mb{F}^\prime$. For the other components in group public key $\ms{gpk}$, group manager secret key $\ms{gsk}$ and opening authority key $\ms{osk}$, $\mc{C}$ sets them up faithfully following system's description in Section~\ref{section:lattice-ts-description}. Note that, the distribution of $\ms{gpk}$ in this case is identical to that of the real system. In addition, let $q_\ms{pJoin}$ be the maximum number of queries to $\mc{O}_{\ms{pJoin}}$. The challenger $\mc{C}$ guesses uniformly at random an index $j \in \{1, \ldots, q_\ms{pJoin}\}$.  \\

\noindent
\noindent\textbf{Handling Queries:} The queries are handled exactly like in the real experiment $\ms{Exp}_{\mc{TS},\mc{A}}^{\ms{trace}}(\lambda)$ (see Figure~\ref{fig:exp-trace}) except for the $j$-th query of $\mc{A}$ to $\mc{O}_\ms{pJoin}$. Instead of simulating $\mc{O}_\ms{pJoin}$ in private faithfully, in the $j$-th query, $\mc{C}$ retrieves $\mb{x}^\prime$ from \textbf{Setup} phase, samples $\mb{e}^\prime \leftarrow D_\ms{LWE}^{m_\mb{B}}$, computes $\mb{y}^\prime = \mb{B}^\top \cdot \mb{x}^\prime + \mb{e}^\prime \bmod q^\prime$ and certifies $\mb{y}^\prime$ as group manager. After $\ms{Join}$ terminates, $\mc{C}$ stores the relevant information in $\ms{info}$, but set the corresponding secret $\ms{usk}$ of user with issued identifier $\ms{id}^\prime$ to be $\bot$. We remark that, the distribution of $\mb{x}^\prime$ in this case is uniform. In comparison, if $\mc{C}$ simulates $\ms{Join}$ faithfully by sampling $\mb{z} \lr \{0, 1\}^{m_\mb{F}}$ and setting the user's tracing key as $\mb{x} = \mb{F} \cdot \mb{z} \bmod q^\prime$, then the distribution of $\mb{x}$ is statistically close to uniform (by Lemma~\ref{lem:lhl}).

If $\mc{A}$ queries $\ms{id}^\prime$ to $\mc{O}_\ms{Corrupt}$ then $\mc{C}$ aborts. If $\mc{A}$ queries $\ms{id}^\prime$ to $\mc{O}_\ms{Sign}$, it generates the components in the signature $\Sigma$ faithfully except for the NIZK proof $\pi$, which is now simulated by the NIZK simulator. In $\mc{A}$'s view, this change is noticeable up to a negligible quantity, since the NIZK is zero-knowledge. 

It follows that $\mc{C}$ successfully simulates the view of $\mc{A}$ in the real experiment $\ms{Exp}_{\mc{TS},\mc{A}}^{\ms{trace}}(\lambda)$, and $\mc{A}$ can only notice with a negligible advantage.\\

\noindent
\noindent\textbf{Exploiting Forgery: } After $\mc{A}$ submits its forgery $(M^\star, \Sigma^\star)$, the challenger runs the NIZK extractor to obtain a witness $\big(\ms{id}^\star, \mb{z}^\star, \mb{x}^\star, \mb{e}^\star, \mathsf{cert}_{\ms{id}}^\star = (\mb{y}^\star, \mb{v}_1^\star, \mb{v}_2^\star), \mb{r}^\star, \mb{e}_\mb{c}^\star, \mb{e}_\mb{t}^\star\big)\Big)$. As the NIZK system is simulation-sound online-extractable, the extracted witness is valid with overwhelming probability.

Now $\mc{C}$ checks if $(\ms{id}^\prime, \mb{y}^\prime) = (\ms{id}^\star, \mb{y}^\star)$. If this is not the case then $\mc{C}$ aborts. As assumed before, $(\ms{id}^\star, \mb{y}^\star)$ was involved prior in some query of $\mc{O}_\ms{pJoin}$. Since the guess of $\mc{C}$ is independent of $\mc{A}$'s view, $\mc{C}$ succeeds with probability at least $1/q_\ms{pJoin}$. Finally, $\mc{C}$ recovers $\mb{z}^\star$ and outputs $(\mb{z}^\star, 1) \in \{0, 1\}^{m_\mb{F}+1}$ as the SIS solution to $\mb{A}_\ms{SIS}$. It is easy to see that $(\mb{z}^\star, 1)$ a valid solution to the SIS instance $\mb{A}_\ms{SIS}$. \qed
\smallskip
\end{pf}
We note that the unforgeability of the signature scheme by Jeudy \textit{et al.} \cite{JRS23} is based on SIS assumptions.

\begin{lem}[Non-frameability]\label{lem:ts-lattice-frame}
The TS scheme in Section~\ref{section:lattice-ts-description} is non-frameable in QROM, assuming that:
(i) the underlying NIZK argument is simulation-sound online-extractable; 
and (ii) the $\ms{SIS}_{n, m_\mb{F}, q, 1}^\infty$ problem is hard.
\end{lem}
\begin{pf}
Consider the experiment $\ms{Exp}_{\mc{TS},\mc{A}}^{\ms{frame}}(\lambda)$ presented in Figure~\ref{fig:exp-frame}. A challenger $\mc{C}$ sets up the public key and provides the adversary $\mc{A}$ with necessary inputs and oracle access. In addition, it programs the random oracles $\mc{H}_\ms{Sign}^{(1)}$, $\mc{H}_\ms{Sign}^{(2)}$, $\mc{H}_\ms{Claim}^{(1)}$, $\mc{H}_\ms{Claim}^{(2)}$ using Proposition~\ref{prop:qrom-simulation}. As simulation of oracles is perfect, the view of $\mc{A}$ remains unchanged.

Whenever $\mc{A}$ outputs a valid signature $(M^\star, \Sigma^\star)$, the challenger $\mc{C}$ parses $\Sigma^\star= (vk^\star, \rho^\star, \mb{c}^\star, \mb{t}^\star, \pi^\star, sig^
\star)$ as in \eqref{eq:ts-sig}. Then it reconstructs the statement of relation $\mc{R}_\ms{Sign}$ in Definition~\ref{def:ts-relation-sign} and invokes the NIZK extractor on the \textsf{NIZKAoK} $\pi^\star$ with the oracle description of $\mc{H}_\ms{Sign}^{(1)}$, $\mc{H}_\ms{Sign}^{(2)}$. Let $\zeta = \big(\ms{id}^\star, \mb{z}^\star, \mb{x}^\star, \mb{e}^\star, \mathsf{cert}_{\ms{id}}^\star = (\mb{y}^\star, \mb{v}_1^\star, \mb{v}_2^\star), \mb{r}^\star, \mb{e}^\star_\mb{c}, \mb{e}^\star_\mb{t}\big)\Big)$ be the output of NIZK extractor. As the NIZK system is online-extractable in QROM, $\zeta$ is a valid witness of $\mc{R}_{\ms{Sign}}$ with overwhelming probability. In particular:
\begin{itemize}
\item $\mb{x}^\star = \mb{F} \cdot \mb{z}^\star \bmod {q^\prime}$;
\item $\big(\mb{A} \mid \ms{id}^\star \cdot \mb{G}_n + \mb{A}^\prime\big)\cdot (\mb{v}_1^\star, \mb{v}_2^\star) = \mb{u} + \mb{D} \cdot \ms{bin}(\mb{y}^\star) \bmod q$; 
\item $\mb{y}^\star = \mb{B}^\top \cdot \mb{x}^\star + \mb{e}^\star \bmod q^\prime$;
\item $\mb{c}^\star = \begin{pmatrix}
        \mb{B}^\top \\
        \mb{v}^\top
    \end{pmatrix} \cdot \mb{r}^\star + \mb{e}_\mb{c}^\star + \begin{pmatrix}
        \mb{0}^{m_\mb{B}} \\
        \lceil q/(2(N+1)) \rfloor \cdot \ms{id}^\star
    \end{pmatrix} \bmod q^\prime$;
\item$\mb{t}^\star = \mb{M}^\star \cdot \mb{x}^\star + \mb{e}_\mb{t}^\star \bmod q^\prime$;
\end{itemize}
where $\mb{M}^\star = \mc{H}_\ms{PRF}(\rho^\star)$. Note that the above imply that the signature is opened to a user with identifier $\ms{id}^\star$, or can be traced by $\mb{x}^\star$, which is obtained by running $\ms{Reveal}$ on input $\ms{id}^\star$. Therefore, if $\mc{A}$ wins by passing the check in line 5 of Figure~\ref{fig:exp-frame}, it must be the case that $(\ms{id}^\star, \mb{v}_1^\star, \mb{v}_2^\star, \mb{y}^\star)$ was issued in an execution of $\mc{O}_\ms{pJoin}$ and $\ms{id}^\star$ was not queried to $\mc{O}_\ms{Corrupt}$. In such case, similar to the proof of Lemma~\ref{lem:ts-lattice-trace} we can construct an algorithm solving $\ms{SIS}_{n, m_\mb{F} + 1, q^\prime, 1}^\infty$, using $\mc{A}$ as a subroutine.

In the case that $\mc{A}$ forges a claim $\chi^\star$ on a signature $(M^\star, \Sigma^\star)$ created by some honest user with identifier $\ms{id}^\star$, $\mc{C}$ can solve $\ms{SIS}_{n, m_\mb{F} + 1, q^\prime, 1}^\infty$ using $\mc{A}$ as a subroutine. The reduction is nearly identical: $\mc{C}$ guesses the targeted $\ms{id}^\star$ and embeds the SIS instance; when exploiting the forgery, it runs the extractor on the forged claim $\chi^\star$ to obtain an SIS solution. \qed
\end{pf}

Combining the results of Lemma~\ref{lem:ts-lattice-anon}, Lemma~\ref{lem:ts-lattice-trace} and Lemma~\ref{lem:ts-lattice-frame}, Theorem~\ref{thm:lattices-TS} then follows.



\bibliographystyle{elsarticle-num}
\bibliography{main}

\begin{thebibliography}{10}
\expandafter\ifx\csname url\endcsname\relax
  \def\url#1{\texttt{#1}}\fi
\expandafter\ifx\csname urlprefix\endcsname\relax\def\urlprefix{URL }\fi
\expandafter\ifx\csname href\endcsname\relax
  \def\href#1#2{#2} \def\path#1{#1}\fi

\bibitem{CV91}
D.~Chaum, E.~van Heyst, Group signatures, in: {EUROCRYPT} '91, Vol. 547 of
  LNCS, Springer, 1991, pp. 257--265.

\bibitem{ACJT00}
G.~Ateniese, J.~Camenisch, M.~Joye, G.~Tsudik, A practical and provably secure
  coalition-resistant group signature scheme, in: {CRYPTO} 2000, Vol. 1880 of
  LNCS, Springer, 2000, pp. 255--270.

\bibitem{BBS04}
D.~Boneh, X.~Boyen, H.~Shacham, Short group signatures, in: {CRYPTO} 2004, Vol.
  3152 of LNCS, Springer, 2004, pp. 41--55.

\bibitem{KY05}
A.~Kiayias, M.~Yung, Group signatures with efficient concurrent join, in:
  {EUROCRYPT} 2005, Vol. 3494 of LNCS, Springer, 2005, pp. 198--214.

\bibitem{BW06}
X.~Boyen, B.~Waters, Compact group signatures without random oracles, in:
  {EUROCRYPT} 2006, Vol. 4004 of LNCS, Springer, 2006, pp. 427--444.

\bibitem{Groth07}
J.~Groth, Fully anonymous group signatures without random oracles, in:
  {ASIACRYPT} 2007, Vol. 4833 of LNCS, Springer, 2007, pp. 164--180.

\bibitem{GKV10}
S.~D. Gordon, J.~Katz, V.~Vaikuntanathan, A group signature scheme from lattice
  assumptions, in: {ASIACRYPT} 2010, Vol. 6477 of LNCS, Springer, 2010, pp.
  395--412.

\bibitem{LLLS13}
F.~Laguillaumie, A.~Langlois, B.~Libert, D.~Stehl{\'{e}}, Lattice-based group
  signatures with logarithmic signature size, in: {ASIACRYPT} 2013, Vol. 8270
  of LNCS, Springer, 2013, pp. 41--61.

\bibitem{LLNW16}
B.~Libert, S.~Ling, K.~Nguyen, H.~Wang, Zero-knowledge arguments for
  lattice-based accumulators: Logarithmic-size ring signatures and group
  signatures without trapdoors, in: {EUROCRYPT} 2016, Vol. 9666 of LNCS,
  Springer, 2016, pp. 1--31.

\bibitem{LNWX18}
S.~Ling, K.~Nguyen, H.~Wang, Y.~Xu, Constant-size group signatures from
  lattices, in: {PKC} 2018, Vol. 10770 of LNCS, Springer, 2018, pp. 58--88.

\bibitem{PLS18}
R.~del Pino, V.~Lyubashevsky, G.~Seiler, Lattice-based group signatures and
  zero-knowledge proofs of automorphism stability, in: {CCS} 2018, {ACM}, 2018,
  pp. 574--591.

\bibitem{LNPS21}
V.~Lyubashevsky, N.~K. Nguyen, M.~Plan{\c{c}}on, G.~Seiler, Shorter
  lattice-based group signatures via "almost free" encryption and other
  optimizations, in: {ASIACRYPT} 2021, Vol. 13093 of LNCS, Springer, 2021, pp.
  218--248.

\bibitem{LNP22}
V.~Lyubashevsky, N.~K. Nguyen, M.~Plan{\c{c}}on, Lattice-based zero-knowledge
  proofs and applications: Shorter, simpler, and more general, in: {CRYPTO}
  2022, Vol. 13508 of LNCS, Springer, 2022, pp. 71--101.

\bibitem{BDKLP23}
W.~Beullens, S.~Dobson, S.~Katsumata, Y.~Lai, F.~Pintore, Group signatures and
  more from isogenies and lattices: generic, simple, and efficient, Des. Codes
  Cryptogr. 91~(6) (2023) 2141--2200.

\bibitem{BM18}
R.~E. Bansarkhani, R.~Misoczki, G-merkle: {A} hash-based group signature scheme
  from standard assumptions, in: PQCrypto 2018, Vol. 10786 of LNCS, Springer,
  2018, pp. 441--463.

\bibitem{ELLNW15}
M.~F. Ezerman, H.~T. Lee, S.~Ling, K.~Nguyen, H.~Wang, A provably secure group
  signature scheme from code-based assumptions, in: {ASIACRYPT} 2015, Vol. 9452
  of LNCS, Springer, 2015, pp. 260--285.

\bibitem{NTWZ19}
K.~Nguyen, H.~Tang, H.~Wang, N.~Zeng, New code-based privacy-preserving
  cryptographic constructions, in: {ASIACRYPT} 2019, Vol. 11922 of LNCS,
  Springer, 2019, pp. 25--55.

\bibitem{BGM21}
O.~Blazy, P.~Gaborit, D.~T. Mac, A rank metric code-based group signature
  scheme, in: CBCrypto 2021, Vol. 13150 of LNCS, Springer, 2021, pp. 1--21.

\bibitem{OTX24}
Y.~Ouyang, D.~Tang, Y.~Xu, Code-based zero-knowledge from vole-in-the-head and
  their applications: Simpler, faster, and smaller, in: K.~Chung, Y.~Sasaki
  (Eds.), {ASIACRYPT} 2024, Vol. 15488 of LNCS, Springer, 2024, pp. 436--470.

\bibitem{BS04}
D.~Boneh, H.~Shacham, Group signatures with verifier-local revocation, in:
  {CCS} 2004, {ACM}, 2004, pp. 168--177.

\bibitem{KohlweissM15}
M.~Kohlweiss, I.~Miers, Accountable metadata-hiding escrow: {A} group signature
  case study, Proc. Priv. Enhancing Technol. 2015~(2) (2015) 206--221.

\bibitem{SEHKO12}
Y.~Sakai, K.~Emura, G.~Hanaoka, Y.~Kawai, T.~Matsuda, K.~Omote, Group
  signatures with message-dependent opening, in: Pairing-Based Cryptography
  2012, Vol. 7708 of LNCS, Springer, 2012, pp. 270--294.

\bibitem{KGK14}
A.~{El Kaafarani}, E.~Ghadafi, D.~Khader, Decentralized traceable
  attribute-based signatures, in: {CT-RSA} 2014, Vol. 8366 of LNCS, Springer,
  2014, pp. 327--348.

\bibitem{LNPY21}
B.~Libert, K.~Nguyen, T.~Peters, M.~Yung, Bifurcated signatures: Folding the
  accountability vs. anonymity dilemma into a single private signing scheme,
  in: {EUROCRYPT} 2021, Vol. 12698 of LNCS, Springer, 2021, pp. 521--552.

\bibitem{NGSY22}
K.~Nguyen, F.~Guo, W.~Susilo, G.~Yang, Multimodal private signatures, in:
  {CRYPTO} 2022, Vol. 13508 of LNCS, Springer, 2022, pp. 792--822.

\bibitem{LMN16}
B.~Libert, F.~Mouhartem, K.~Nguyen, A lattice-based group signature scheme with
  message-dependent opening, in: {ACNS} 2016, Vol. 9696 of LNCS, Springer,
  2016, pp. 137--155.

\bibitem{CNW16}
S.~Cheng, K.~Nguyen, H.~Wang, Policy-based signature scheme from lattices, Des.
  Codes Cryptogr. 81~(1) (2016) 43--74.

\bibitem{LNRW18}
S.~Ling, K.~Nguyen, A.~Roux{-}Langlois, H.~Wang, A lattice-based group
  signature scheme with verifier-local revocation, Theor. Comput. Sci. 730
  (2018) 1--20.

\bibitem{LNWX19}
S.~Ling, K.~Nguyen, H.~Wang, Y.~Xu, Accountable tracing signatures from
  lattices, in: {CT-RSA} 2019, Vol. 11405 of LNCS, Springer, 2019, pp.
  556--576.

\bibitem{KTY04}
A.~Kiayias, Y.~Tsiounis, M.~Yung, Traceable signatures, in: {EUROCRYPT} 2004,
  2004, pp. 571--589.

\bibitem{IEHT18}
A.~Ishida, Y.~Sakai, K.~Emura, G.~Hanaoka, K.~Tanaka, Fully anonymous group
  signature with verifier-local revocation, in: {SCN} 2018, Vol. 11035 of LNCS,
  Springer, 2018, pp. 23--42.

\bibitem{ZLHZJ19}
Y.~Zhang, X.~Liu, Y.~Hu, Q.~Zhang, H.~Jia, Lattice-based group signatures with
  verifier-local revocation: Achieving shorter key-sizes and explicit
  traceability with ease, in: {CANS} 2019, Vol. 11829 of LNCS, Springer, 2019,
  pp. 120--140.

\bibitem{CCMC23}
S.~Chen, J.~Chen, A.~Miyaji, K.~Chen, Constant-size group signatures with
  message-dependent opening from lattices, in: {P}rov{S}ec 2023, Vol. 14217 of
  LNCS, Springer, 2023, pp. 166--185.

\bibitem{LY09}
B.~Libert, M.~Yung, Efficient traceable signatures in the standard model, in:
  Pairing-Based Cryptography, 2009, pp. 187--205.

\bibitem{BP12}
O.~Blazy, D.~Pointcheval, Traceable signature with stepping capabilities, in:
  Cryptography and Security: From Theory to Applications, Vol. 6805 of LNCS,
  Springer, 2012, pp. 108--131.

\bibitem{PA20}
T.~Preethi, B.~Amberker, Traceable signatures using lattices, Int. Arab J. Inf.
  Technol. 17~(6) (2020) 965--975.

\bibitem{CG04}
J.~Camenisch, J.~Groth, Group signatures: Better efficiency and new theoretical
  aspects, in: {SCN} 2004, Vol. 3352 of LNCS, Springer, 2004, pp. 120--133.

\bibitem{BSZ05}
M.~Bellare, H.~Shi, C.~Zhang, Foundations of group signatures: The case of
  dynamic groups, in: {CT-RSA} 2005, Vol. 3376 of LNCS, Springer, 2005, pp.
  136--153.

\bibitem{CPY06}
S.~G. Choi, K.~Park, M.~Yung, Short traceable signatures based on bilinear
  pairings, in: {IWSEC} 2006, 2006, pp. 88--103.

\bibitem{DChow09}
S.~S.~M. Chow, Real traceable signatures, in: {SAC} 2009, 2009, pp. 92--107.

\bibitem{DY05}
Y.~Dodis, A.~Yampolskiy, A verifiable random function with short proofs and
  keys, in: S.~Vaudenay (Ed.), {PKC} 2005,, Vol. 3386 of LNCS, Springer, 2005,
  pp. 416--431.

\bibitem{FS86C}
A.~Fiat, A.~Shamir, How to prove yourself: Practical solutions to
  identification and signature problems, in: {CRYPTO} 1986, Vol. 263 of LNCS,
  Springer, 1986, pp. 186--194.

\bibitem{GS08}
J.~Groth, A.~Sahai, Efficient non-interactive proof systems for bilinear
  groups, in: {EUROCRYPT} 2008, Vol. 4965 of LNCS, Springer, 2008, pp.
  415--432.

\bibitem{JRS23}
C.~Jeudy, A.~Roux{-}Langlois, O.~Sanders, Lattice signature with efficient
  protocols, application to anonymous credentials, in: {CRYPTO} 2023, Vol.
  14082 of LNCS, Springer, 2023, pp. 351--383.

\bibitem{GPV08}
C.~Gentry, C.~Peikert, V.~Vaikuntanathan, Trapdoors for hard lattices and new
  cryptographic constructions, in: STOC 2008, {ACM}, 2008, pp. 197--206.

\bibitem{YAZ19}
R.~Yang, M.~H. Au, Z.~Zhang, Q.~Xu, Z.~Yu, W.~Whyte, Efficient lattice-based
  zero-knowledge arguments with standard soundness: Construction and
  applications, in: {CRYPTO} 2019, Vol. 11692 of LNCS, Springer, 2019, pp.
  147--175.

\bibitem{Unruh15}
D.~Unruh, Non-interactive zero-knowledge proofs in the quantum random oracle
  model, in: {EUROCRYPT} 2015, Vol. 9057 of LNCS, Springer, 2015, pp. 755--784.

\bibitem{FLW19}
H.~Feng, J.~Liu, Q.~Wu, Secure stern signatures in quantum random oracle model,
  in: {ISC} 2019, Vol. 11723 of LNCS, Springer, 2019, pp. 425--444.

\bibitem{KYY18}
S.~Katsumata, S.~Yamada, T.~Yamakawa, Tighter security proofs for {GPV-IBE} in
  the quantum random oracle model, in: {ASIACRYPT} 2018, Vol. 11273 of LNCS,
  Springer, 2018, pp. 253--282.

\bibitem{Ajtai96}
M.~Ajtai, Generating hard instances of lattice problems (extended abstract),
  in: STOC 1996, {ACM}, 1996, pp. 99--108.

\bibitem{CHK04}
R.~Canetti, S.~Halevi, J.~Katz, Chosen-ciphertext security from identity-based
  encryption, in: {EUROCRYPT} 2004, Vol. 3027 of LNCS, Springer, 2004, pp.
  207--222.

\bibitem{Stern96}
J.~Stern, A new paradigm for public key identification, {IEEE} Trans. Inf.
  Theory 42~(6) (1996) 1757--1768.

\bibitem{ACHO11}
M.~Abe, S.~S.~M. Chow, K.~Haralambiev, M.~Ohkubo, Double-trapdoor anonymous
  tags for traceable signatures, in: {ACNS} 2011, Vol. 6715 of LNCS, 2011, pp.
  183--200.

\bibitem{PS19}
S.~Park, A.~Sealfon, It wasn't me! repudiability and claimability of ring
  signatures, in: {CRYPTO} 2019, Vol. 110694 of LNCS, Springer, 2019, pp.
  159--190.

\bibitem{KYY21}
S.~Katsumata, S.~Yamada, T.~Yamakawa, Tighter security proofs for {GPV-IBE} in
  the quantum random oracle model, Journal of Cryptology 34~(1) (2021) 5.

\bibitem{LLNW14}
A.~Langlois, S.~Ling, K.~Nguyen, H.~Wang, Lattice-based group signature scheme
  with verifier-local revocation, in: {PKC} 2014, Vol. 8383 of LNCS, Springer,
  2014, pp. 345--361.

\bibitem{GKPV10}
S.~Goldwasser, Y.~T. Kalai, C.~Peikert, V.~Vaikuntanathan, Robustness of the
  learning with errors assumption, in: {ICS} 2010, 2010, pp. 230--240.

\bibitem{Ban93}
W.~Banaszczyk, New bounds in some transference theorems in the geometry of
  numbers, Mathematische Annalen 296 (1993) 625--635.

\bibitem{Regev05}
O.~Regev, On lattices, learning with errors, random linear codes, and
  cryptography, in: {STOC} 2005, {ACM}, 2005, pp. 84--93.

\bibitem{MP12}
D.~Micciancio, C.~Peikert, Trapdoors for lattices: Simpler, tighter, faster,
  smaller, in: {EUROCRYPT} 2012, Vol. 7237 of LNCS, Springer, 2012, pp.
  700--718.

\bibitem{LNSW13}
S.~Ling, K.~Nguyen, D.~Stehl{\'{e}}, H.~Wang, Improved zero-knowledge proofs of
  knowledge for the {ISIS} problem, and applications, in: {PKC} 2013, Vol. 7778
  of LNCS, Springer, 2013, pp. 107--124.

\bibitem{BDLOP18}
C.~Baum, I.~Damg{\aa}rd, V.~Lyubashevsky, S.~Oechsner, C.~Peikert, More
  efficient commitments from structured lattice assumptions, in: SCN 2018, Vol.
  11035 of LNCS, Springer, 2018, pp. 368--385.

\bibitem{Mohassel10}
P.~Mohassel, One-time signatures and chameleon hash functions, in: {SAC} 2010,
  Vol. 6544 of LNCS, Springer, 2010, pp. 302--319.

\bibitem{Zhandry2015}
M.~Zhandry, Secure identity-based encryption in the quantum random oracle
  model, in: {CRYPTO} 2012, Vol. 7417 of LNCS, Springer, 2012, pp. 758--775.

\bibitem{KTX08}
A.~Kawachi, K.~Tanaka, K.~Xagawa, Concurrently secure identification schemes
  based on the worst-case hardness of lattice problems, in: {ASIACRYPT} 2008,
  Vol. 5350 of LNCS, Springer, 2008, pp. 372--389.

\bibitem{TNLPS24}
N.~Tran, K.~Nguyen, D.~Liu, J.~Pieprzyk, W.~Susilo, Improved multimodal private
  signatures from lattices, in: {ACISP} 2024, 2024, pp. 3--23.

\end{thebibliography}

\appendix
\section{Security Requirement for Traceable Signatures} \label{appendix:security-requirement}
We follow the BP model \cite{BP12}. The model allows an adversary access to numerous oracles. Whenever some of the oracles are invoked, they make changes to databases that are empty from the start. They are defined as below:
\begin{itemize}
\item $\ms{info}$: contains information obtainable from the system point of view. 

\item $\ms{Control}$: contains identities of users controlled by adversary.

\item $\ms{Honest}$: contains identities of users that adversary does not know the secrets.

\item $\ms{SIGN}$: contains signatures created by honest users. 

\item $\ms{REV}$: contains revealed tracing trapdoors.

\item $\ms{CLAIM}$: contains claims created by honest users.
\end{itemize}

The oracles are nearly identical to those described in \cite{LY09,BP12}. First, the adversary can introduce users to the group passively (only observing the communication) or actively (playing the role of a user). These are allowed via access to the following oracles:
\begin{itemize}    
    \item $\mc{O}_\ms{pJoin}$: adds new honest users to the group. It simulates $\ms{Join}$ to obtain $(\ms{id}, \ms{usk}_\ms{id}, \ms{cert}_\ms{id})$ and $\ms{transcript}_\ms{id}$, then adds $\ms{id}$ to $\ms{Honest}$ and updates $\ms{info}$ by adding an entry of the form $(\ms{id}, \ms{usk}_\ms{id}, \ms{cert}_\ms{id}, \ms{transcript}_\ms{id})$.
    
    \item $\mc{O}_{\ms{aJoin}}$: adds controlled users to the group. The oracle, acting as $\mc{GM}$, interacts with the adversary (the controlled user) via $\ms{Join}$. If $\ms{Join}$ terminates, $\mathcal{O}_{\ms{aJoin}}$ adds $\ms{id}$ to $\ms{Control}$ and updates $\ms{info}$ by adding an entry of the form $(\ms{id}, \bot, \ms{cert}_\ms{id}, \ms{transcript}_\ms{id})$.
\end{itemize}
In framing attacks, adversary can corrupt group manager $\mc{GM}$ (obtaining $\ms{gsk}$). In such case, it can run $\mc{O}_\ms{aJoin}$ by itself to introduce new controlled users to the group.

After users are added, the adversary can access the following oracles:
\begin{itemize}
    \item $\mc{O}_{\ms{Corrupt}}$: takes as input an identifier $\ms{id}$. The oracle returns $\bot$ if $\ms{id} \not \in \ms{Honest}$. Else, it returns $\ms{usk}_\ms{id}$ in the entry of $\ms{info}$ containing $\ms{id}$. Finally, it removes this entry from $\ms{Honest}$ and adds it to $\ms{Control}$.

    \item $\mc{O}_{\ms{Sign}}$: receives query on an identifier $\ms{id}$ and message $M$, $\mathcal{O}_{\ms{Sign}}$ if $\ms{id} \in \ms{Honest}$. It returns $\bot$ if such $\ms{id}$ does not exist. Else, it retrieves $(\ms{usk}_\ms{id}, \ms{cert}_\ms{id})$ from the entry containing $\ms{id}$ in $\ms{info}$, returns $\Sigma \leftarrow \ms{Sign}(\ms{gpk}, (\ms{usk}_\ms{id}, \ms{cert}_\ms{id}), M)$ and sets $\ms{SIGN} \leftarrow \ms{SIGN}\left\|(\ms{id}, M, \Sigma)\right.$.

    \item $\mc{O}_{\ms{Reveal}}$: receives query on a user identifier $\ms{id}$, $\mathcal{O}_{\ms{Reveal}}$ checks if $\ms{id} \in \ms{Honest}$. It returns $\bot$ if such $\ms{id}$ does not exist. Else, it returns the output of $\ms{Reveal}(\ms{gpk}, \ms{osk}, \ms{id})$ and adds $\ms{id}$ to $\ms{REV}$.
    
    \item $\mc{O}_{\ms{Claim}}$: receives query on an identifier $\ms{id}$, a message $M$ and a signature $\Sigma$, this oracle first checks whether $\ms{id} \in \ms{Honest}$ and whether there is an entry $(\ms{id}, M, \Sigma)$ in $\ms{SIGN}$. If one of these conditions does not hold, it returns $\bot$. Else, the oracle probes $\ms{info}$ for an entry containing $(\ms{usk}_\ms{id}, \ms{cert}_\ms{id})$ of user $\ms{id}$, and returns $\chi \leftarrow \ms{Claim}(\ms{gpk}, (\ms{usk}_\ms{id}, \ms{cert}_\ms{id}), (M, \Sigma))$. It then sets $\ms{CLAIM} \leftarrow \ms{CLAIM}\left\|(\ms{id}, (M, \Sigma), \chi)\right.$. 
        
    \item $\mc{O}_{\ms{Open}}$: receives query on a message $M$ and a signature $\Sigma$, the oracle verifies if $(M, \Sigma)$ is valid. If not, it returns $\bot$. Otherwise, it returns the output of $\ms{Open}(\ms{gpk}, \ms{osk}, (M, \Sigma))$. 
    
\end{itemize}
For an oracle $\mc{O}$, we denote $\mc{O}^{\neg S}$ to indicate that $\mc{O}$ rejects queries on any input in the set $S$. We remark that from the above description, an adversary can run $\mc{O}_\ms{Sign}$, $\mc{O}_\ms{Reveal}$ and $\mc{O}_\ms{Claim}$ by itself on any controlled users. For such reason, these oracles only accept queries on identifiers that belong to honest users. 

For security notion, a traceable signature scheme should be traceable, non-frameable and anonymous. \\

\noindent \textit{Traceability.} This is captured by \textit{misidentification attacks}, in which an adversary attempts to create a non-trivial valid signature that is not opened to any of controlled users, or cannot be traced back to one of them. The adversary can control some group members and even some tracing agents. 

\begin{defn}[Traceability \protect{\cite{KTY04,LY09}}]
A TS scheme $\mc{TS}$ is said to be traceable if for all polynomial $\lambda(\cdot)$ and all probabilistic, polynomial time (PPT) adversaries $\mathcal{A}$, its advantage $\ms{Adv}_{\ms{TS},\mathcal{A}}^{\ms{trace}}(\lambda): = \Pr[\ms{Exp}_{\mc{TS},\mathcal{A}}^{\ms{trace}}(\lambda)=1]$ is negligible in the security parameter $\lambda$. The experiment $\ms{Exp}_{\mc{TS},\mathcal{A}}^{\ms{trace}}$, formalizing misidentification attacks, is depicted as in Figure~\ref{fig:exp-trace}.
\begin{center}
\begin{figure}[H]
\centering
\begin{algorithm}[H]
$\ms{pp} \leftarrow \ms{Setup}(1^\lambda, N)$\;
$\ms{(gpk, gsk, osk)} \leftarrow \ms{KeyGen}(1^\lambda)$\;
$(M^\star, \Sigma^\star) \leftarrow \mathcal{A}^{\mc{O}_\ms{pJoin}, \mc{O}_\ms{aJoin}, \mc{O}_\ms{Corrupt}, \mc{O}_\ms{Sign}, \mc{O}_\ms{Reveal}}(\ms{gpk})$\;
\If{$\ms{Verify}(\ms{gpk}, M^\star, \Sigma^\star) = 0$}{\Return 0\;}
\If{$\exists \ms{id} \in \ms{Honest}: (\ms{id}, M^\star, \Sigma^\star) \in \ms{SIGN}$}{\Return 0\;}
\If{$\ms{Open}(\ms{gpk}, \ms{osk}, (M^\star, \Sigma^\star)) \not \in \ms{Control}$ \Or $\forall \ms{id} \in \ms{Control}: \ms{Trace}(\ms{gpk}, (M^\star, \Sigma^\star), \ms{Reveal}(\ms{gpk}, \ms{osk}, \ms{id})) = 0$}{\Return 1\;}
\Return 0\;
\end{algorithm}
\caption{Experiment $\ms{Exp}_{\mc{TS},\mathcal{A}}^{\ms{trace}}(\lambda)$ defining traceability.}
\label{fig:exp-trace}
\end{figure}
\end{center}
\end{defn}

\noindent \textit{Non-frameability.} This is captured by \textit{framing attacks}, in which an adversary aims to 1) produce a valid signature that is opened or traced to an honest user; or 2) forge a claim of a signature generated by honest user. In this type of attacks, adversary can corrupt both of the $\mc{GM}$ and $\mc{OA}$.

\begin{defn}[Non-frameability \protect{\cite{KTY04,CPY06}}]
A TS scheme $\mc{TS}$ is said to be non-frameable if for all polynomial $\lambda(\cdot)$ and all PPT adversaries $\mathcal{A}$, its advantage $\ms{Adv}_{\mc{TS},\mathcal{A}}^{\ms{frame}}(\lambda): = \Pr[\ms{Exp}_{\mc{TS},\mathcal{A}}^{\ms{frame}}(\lambda)=1]$ is negligible in the security parameter $\lambda$. The experiment $\ms{Exp}_{\mc{TS},\mathcal{A}}^{\ms{frame}}$, formalizing framing attacks, is depicted as in Figure~\ref{fig:exp-frame}.
\begin{center}
\begin{figure}[H]
\centering
\begin{algorithm}[H]
$\ms{pp} \leftarrow \ms{Setup}(1^\lambda, N)$\;
$\ms{(gpk, gsk, osk)} \leftarrow \ms{KeyGen}(1^\lambda)$\;
$(M^\star, \Sigma^\star, \chi^\star) \leftarrow \mathcal{A}^{\mc{O}_\ms{pJoin}, \mc{O}_\ms{Corrupt}, \mc{O}_\ms{Sign}}(\ms{gpk},\ms{gsk},\ms{osk})$\;
\If{$\ms{Verify}(\ms{gpk}, M^\star, \Sigma^\star) = 0$}{\Return 0\;}
\If{$\left(\ms{Open}(\ms{gpk}, \ms{osk}, (M^\star, \Sigma^\star)) \in \ms{Honest}\right) \lor \left(\exists \ms{id} \in \ms{Honest}: \ms{Trace}(\ms{gpk}, (M^\star, \Sigma^\star), \ms{Reveal}(\ms{gpk}, \ms{osk}, \ms{id})) = 1\right)$ \And $(\cdot, M^\star, \Sigma^\star) \not \in \ms{SIGN}$}{\Return 1\;}
\If{$\exists \ms{id}^\prime \in \ms{Honest}: (\ms{id}^\prime, M^\star, \Sigma^\star) \in \ms{SIGN}$ \And $\ms{ClaimVerify}(\ms{gpk}, (M^\star, \Sigma^\star), \chi^\star) = 1$}{\Return 1\;}
\Return 0\;
\end{algorithm}
\caption{Experiment $\ms{Exp}_{\mc{TS},\mathcal{A}}^{\ms{frame}}(\lambda)$ defining non-frameability.}
\label{fig:exp-frame}
\end{figure}
\end{center}
\end{defn}
In Figure~\ref{fig:exp-frame}, line 8 captures the winning condition when the adversary successfully creates a claim of a signature created by an honest user $\ms{id}^\prime$, without knowing $\ms{id}^\prime$'s secret. \\

\noindent \textit{Anonymity.} This is captured via a 2-phase game. During the first phase, an adversary is allowed to introduce honest and controlled users to the system, observe generated signatures and control some of the tracing agents. Then it chooses a challenge message $M^\star$ and two honest users $\ms{id}_0^\star$ and $\ms{id}_1^\star$ of which the tracing trapdoors are not revealed. The phase concludes when adversary receives a signature $\Sigma^\star$ which is signed by either $\ms{id}_0^\star$ or $\ms{id}_1^\star$. The adversary continues its interaction with the oracles in second phase and attempts to guess which of the targeted users creates $\Sigma^\star$. 

Naturally, Definition~\ref{def:ts-anonymity} captures anonymity in the strongest sense (CCA-anonymity) by providing the adversary access to opening oracle. If this is not the case, we obtain a weaker notion of anonymity as in~\cite{KTY04,LY09}.

In the experiment defining anonymity, unlike \cite{BP12} we do not provide adversary with a tracing oracle. This is because tracing can be done implicitly if access to opening oracle is allowed. In addition, $\mc{O}_\ms{Reveal}$ already models the threat of corrupted tracing agents (as discussed in \cite{LY09}). 
\begin{defn}[Anonymity \protect{\cite{KTY04,LY09,BP12}}] \label{def:ts-anonymity} A TS scheme $\mc{TS}$ is CCA-anonymous if for all polynomial $\lambda(\cdot)$ and all PPT adversaries $\mathcal{A}$, its advantage \[\ms{Adv}_{\mc{TS},\mathcal{A}}^{\ms{anon}}(\lambda): = \left| \Pr[\ms{Exp}_{\mc{TS},\mathcal{A}}^{\ms{anon}}(\lambda)=1] - 1/2 \right|\] is negligible in the security parameter $\lambda$. The experiment $\ms{Exp}_{\mc{TS},\mathcal{A}}^{\ms{anon}}$ is depicted as in Figure~\ref{fig:exp-anon}.
\begin{center}
\begin{figure}[H]
\centering
\begin{algorithm}[H]
$\ms{pp} \leftarrow \ms{Setup}(1^\lambda, N)$\;
$\ms{(gpk, gsk, osk)} \leftarrow \ms{Setup}(1^\lambda)$\;
$({st}, M^\star, \ms{id}^\star_0, \ms{id}^\star_1) \leftarrow \mathcal{A}^{\mc{O}_\ms{pJoin}, \mc{O}_\ms{Corrupt}, \mc{O}_\ms{Sign}, \mc{O}_\ms{Reveal}, \mc{O}_\ms{Claim}, \mc{O}_\ms{Open}}(\ms{gpk}, \ms{gsk})$\;
\If{$\ms{id}_0^\star \not \in \ms{Honest}$ \Or $\ms{id}_0^\star \in \ms{REV}$ \Or $\ms{id}_1^\star \not \in \ms{Honest}$ \Or $\ms{id}_1^\star \in \ms{REV}$}{\Return 0\;}
$b \xleftarrow{\$} \{0, 1\}$ \;
$\Sigma^\star\leftarrow \ms{Sign}(\ms{gpk}, (\ms{sk}_{\ms{id}_{b}^\star}, \ms{cert}_{\ms{id}_{b}^\star}), M^\star)$\;
$b^\prime \leftarrow \mathcal{A}^{\mc{O}_\ms{pJoin}, \mc{O}_\ms{Corrupt}^{\neg{\{\ms{id}_0^\star, \ms{id}_1^\star\}}}, \mc{O}_\ms{Sign}, \mc{O}_\ms{Reveal}^{\neg{\{\ms{id}_0^\star, \ms{id}_1^\star\}}}, \mc{O}_\ms{Claim}^{{\neg\{(\cdot, M^\star, \Sigma^\star)\}}}, \mc{O}_\ms{Open}^{\neg(M^\star, \Sigma^\star)}}(\ms{gpk}, \ms{gsk}, {st})$\;
\If{$b = b^\prime$}{\Return 1\;}
\Return 0\;
\end{algorithm}
\caption{Experiment $\ms{Exp}_{\mc{TS},\mathcal{A}}^{\ms{anon}}(\lambda)$ defining CCA-anonymity, the adversary is not allowed to make trivial query to reveal/claiming/opening oracles.}
\label{fig:exp-anon}
\end{figure}
\end{center}
\end{defn}

\section{NIZK in QROM} \label{appendix:nizk-qrom}
We recall some results regarding NIZK proofs/arguments in QROM in \cite{FLW19}. 

\subsection{Sigma Protocol}
A Sigma protocol $\Sigma$ for a relation $\mc{R}$ is a tuple $\left(D_\textrm{com}, D_\textrm{ch}, D_\textrm{rsp}, P_\Sigma^1, P_\Sigma^2, V_\Sigma \right)$; in which $D_\textrm{com}, D_\textrm{ch}, D_\textrm{rsp}$ are domains of first, second and third message respectively; $(P_\Sigma^1, P_\Sigma^2)$ are two PPT algorithms that share a shame state and together form a prover $\mc{P}$; $V_\Sigma$ is a deterministic, PT algorithm constituting a verifier $\mc{V}$ and outputting a single bit. On input $(x, w) \in \mc{R}$ where $x$ is a statement and $w$ is a corresponding witness, $P_\Sigma^1(x, w) \rightarrow \ms{com}$, where the first message $\ms{com}$ is called \textit{commitment} and is sent to $\mc{V}$. The second message is called \textit{challenge}, denoted by $\ms{ch}$ and is uniformly chosen from $D_\textrm{ch}$ by $\mc{V}$ and is sent to the $\mc{P}$. The final message is called \textit{response}, sent by $\mc{P}$ to $\mc{V}$ as $\ms{rsp} \leftarrow P_\Sigma^2(x, w, \ms{com}, \ms{ch})$. Finally, $\mc{V}$ runs $V_\Sigma$ on input $(x, \ms{com}, \ms{ch}, \ms{rsp})$ and accepts iff $V_\Sigma$ outputs 1. The tuple $(\ms{com}, \ms{ch}, \ms{rsp})$ is called a transcript of $\Sigma$. \\

A sigma protocol should satisfy the following properties:
\begin{enumerate}
\item \textbf{Completeness:} If $\mc{P}$ and $\mc{V}$ follow the protocol with statement $x$ as the common input and witness $w$ as the private input to $\mc{P}$, such that $(x, w) \in \mc{R}$; then $\mc{V}$ accepts with overwhelming probability.
\item \textbf{(Computational) $k$-special soundness:} Given a positive integer $k \geq 2$ and let $(\ms{com}, \ms{ch}_1, \ms{rsp}_1)$, $\ldots$, $(\ms{com}, \ms{ch}_k, \ms{rsp}_k)$ be $k$ valid transcripts of $\Sigma$ for a statement $x$, produced by some quantum PT algorithm $\mc{A}$, where $\ms{ch}_1, \ldots, \ms{ch}_k$ are pairwise distinct. Then there exists a PPT algorithm $E_\Sigma$ called \textit{knowledge extractor}, taking $k$ valid transcripts as input and return a witness $w$ such that $(x, w) \in \mc{R}$ with overwhelming probability.
\item \textbf{Honest verifier zero-knowledge (HVZK):} There exists a PPT algorithm $S_\Sigma$, called \textit{simulator}, such that for all PPT quantum algorithm $\mc{A}$ and all $(x, w) \in \mc{R}$, the following quantity is negligible \begin{align*}
&\vert\Pr\left[b = 1: (\ms{com}, \ms{ch}, \ms{rsp}) \leftarrow (\mc{P}(x, w) \leftrightarrow \mc{V}(x)), b \leftarrow \mc{A}(\ms{com}, \ms{ch}, \ms{rsp}) \right] - \\
&\Pr\left[b = 1: (\ms{com}, \ms{ch}, \ms{rsp}) \leftarrow S_\Sigma(x), b \leftarrow \mc{A}(\ms{com}, \ms{ch}, \ms{rsp}) \right] \vert.  
\end{align*}  Here $(\ms{com}, \ms{ch}, \ms{rsp}) \leftarrow (\mc{P}(x, w) \leftrightarrow \mc{V}(x))$ denotes the transcript obtained by executing protocol honestly with prover $\mc{P}$ and verifier $\mc{V}$.
\end{enumerate}

\subsection{NIZK protocol in QROM}
A non-interactive proof protocol $\Psi = (\mc{P}, \mc{V})$ for a relation $\mc{R}$ can be described as $(\pi \leftarrow \mc{P}^\mc{H}(x, w), b \leftarrow \mc{V}^\mc{H}(x, \pi))$. The prover $\mc{P}$ and the verifier $\mc{V}$ are PPT algorithms, $\mc{H}$ is a random oracle, $x$ is the statement, $w$ is the witness such that $(x, w) \in \mc{R}$ and $b \in \{0, 1\}$. Denote by $R$ the uniform distribution of random oracles.

For a non-interactive protocol $\Psi$, the following should be satisfied:
\begin{enumerate}
\item \textbf{Completeness:} If $\mc{P}$ and $\mc{V}$ follow the protocol with statement $x$ as the common input and witness $w$ as the private input to $\mc{P}$, such that $(x, w) \in \mc{R}$; then $\mc{V}$ outputs with overwhelming probability.
\item \textbf{Zero-knowledge:} There is a pair of algorithms $(S_P , S_O)$ (called simulator), such that for any quantum polynomial-time algorithm $\mc{A}$, we have that
\[ \left| \Pr[b = 1: \mc{H} \leftarrow R, b \leftarrow \mc{A}^{\mc{H},\mc{P}}()] - \Pr[b = 1: \mc{H} \leftarrow S_O, b \leftarrow A^{\mc{H}, S_P}() \right| \]
is negligible, where $\mc{A}^{\mc{H}, \mc{P}}$ denotes the quantum PT algorithm $\mc{A}$ with oracle access to $\mc{H}$ and $\mc{P}$ (similar for $A^{\mc{H}, S_P}$).
\end{enumerate}

Another property is \textit{online-extractability}, which is a special case of \textit{special soundness}. Roughly speaking, online-extractability captures the ability to extract a witness without rewinding a prover. In QROM, an extractor against this property is allowed to get the circuit description of random oracles, which are generated by the random oracle simulator $S_O$ of zero-knowledge property.

\begin{defn}
$\Psi$ is online-extractable with respect to a simulator $S_O$ if there exist a PPT extractor $\mc{E}$ such that for any quantum, PT algorithm $\mc{A}$, the following probability 
\[\Pr[b = 1 \land (x, w) \not \in \mc{R}: \mc{H} \leftarrow S_O, (x, \pi) \leftarrow \mc{A}^\mc{H}, b \leftarrow \mc{P}^\mc{H}(x, \pi), w \leftarrow \mc{E}(\mc{H}, x, \pi)] \] is negligible.
\end{defn}

\textit{Simulation-sound online-extractability} is a special case of \textit{online-extractability}, which essentially states a witness extractor succeeds with overwhelming probability, even against an adversary that adaptively gets many simulated proofs.

\begin{defn}
$\Psi$ is simulation-sound online-extractable, if there exist a PPT extractor $\mc{E}$ such that for any quantum, PT algorithm $\mc{A}$, the following probability 
\begin{align*}
\Pr[&b = 1 \land (x, \pi) \not \in \mathbb{X} \land (x, w) \not \in \mc{R}: \\
&\mc{H} \leftarrow S_O, (x, \pi) \leftarrow \mc{A}^{\mc{H}, S_P}, b \leftarrow \mc{P}^\mc{H}(x, \pi), w \leftarrow \mc{E}(\mc{H}, x, \pi)] \end{align*} is negligible. Here $\mathbb{X}$ denotes the set of proofs produced by $S_P$.
\end{defn}

\subsection{NIZK from Generalized Unruh Transformation}
Let $\Sigma = \left(D_\textrm{com}, D_\textrm{ch}, D_\textrm{rsp}, P_\Sigma^1, P_\Sigma^2, V_\Sigma \right)$ be a sigma protocol for a relation $\mc{R}$ with completeness, $k$-special soundness and HVZK properties. The generalized Unruh transformation \cite{FLW19} transform $\Sigma$ into a non-interactive protocol $\Psi$ between prover $\mc{P}$ and verifier $\mc{V}$ by utilizing two hash functions $\mc{H}_1$ and $\mc{H}_2$ modeled as random oracles, where $\mc{H}_1: \{0, 1\}^* \rightarrow \{1, 2, \ldots, m\}^t$ and $\mc{H}_2: D_\textrm{rsp} \rightarrow D_\textrm{rsp}$. Here $t$ is the number of times prover $\mc{P}$ needs to execute  $P_\Sigma^1$ and $m$ is a positive integer such that $m > k$. The algorithms of prover $\mc{P}$ and verifier $\mc{V}$ is depicted as in Figure~\ref{fig:unruh-prove} and Figure~\ref{fig:unruh-verif} respectively. Here, we let  $[n]$ be the set $\{1, 2, \ldots, n\}$.

\begin{figure}
\centering
\begin{algorithm}[H] \label{fig:unruh-prove}
    \LinesNumberedHidden
    \For{$i = 1, \ldots, t$}{
        $\ms{com}_i \leftarrow P_\Sigma^1(x, w)$ \;
        \For{$j = 1, \ldots, m$}{
            $\ms{ch}_{i, j} \lr D_\textrm{ch} \backslash \{\ms{ch}_{i, 1}, \ldots, \ms{ch}_{i, j-1} \}$\;
            $\ms{rsp}_{i, j} \leftarrow P_\Sigma^2(x, w, \ms{com}_i, \ms{ch}_{i, j})$\;
        }
    }
    \For{$i = 1, \ldots, t$}{
        \For{$j = 1, \ldots, m$}{
            $h_{i, j} = \mc{H}_2(\ms{rsp}_{i, j})$\;
        }
    }
    Parse $J_1 | \cdots | J_t := \mc{H}_1\left(x, (\ms{com}_i)_{[t]}, (\ms{ch}_{i, j})_{[t] \times [m]}, (h_{i, j})_{[t] \times [m]}\right)$, where $J_i \in [m]$ \;
    \For{$i = 1, \ldots, t$}{
        $\ms{rsp}_i := \ms{rsp}_{i, J_i}$ \;
    }   
    $\pi := \left((\ms{com}_i)_{[t]}, (\ms{ch}_{i, j})_{[t] \times [m]}, (h_{i, j})_{[t] \times [m]}, (\ms{rsp}_i)_{[t]} \right)$ \;
    \KwRet{$\pi$}
\end{algorithm}
\caption{A prover $\mc{P}$, generates an NIZK proof $\pi$ for $(x, w) \in \mc{R}$.}
\end{figure}

\smallskip

\begin{figure}
\begin{algorithm}[H] \label{fig:unruh-verif} 
    Parse $J_1 | \cdots | J_t := \mc{H}_1\left(x, (\ms{com}_i)_{[t]}, (\ms{ch}_{i, j})_{[t] \times [m]}, (h_{i, j})_{[t] \times [m]}\right)$, where $J_i \in [m]$ \;
    \For{$i = 1, \ldots, t$}{
        Verify that $\ms{ch}_{i, 1}, \ldots, \ms{ch}_{i, t}$ pairwise distinct \;
    }
    \For{$i = 1, \ldots, t$}{
        Verify that $V_\Sigma(x, \ms{com}_i, \ms{ch}_{i, j}, \ms{rsp}_i) = 1$\;
    }
    \For{$i = 1, \ldots, t$}{
        Verify that $h_{i, J_i} = \mc{H}_2(\ms{rsp}_{i, j})$ \;
    }   
    \If{all checks succeed}{\Return 1\;}
    \Else{\Return 0\;}
\end{algorithm}
\caption{Verifier $\mc{V}$ checks the validity of a proof $\pi$ on statement $x$.}
\end{figure}

\begin{prop}[\cite{FLW19}]
Let $\Sigma$ be a Sigma protocol that is complete, $k$-special soundness and HVZK. Then the non-interactive protocol $\Psi$ derived from $\Sigma$ as described in \textbf{Fig.}~\ref{fig:unruh-prove} and \textbf{Fig.}~\ref{fig:unruh-verif} is complete, zero-knowledge and simulation-sound online-extractable. 
\end{prop}

To turn $\Psi$ into a signature of knowledge on a message $M$, we simply include $M$ as an input to $\mc{H}_1$. That is, in the prover's algorithm in \textbf{Fig.}~\ref{fig:unruh-prove}, $\mc{P}$ computes \[\mc{H}_1\left(M, x, (\ms{com}_i)_{[t]}, (\ms{ch}_{i, j})_{[t] \times [m]}, (h_{i, j})_{[t] \times [m]}  \right).\] In the case $\Sigma$ has a common reference string $\ms{crs}$, the NIZK prover computes \[\mc{H}_1\left(\ms{crs}, M,  x, (\ms{com}_i)_{[t]}, (\ms{ch}_{i, j})_{[t] \times [m]}, (h_{i, j})_{[t] \times [m]}  \right).\]

\section{Lattice-Based ZK Argument for Quadratic Relation} \label{appendix:lattice-based-zk}
We recall the ZK argument system from \cite{YAZ19}. Let $n, m, \ell$ be positive integer. Let $\mathbf{A} \in \mathbb{Z}_q^{m \times n}$, $\mathbf{x} \in \mathbb{Z}_q^n$ and $\mathbf{y} \in \mathbb{Z}_q^m$. Let $\mathcal{S}$ be a set of tuples $(h, i, j)$ of integers in $[1, n]$ such that $\vert \mc{S} \vert = \ell$. The ZK argument system from \cite{YAZ19} is a proof-of-knowledge protocol for the relation \begin{equation} \label{eq:quadratic-relation}
\mathcal{R} = \left\{ \left(\left(\mathbf{A}, \mathbf{y}, \mathcal{S}\right), \mathbf{x}\right): \mathbf{A} \cdot \mathbf{x} = \mathbf{y} \bmod{q} \land \forall (h, i, j) \in \mathcal{M}: \mathbf{x}[h] = \mathbf{x}[i] \cdot \mathbf{x}[j] \bmod{q} \right\},   
\end{equation}

\noindent
\textbf{The basic protocol} Let $\lambda$ be a security parameter and \textsf{AuxCom} be an arbitrary secure string commitment scheme with randomness space $\{0, 1\}^\kappa$. Let $l_1, l_2 = \Theta(\lambda)$ be positive integers; $\mb{B}_1$ and $\mb{B}_2$ are two random matrices in the form 
\begin{eqnarray*}
\mb{B}_1 = \left(\begin{array}{ c | c }
    \mb{I}_{l_1} & \mb{B}_{1, 1} \\[5pt]
    \mb{0}^{n \times l_1} & \mb{I}_n~~~~\mb{B}_{1, 2}
  \end{array}\right), ~~~~~\mb{B}_2 = \left(\begin{array}{ c | c }
    \mb{I}_{l_1} & \mb{B}_{2, 1} \\[5pt]
    \mb{0}^{\ell \times l_1} & \mb{I}_\ell~~~~\mb{B}_{2, 2}
  \end{array}\right),
\end{eqnarray*}
where $\mb{B}_{1,1}, \mb{B}_{1,2}, \mb{B}_{2,1}, \mb{B}_{2,2}$ are uniformly random matrices sampled from $\mathbb{Z}_q^{l_1 \times (l_2 + n)}, \mathbb{Z}_q^{n \times l_2}, \mathbb{Z}_q^{l_1 \times (l_2 + \ell)}$ and $\mathbb{Z}_q^{\ell \times l_2}$ respectively. Looking forward, $\mb{B}_1$ and $\mb{B}_2$ are public parameters of the BDLOP commitment scheme~\cite{BDLOP18}.

In addition, the system choose a small integer $p = \textrm{poly}(\lambda)$ defining the soundness error. The randomnesses for BDLOP commitment scheme are sampled from two discrete Gaussian distributions, with width $\sigma_1 \geq \sqrt{2l_2/\pi}$ and $\sigma_2 = 2 \cdot p \cdot l \cdot \log l \cdot \sigma_1$, where $l = 2l_1 + 2l_2 + n + \ell$. The system modulus $q$ is a power of a prime $q_0$, i.e. $q = q_0^e$ and is chosen such that $q \geq 16p \cdot \max\{l_1 + l_2 + n, l_1 + l_2 + \ell\} \cdot (\sigma_2 + p \cdot \sigma_1) \cdot \Tilde{\mc{O}}(\sqrt{l_1})$ and $q/\sigma_1$ is a polynomial. The common reference string is defined as \[
\ms{crs} = \left(\ms{AuxCom}, \mb{B}_1, \mb{B}_2, \sigma_1, \sigma_2, p\right).
\]

In the following protocol, let $\mathfrak{p}(\mb{v}, \mb{z}) = \min \left(1, \dfrac{D_{\sigma_2}^l(\mb{z})}{M \cdot D_{\sigma_2, \mb{v}}^l(\mb{z})}\right)$ for any vectors $\mb{v}, \mb{z} \in \mathbb{Z}^l$, where $M = e^{\mc{O}(1/ \log^2 l)}$.

\begin{tcolorbox}[enhanced,breakable,title={$\Sigma$-protocol for relation $\mc{R}$}]
\begin{enumerate}
  \item \textbf{Commitment:} 
  
    - $\mathcal{P}$ samples uniformly random $\mb{r} \in \mathbb{Z}_q^n$, computes $\mb{t} \leftarrow \mb{A} \cdot \mb{r} \bmod q$;

    - $\mathcal{P}$ samples randomnesses $\mb{s}_1 \leftarrow D_{\sigma_1}^{l_2 + n + l_1}$, $\mb{s}_2 \leftarrow D_{\sigma_2}^{l_2 + n + l_1}$, $\mb{s}_3 \leftarrow D_{\sigma_1}^{l_2 + \ell + l_1}$, $\mb{s}_4 \leftarrow D_{\sigma_2}^{l_2 + \ell + l_1}$;

    - $\mb{c}_1 \leftarrow \mb{B}_1 \cdot \mb{s}_1 + \left( \mb{0}^T | \mb{x}^T \right)^T \bmod q$, $\mb{c}_2 \leftarrow \mb{B}_1 \cdot \mb{s}_2 + \left( \mb{0}^T | \mb{r}^T \right)^T \bmod q$;

    - Let $\mb{a}, \mb{b}$ be two vectors in $\mathbb{Z}_q^\ell$;

    - For $k \in [1, \ell]$, let $(h, i, j)$ be the $k$-th element of $\mc{S}$ and compute 
    \begin{eqnarray*}
        \mb{a}[k] = \mb{r}[h] - \mb{r}[i] \cdot \mb{r}[j] - \mb{r}[j] \cdot \mb{x}[i],~~~\mb{b}[k] = \mb{r}[i] \cdot \mb{r}[j];
    \end{eqnarray*}

    - $\mb{c}_3 \leftarrow \mb{B}_2 \cdot \mb{s}_3 + \left( \mb{0}^T | \mb{a}^T \right)^T$, $\mb{c}_4 \leftarrow \mb{B}_2 \cdot \mb{s}_4 + \left( \mb{0}^T | \mb{b}^T \right)^T$;

    - Samples randomness $\rho \lr \{0, 1\}^\kappa$, commits \[\ms{com}_\ms{aux} \leftarrow \ms{AuxCom}(\mb{t} \mid \mb{c}_1 \mid \mb{c}_2 \mid \mb{c}_3 \mid \mb{c}_4; \rho)\] and sends $\ms{com}_\ms{aux}$ to $\mc{V}$.

  \item \textbf{Challenge:} $\mc{V}$ sends a uniformly random challenge $\ms{ch} \in [-p, p]$ to $\mathcal{P}$.
  
  \item \textbf{Response:} $\mathcal{P}$ computes
  \begin{eqnarray*}
      \mb{z}_0 \leftarrow \ms{ch} \cdot \mb{x} + \mb{r},~~~\mb{z}_1 \leftarrow \ms{ch} \cdot \mb{s}_1 + \mb{s}_2,~~~\mb{z}_2 \leftarrow \ms{ch} \cdot \mb{s}_3 - \mb{s}_4;
  \end{eqnarray*}
  aborts with probability $1 - \mathfrak{p}\left( (\ms{ch} \cdot \mb{s}_1^T | \ms{ch} \cdot \mb{s}_2^T)^T, (\mb{z}_1^T | \mb{z}_2^T)^T \right)$, and sends the response as $\ms{rsp} = \left(\mb{c}_1, \mb{c}_3, \rho, \mb{z}_0, \mb{z}_1, \mb{z}_2 \right)$.
\end{enumerate}
\smallskip
\textbf{Verification:}  Receiving $\ms{rsp}$, $\mathcal{V}$ parses $\ms{rsp} = \left(\mb{c}_1, \mb{c}_3, \rho, \mb{z}_0, \mb{z}_1, \mb{z}_2 \right)$ and proceeds as follows:
\begin{itemize} 
\item Let $\mb{d} \in \mathbb{Z}^\ell$. For $k \in [1, \ell]$, let $(h, i, j)$ be the $k$-th element of $\mc{S}$ and set \begin{align*}
\mb{d}[k] \leftarrow \ms{ch} \cdot \mb{z}_0[h] - \mb{z}_0[i] \cdot \mb{z}_0[j];
\end{align*}
\item Set \begin{align*}
\mb{t} &\leftarrow \mb{A} \cdot \mb{z}_0 - \ms{ch} \cdot \mb{y}; \\
\mb{c}_2 &\leftarrow \mb{B}_1 \cdot \mb{z}_1 + \left(\mb{0}^T | \mb{z}_0^T \right)^T - \ms{ch} \cdot \mb{c}_1; \\
\mb{c}_4 &\leftarrow \ms{ch} \cdot \mb{c}_3 - \mb{B}_2 \cdot \mb{z}_2 + \left(\mb{0}^T | \mb{d}_0^T \right)^T; 
\end{align*}
\item Verify that:

- $\ms{com}_\ms{aux} = \mathsf{AuxCom}(\mb{t} | \mb{c}_1 | \mb{c}_2 | \mb{c}_3 | \mb{c}_4; \rho)$;

- $\left\| \mb{z}_1 \right\| \leq 2 \sqrt{l_1 + l_2 + n} \cdot (\sigma_2 + p \cdot \sigma_1)$;

- $\left\| \mb{z}_2 \right\| \leq 2 \sqrt{l_1 + l_2 + \ell} \cdot (\sigma_2 + p \cdot \sigma_1)$.
\end{itemize}
$\mathcal{V}$ outputs $1$ if and only if all the conditions hold.
\end{tcolorbox}

In the above protocol, response and verification algorithms slightly differ from the original one. In fact, these are the tweaks proposed in \cite{YAZ19} for the non-interactive version. They help reduce the communication cost and do not affect the overall security.

\begin{prop}[Adapted from {\cite{YAZ19}}] \label{prop:yaz-protocol} Suppose that $p < q_0/2$, then the above protocol is a ZK argument protocol for relation $\mathcal{R}$ with completeness error $1-1/M$, soundness error at most $2/(2p+1)$ and is honest-verifier zero knowledge assuming the hardness of $\ms{SIS}_{l_1, l_1 + l_2 + n, q, \beta_1}^{(2)}$, $\ms{SIS}_{l_1, l_1 + l_2 + \ell, q, \beta_2}^{(2)}$ and $\ms{LWE}_{l_2, q, \mc{D}_{\sigma_1}}$, where $\beta_1 = 16p\cdot\sqrt{l_1 + l_2 + n} \cdot (\sigma_2 + p \cdot \sigma_1)$ and $\beta_2 = 16p\cdot\sqrt{l_1 + l_2 + \ell} \cdot (\sigma_2 + p \cdot \sigma_1)$. In particular, the prover's response has bit size at most $\kappa + (6l_1 + 2l_2 + 3n + 2\ell)\log q$ and the protocol is 3-soundness: given 3 valid pairs of challenge-response with the same commitment, there is a PPT extractor for a witness of $(\mb{A}, \mb{y}, \mc{S})$.
\end{prop}

We remark that the protocol has non-negligible completeness error due to aborting, therefore when apply the algorithm in \textbf{Fig.}~\ref{fig:unruh-prove} one keeps repeating the protocol until all responses are generated successfully. The number repetitions can be reduced by employing rejection sampling technique. Additionally, \textsf{AuxCom} should be a binding and hiding string commitment scheme. In lattice settings, it can be instantiated by the KTX commitment scheme \cite{KTX08}, which is computationally binding (under SIS assumption) and statistically hiding.
\section{Detailed NIZK Proofs for Lattice-Based Traceable Signature Scheme}\label{appendix:lattice-ts-nizk}
We explain how to prove the relations $\mc{R}_\ms{Sign}$ in Definition~\ref{def:ts-relation-sign} and $\mc{R}_\ms{Claim}$ in Definition~\ref{def:ts-relation-claim}. First, we consider a generic relation that captures all the relations defining $\mc{R}_\ms{Sign}$ and $\mc{R}_\ms{Claim}$. Then we give the detail transformation of each of those relations to a case of \eqref{eq:quadratic-relation}.
\subsection{Proving a Generic Relation}
Recall that the TS scheme in Section~\ref{section:lattice-ts-description} works with two moduli $q$ and $q^\prime$, where $q > q^\prime$. As the relations defining our TS scheme involves equations modulo $q^\prime$, we need to appropriately transform them into sets of equations modulo $q$.

Following the observation in \cite{TNLPS24}, statements \ref{item-1}, \ref{item-3} and \ref{item-4} are special cases of the following relation
\begin{align} \label{eq:basic-relation}
\mathcal{R}_{m, n, q^\prime, \beta} &= \{(\mb{A}, \mb{y}, \mb{x}) \in \mathbb{Z}_{q^\prime}^{m \times n} \times \mathbb{Z}_{q^\prime}^m \times \mathbb{Z}_{q^\prime}^n : \mb{y} = \mb{A} \cdot \mb{x} \bmod{{q^\prime}}; \left\|\mb{x}\right\|_\infty \leq \beta\}    
\end{align}  
where $1 \leq \beta \leq ({q^\prime}-1)/2$. We show that relation \eqref{eq:basic-relation} can be transformed into a case of relation \eqref{eq:quadratic-relation}. Rewrite the equation modulo $q^\prime$ as $\mb{y} = \mb{A} \cdot \mb{x} + q^\prime \cdot \mb{v}$ for some $\mb{v} \in \mathbb{Z}^m$. Observe that \[q^\prime \cdot \left\| \mb{v} \right\|_\infty \leq  \left(\left\|\mb{y} \right\|_\infty + \left\|\mb{A} \cdot \mb{x} \right\|_\infty \right) \leq q^\prime \cdot \frac{n\beta+1}{2}, \] since the entries of $\mb{A}$ have absolute values at most $(q^\prime-1)/2$.
Thus, for sufficiently large $q$, namely $q > q^\prime (n\beta+1)$, if we let  $\mb{A}^\prime = \left[\mb{A} | q^\prime \mb{I}_m \right] \in \mathbb{Z}_q^{m \times (n+m)}$, $\mb{x}^\prime =  \left[\mb{x} | \mb{v} \right] \in \mathbb{Z}_q^{n+m}$ and $\beta^\prime = (n\beta+1)/2$, then $\mathcal{R}_{m, n, q^\prime, \beta}$ is equivalent to the following relation 
\begin{align} \label{eq:basic-relation-transformed}
\left.\begin{array}{rl}
\mathcal{R}_{n, m, q, \beta, \beta^\prime} &=\{((\mb{A}^\prime, \mb{y}), \mb{x}^\prime) \in \mathbb{Z}_q^{m \times (n+m)} \times \mathbb{Z}_q^m \times \mathbb{Z}_q^{n+m} : \mb{y} = \mb{A}^\prime \cdot \mb{x}^\prime \bmod{q}; \\
&\mb{x}^\prime =  \left[\mb{x} | \mb{v} \right] \in \mathbb{Z}_q^{n+m}; \left\|\mb{x}\right\|_\infty \leq \beta;
\left\| \mb{v} \right\|_\infty \leq \beta^\prime \},
\end{array} \right.
\end{align}
since $\left\|\mb{A}^\prime \mb{x}^\prime \right\|_\infty < q/2$. 
Now we transform the relation $\mathcal{R}_{n, m, q, \beta, \beta^\prime}$ to a case of relation $\mc{R}$ in \eqref{eq:quadratic-relation}, by applying the binary decomposition technique from \cite{LNSW13}. Formally, let $\mb{b} = \begin{pmatrix} \beta  \cdots  \beta \end{pmatrix}^T \in \mathbb{Z}_q^{n}$, $\mb{b}^\prime = \begin{pmatrix} \beta^\prime & \cdots & \beta^\prime \end{pmatrix}^T \in \mathbb{Z}_q^{m}$ and let $\mb{x}_1 = \mb{x} + \mb{b}, \mb{x}_2 = \mb{v} + \mb{b}^\prime$. Let $k =\lceil \log(2\beta) \rceil + 1$ and \[\mb{g}_1 = \begin{pmatrix}
 \lfloor \dfrac{2\beta + 1}{2} \rfloor  &  & \lfloor \dfrac{2\beta + 2}{4} \rfloor &  & \cdots & & \lfloor \dfrac{2\beta + 2^{k-1}}{2^k} \rfloor \end{pmatrix} \in \mathbb{Z}^{k} \] be a row vector. In \cite{LNSW13}, it is proven that an integer $a \in [0, 2\beta]$ if and only if there exists a binary vector $\mb{a} \in \{0, 1\}^k$ such that $\mb{g}_1 \cdot \mb{a} = a$. Using this fact, if we let $\mb{G}_1 = \mb{I}_n \otimes \mb{g}_1$, then there exists a binary vector $\overline{\mb{x}_1} \in \{0, 1\}^{kn}$ such that $\mb{G}_1 \cdot \overline{\mb{x}_1} = \mb{x}_1$. Similarly, let \begin{align*}
 k^\prime &= \lceil \log(2\beta^\prime) \rceil + 1; \\
\mb{g}_2 &= \begin{pmatrix}
 \lfloor \dfrac{2\beta^\prime + 1}{2} \rfloor  & &  \lfloor \dfrac{2\beta^\prime + 2}{4} \rfloor & &\cdots & &\lfloor \dfrac{2\beta^\prime + 2^{k^\prime-1}}{2^{k^\prime}} \rfloor \end{pmatrix}  \in \mathbb{Z}^{k^\prime}; \\
 \mb{G}_2 &= \mb{I}_m \otimes \mb{g}_2 \in \mathbb{Z}^{m \times k^\prime m};
\end{align*}
then then there exists a binary vector $\overline{\mb{x}_2} \in \{0, 1\}^{k^\prime m}$ such that $\mb{G}_2 \cdot \overline{\mb{x}_2} = \mb{x}_2$. Now, form the matrix $\overline{\mb{A}^\prime} = \left[\mb{A} \cdot \mb{G}_1 | q^\prime \mb{I}_m \cdot \mb{G}_2\right]  \in \mathbb{Z}_q^{m \times (kn + k^\prime m)}$ and form the column vector $\overline{\mb{x}^\prime} = \left[\overline{\mb{x}_1}^T | \overline{\mb{x}_2}^T \right]^T \in \mathbb \{0, 1\}^{kn + k^\prime m}$. The relation $\mathcal{R}_{n, m, q, \beta, \beta^\prime}$ is transformed into \begin{eqnarray} \label{eq:binary-relation}
\{((\overline{\mb{A}^\prime}, \mb{y}), \overline{\mb{x}^\prime}) \in \mathbb{Z}_q^{m \times (kn+k^\prime m)} \times \mathbb{Z}_q^m \times \{0, 1\}^{kn+k^\prime m} : \mb{y} = \overline{\mb{A}^\prime} \cdot \overline{\mb{x}^\prime} \bmod{q}\}.    
\end{eqnarray}
This is a linear relation modulo $q$, with a binary constraint over the witness $\overline{\mb{x}^\prime}$. Note that when $q$ is prime and $x \in \mathbb{Z}_q$, then we have $x \in \{0, 1\}$ iff $x = x^2 \bmod{q}$. Hence, if we let $\mc{S} = \left\{(i, i, i): i \in \{1, 2, \ldots, kn + k^\prime m\} \right\}$, then the above relation can be rewritten into the following \[ 
\{((\overline{\mb{A}^\prime}, \mb{y}, \mc{S}), \overline{\mb{x}^\prime}): \mb{y} = \overline{\mb{A}^\prime} \cdot \overline{\mb{x}^\prime} \bmod{q} \land \forall(i, i, i) \in \mc{S}: \overline{\mb{x}^\prime}[i] =\overline{\mb{x}^\prime}[i] \cdot \overline{\mb{x}^\prime}[i] \}.
\] which is a case of the quadratic relation in \eqref{eq:quadratic-relation}. It follows that one can prove relation \eqref{eq:basic-relation} by Yang \textit{et al} argument \cite{YAZ19}.

\subsection{Proving the Defining Statements of Lattice-Based Traceable Signature Scheme} \label{appendix:proof-statements}
In this section we describe the transformations of the statements \ref{item-1}, \ref{item-3}, \ref{item-4} and \ref{item-5} into a case of relation \eqref{eq:basic-relation}. 

\noindent\textbf{Proof of Image-Preimage of SIS One-way Function. }
Let $\mathbf{F} \in \mathbb{Z}_{q^\prime}^{n \times m_\mb{F}}$ define an SIS-based one-way function over domain $\{0, 1\}^{m_\mb{F}}$ and range $\mathbb{Z}_{q^\prime}^n$. We want to prove in zero-knowledge the following relation \[\mathcal{R}_{\ms{SIS}} = \left\{(\mathbf{F}, (\mathbf{x}, \mathbf{z})) \in \mathbb{Z}_{q^\prime}^{n \times {m_\mb{F}}} \times \mathbb{Z}_q^n \times \{0, 1\}^{m_\mb{F}} : \mathbf{x} = \mathbf{F} \cdot \mathbf{z} \bmod{q^\prime} \right\}.\]
Let $\mb{w} = \ms{bin}(\mb{x}) \in \{0, 1\}^{k}$, where $k = \lceil \log q^\prime \rceil$ and $\mb{G}_n \in \mathbb{Z}_q^{n \times k}$ be a gadget matrix. It follows that $\mb{G}_n \cdot \mb{w} = \mb{x}$ and that $\mb{G}_n \cdot \mb{w} = \mb{F} \cdot \mb{s}  - {q^\prime} \cdot \mb{a}$ for $\mb{a} \in \mathbb{Z}^{m_\mb{F}}$. In addition \[q^\prime \left\|\mb{a}\right\|_\infty \leq (\left\|\mb{t}\right\|_\infty + \left\|\mb{F} \cdot \mb{s} \right\|_\infty) \leq q^\prime (m_\mb{F}+1)/2.\] By choosing $q > {q^\prime}(m_\mb{F} + 1)$, and letting 
\begin{align*}
\overline{\mb{A}} &= \begin{pmatrix} \mb{G}_n & & -\mb{F} & & q^\prime\mb{I}_{m_\mb{F}}
\end{pmatrix} \in \mathbb{Z}^{n \times (2m_\mb{F} + k)}; \\
\overline{\mb{x}} &= \left(\mb{w}, \mb{z}, \mb{a} \right) \in \mathbb{Z}^{2m_\mb{F} + k};
\end{align*}
then we have $\overline{\mb{A}} \cdot \overline{\mb{x}} = \mb{0} \bmod{q}$, where the components $\mb{w}, \mb{z}$ in $\overline{\mb{x}}$ are binary vectors and $\mb{a}$ has infinity norm at most $(m_\mb{F} + 1)/2$. Note that, $(\overline{\mb{A}}, \overline{\mb{x}})$ is an instance of relation ~\eqref{eq:basic-relation-transformed}. Hence we can apply the binary decomposition technique to $\mb{a}$ to transform $\mathcal{R}_{\ms{SIS}}$  to a case of relation \eqref{eq:quadratic-relation}. Overall, the final quadratic relation has witness length and number of quadratic constraints approximately $m_\mb{F} + n \log q^\prime + m_\mb{F}\log(m_\mb{F} + 1) \approx 2n \log q^\prime + 3n\log(n \log q^\prime)$. \\

\smallskip
\noindent\textbf{Proof of LWE Sample-Secret.} Let $\mb{y} \in \mathbb{Z}_{q^\prime}^{m_\mb{B}}$, $\mb{x} \in \mathbb{Z}_{q^\prime}^n$, $\mb{e} \in \mathbb{Z}^{m_\mb{B}}$ and $\mb{B} \in \mathbb{Z}_{q^\prime}^{m_\mb{B} \times n}$, we need to prove in ZK the following relation 
 \begin{align*}
     \mc{R}_{\ms{LWE}}^{(1)} = &\left\{(\mb{M}, (\mb{y}, \mb{x}, \mb{e})): \mb{y} = \mb{B}^\top \cdot \mb{x} + \mb{e} \bmod q^\prime \land \left\| \mb{e} \right\|_\infty \leq B_\ms{LWE} \right\}.
 \end{align*}
Let $\mb{G}_{m_\mb{B}}$ be a gadget matrix and $\ms{bin}(\mb{y}) \in \{0, 1\}^{m_\mb{B} \lceil \log q^\prime \rceil}$, we have that $\mb{G}_{m_\mb{B}} \cdot \ms{bin}(\mb{y}) = \mb{y} \bmod q^\prime$. Write $\mb{y} - \mb{B}^\top \cdot \mb{x} - \mb{e} = q^\prime \cdot \mb{a}$ for $\mb{a} \in \mathbb{Z}^{m_\mb{B}}$ then \[
q^\prime \cdot \left\| \mb{a} \right\|_\infty \leq  \left\|\mb{y} - \mb{B}^\top \cdot \mb{x} - \mb{e} \right\|_\infty \leq (q^\prime/2) + (q^\prime/2)^2 + B_\ms{LWE}.
\]
By choosing $q > q^\prime + {q^\prime}^2 + 2B_\ms{LWE}$ and letting 
\begin{align*}
\overline{\mb{A}} &= \begin{pmatrix}\mb{I}_{m_\mb{B}} & & -\mb{B}^\top & & \mb{I}_{m_\mb{M}} & & \mb{0}^{m_\mb{B} \times m_\mb{B}  \lceil \log q^\prime \rceil} & &-q^\prime\mb{I}_{m_\mb{M}} \\
-\mb{I}_{m_\mb{B}} & & \mb{0}^{m_\mb{B} \times n} & & \mb{0}^{m_\mb{B} \times m_\mb{B}}  & & \mb{G}_{m_\mb{B}} & & \mb{0}^{m_\mb{B} \times m_\mb{B}} 
\end{pmatrix} \in \mathbb{Z}^{2m_\mb{B} \times (3m_\mb{B} + n + m_\mb{B} \lceil \log q^\prime \rceil)}_q; \\
\overline{\mb{x}} &= \left(\mb{y}, \mb{x}, \mb{e}, \ms{bin}(\mb{y}), \mb{a} \right) \in \mathbb{Z}^{m_\mb{B}} \times \mathbb{Z}^{n} \times \mathbb{Z}^{m_\mb{B}} \times \{0, 1\}^{m_\mb{B} \lceil \log q^\prime \rceil} \times \mathbb{Z}^{m_\mb{B}};
\end{align*}
we have that $\overline{\mb{A}} \cdot \overline{\mb{x}} = \mb{0} \bmod q$. By using the technique that transforms relation \eqref{eq:basic-relation} to a case of relation \eqref{eq:quadratic-relation}, we can transform $\mc{R}_\ms{LWE}^{(1)}$ to a case of relation \eqref{eq:quadratic-relation} as well. The final quadratic relation has witness length and number of quadratic constraints approximately $3m_\mb{B} \log(q^\prime) + n \log(q^\prime) +  m_\mb{B}\log(2B_\ms{LWE})$. 

We remark here that in addition to proving $\mc{R}_{\ms{LWE}}^{(1)}$, we also prove the well-formedness of the binary message $\ms{bin}(\mb{y})$  certified by the group manager (i.e. $\ms{bin}(\mb{y})$ is a correct binary decomposition of $\mb{y}$), thus provide a link between statements \ref{item-2} and \ref{item-3}. \\

\smallskip
\noindent\textbf{Proof of Plaintext Knowledge in GPV-IBE. }
Let $\mb{c} \in \mathbb{Z}_{q^\prime}^{m_\mb{B} + 1}$, $\mb{B} \in \mathbb{Z}_{q^\prime}^{n \times m_\mb{B}}$, $\mb{v} \in \mathbb{Z}_{q^\prime}^n$, $\mb{r} \in \mathbb{Z}_{q^\prime}^{n}$, $\mb{e}_\mb{c} \in [-B_\ms{GPV}, B_\ms{GPV}]^{m_\mb{B}+1}$ and $\ms{id} \in \{1, \ldots, N = 2^\ell-1\}$. We need to prove in ZK the following relation 
 \begin{align*}
     \mc{R}_{\ms{Enc}} = &\left\{((\mb{c}, \mb{B}, \mb{v}), (\mb{r}, \mb{e}_\mb{c}, \ms{id})): \mb{c} =  \begin{pmatrix}
         \mb{B}^\top \\
         \mb{v}^\top
     \end{pmatrix} \cdot \mb{r} + \mb{e}_\mb{c} +  \begin{pmatrix}
         \mb{0}^{m_\mb{B}} \\
         \lceil q^\prime/(2(N+1))\rfloor \cdot \ms{id}
     \end{pmatrix} \bmod q^\prime \right\}.
 \end{align*}
Let  \[
\mb{C}^\prime = \left(\begin{array}{ccc}
  \begin{pmatrix}
         \mb{B}^\top \\
         \mb{v}^\top
     \end{pmatrix}    &  \mb{I}_{m_\mb{B}+1} & \begin{pmatrix}
    \mb{0}^{m_\mb{B}} \\
    \lceil q^\prime/(2(N+1) \rfloor
\end{pmatrix}
\end{array} \right) \in \mathbb{Z}_{q^\prime}^{(m_\mb{B}+1) \times (n + m_\mb{B} + 1)}
\]
it follows that $\mb{c} = \mb{C}^\prime \cdot (\mb{r}, \mb{e}_\mb{c}, \ms{id}) \bmod q^\prime$. Write $\mb{c} = \mb{C}^\prime \cdot (\mb{r}, \mb{e}_\mb{c}, \ms{id}) - q^\prime \cdot \mb{a}$ for $\mb{a} \in \mathbb{Z}^{m_\mb{B}+1}$, then \[
q^\prime \cdot \left\|\mb{a} \right\|_\infty \leq \left\|\mb{c} \right\|_\infty + \left\|\mb{C}^\prime \cdot (\mb{r}, \mb{e}_\mb{c}, \ms{id}) \right\|_\infty \leq  q^\prime \cdot \dfrac{1+nq^\prime/2+B_\ms{GPV}+N}{2}.
\]
By choosing $q > {q^\prime} \cdot \dfrac{1+nq^\prime/2+B_\ms{GPV}+N}{2}$, and letting 
\begin{align*}
\overline{\mb{A}} &= \begin{pmatrix} \mb{C}^\prime & &  -q^\prime\mb{I}_{m_\mb{B}+1}
\end{pmatrix} \in \mathbb{Z}^{(m_\mb{B} + 1) \times (n + 2m_\mb{B} + 2)}_q; \\
\overline{\mb{x}} &= \left(\mb{r}, \mb{e}_\mb{c}, \ms{id},\mb{a} \right) \in \mathbb{Z}^{n} \times \mathbb{Z}^{m_\mb{B}+1} \times \mathbb{Z} \times \mathbb{Z}^{m_\mb{B}+1};
\end{align*}
we have that $\overline{\mb{A}} \cdot \overline{\mb{x}} = \mb{c} \bmod q$. By using the technique that transforms relation \eqref{eq:basic-relation} to a case of relation \eqref{eq:quadratic-relation}, we can transform $\mc{R}_\ms{Enc}$ to a case of relation \eqref{eq:quadratic-relation} as well. The final quadratic relation has witness length and number of quadratic constraints approximately $n \log(q^\prime) + (m_\mb{B}+1)\log(2B_\ms{GPV}) + \log(N) + (m_\mb{B}+1)\log(1+nq^\prime/2+B_\ms{GPV}+N)$.\\

\smallskip
\noindent\textbf{Proof of LWE Secret. } Let $\mb{t} \in \mathbb{Z}_{q^\prime}^{m_\mb{M}}$, $\mb{x} \in \mathbb{Z}_{q^\prime}^n$, $\mb{e}_\mb{t} \in \mathbb{Z}^{m_\mb{M}}$ and $\mb{M} \in \mathbb{Z}_{q^\prime}^{m_\mb{M} \times n}$, we need to prove in ZK the following relation 
 \begin{align*}
     \mc{R}_{\ms{LWE}}^{(2)} = &\left\{((\mb{t}, \mb{M}), (\mb{x}, \mb{e}_\mb{t})): \mb{t} = \mb{M} \cdot \mb{x} + \mb{e} \bmod q^\prime \land \left\| \mb{e} \right\|_\infty \leq B_\ms{LWE} \right\}.
 \end{align*}
Write $\mb{t} = \mb{M} \cdot \mb{x} + \mb{e} - q^\prime \cdot \mb{a}$ for $\mb{a} \in \mathbb{Z}^{m_\mb{M}}$, then \[
q^\prime \cdot \left\| \mb{a} \right\|_\infty \leq \left\|\mb{t} \right\|_\infty + \left\|\mb{M} \cdot \mb{x} \right\|_\infty + \left\|\mb{e}\right\|_\infty \leq (q^\prime/2) + (q^\prime/2)^2 + B_\ms{PRF}.
\]
By choosing $q > q^\prime + {q^\prime}^2 + 2B_\ms{LWE}$ and letting 
\begin{align*}
\overline{\mb{A}} &= \begin{pmatrix} \mb{M} & & \mb{I}_{m_\mb{M}} & & -q^\prime\mb{I}_{m_\mb{M}}
\end{pmatrix} \in \mathbb{Z}^{m_\mb{M} \times (n + 2m_\mb{M})}_q; \\
\overline{\mb{x}} &= \left(\mb{x}, \mb{e}_t, \mb{a} \right) \in \mathbb{Z}^{n} \times \mathbb{Z}^{m_\mb{M}} \times \mathbb{Z}^{m_\mb{M}};
\end{align*}
we have that $\overline{\mb{A}} \cdot \overline{\mb{x}} = \mb{t} \bmod q$.  By using the technique that transforms relation \eqref{eq:basic-relation} to a case of relation \eqref{eq:quadratic-relation}, we can transform $\mc{R}_\ms{LWE}^{(2)}$ to a case of relation \eqref{eq:binary-relation}. The final quadratic relation has witness length and number of quadratic constraints approximately $n \log q^\prime + m_\mb{M}\log(2B_\ms{LWE}) + m_\mb{M}\log(q^\prime) \approx 4n \log q^\prime$.

\subsection{Putting Everything Together} \label{appendix:asymptotic-size}
To prove relation $\mc{R}_\ms{Sign}$ (Definition~\ref{def:ts-relation-sign}), each of the statements defining $\mc{R}_\ms{Sign}$ is transformed to a case of quadratic relation as in \eqref{eq:quadratic-relation}:
\begin{itemize}
    \item Statement proving knowledge of preimage-image under SIS-based function (item~\ref{item-1}), is transformed into $((\mb{A}_\ms{SIS}, \mb{y}_\ms{SIS}, \mc{S}_\ms{SIS}), \mb{x}_\ms{SIS})$, where $\mb{A}_\ms{SIS} \in \mathbb{Z}_q^{n_\ms{SIS} \times m_\ms{SIS}}$, $\mb{y}_\ms{SIS} = \mb{0}^{n_\ms{SIS}}$, $\#\mc{S}_\ms{SIS} =m_\ms{SIS}$ and $\mb{x}_\ms{SIS} \in \{0, 1\}^{m_\ms{SIS}}$. Here, $n_\ms{SIS} = n$ and $m_\ms{SIS} \approx 2n \log q^\prime + 3n\log(n \log q^\prime)$;
    \item Statement proving message-signature pair in the signature scheme by Jeudy \textit{et al.} \cite{JRS23} (item~\ref{item-2}), is transformed into a case of \ref{eq:quadratic-relation} (see~\cite[Appendix F.2]{JRS23}). The final statement is $((\mb{A}_\ms{sig}, \mb{y}_\ms{sig}, \mc{S}_\ms{sig}), \mb{x}_\ms{sig})$, where $\mb{A}_\ms{sig} \in \mathbb{Z}_q^{n_\ms{sig} \times m_\ms{sig}}$, $\mb{y}_\ms{sig} \in \mathbb{Z}_q^{n_\ms{sig}}$ and $\mb{x}_\ms{sig} \in \mathbb{Z}^{m_\ms{sig}}$. Here, $n_\ms{sig} = 2n + 1$, $m_\ms{sig} \approx 1 + \log N + m_1 \log \beta_1 + m_2 \log \beta_2 + m_\mb{B}\log q^\prime + 2n$;  and $\#\mc{S}_\ms{sig} = m_\ms{sig} - n - 1$;
    \item Statement proving LWE sample and secret (item~\ref{item-3}), is transformed into $((\mb{A}_\ms{LWE}^{(1)}, \mb{y}_\ms{LWE}^{(1)}, \mc{S}_\ms{LWE}^{(1)}), \mb{x}_\ms{LWE}^{(1)})$, where $\mb{A}_\ms{LWE}^{(1)} \in \mathbb{Z}_q^{n_\ms{LWE}^{(1)} \times m_\ms{LWE}^{(1)}}$, $\mb{y}_\ms{LWE}^{(1)} \in \mathbb{Z}_q^{n_\ms{LWE}^{(1)}}$, $\#\mc{S}_\ms{enc}^{(1)} =m_\ms{LWE}^{(1)}$ and $\mb{x}_\ms{LWE}^{(1)} \in \{0, 1\}^{m_\ms{LWE}^{(1)}}$. Here, $n_\ms{LWE}^{(1)} = 2m_\mb{B} $ and $m_\ms{LWE}^{(1)} \approx 3m_\mb{B} \log(q^\prime) + n \log(q^\prime) +  m_\mb{B}\log(2B_\ms{LWE})$;
    \item Statement proving plaintext corresponding to a ciphertext of GPV-IBE \cite{GPV08} (item~\ref{item-4}), is transformed into $((\mb{A}_\ms{enc}, \mb{y}_\ms{enc}, \mc{S}_\ms{enc}), \mb{x}_\ms{enc})$, where $\mb{A}_\ms{enc} \in \mathbb{Z}_q^{n_\ms{enc} \times m_\ms{enc}}$, $\mb{y}_\ms{enc} \in \mathbb{Z}_q^{n_\ms{enc}}$, $\#\mc{S}_\ms{enc} =m_\ms{enc}$ and $\mb{x}_\ms{enc} \in \{0, 1\}^{m_\ms{enc}}$. Here, $n_\ms{enc}= m_\mb{B}+1$ and $m_\ms{enc} \approx n \log(q^\prime) + (m_\mb{B}+1)\log(2B_\ms{GPV}) + \log(N) + (m_\mb{B}+1)\log(1+nq^\prime/2+B_\ms{GPV}+N)$;
    \item Statement proving LWE secret (item~\ref{item-5}), is transformed into $((\mb{A}_\ms{LWE}^{(2)}, \mb{y}_\ms{LWE}^{(2)}, \mc{S}_\ms{LWE}^{(2)}), \mb{x}_\ms{LWE}^{(2)})$, where $\mb{A}_\ms{LWE}^{(2)} \in \mathbb{Z}_q^{n_\ms{LWE}^{(2)} \times m_\ms{LWE}^{(2)}}$, $\mb{y}_\ms{LWE}^{(2)} \in \mathbb{Z}_q^{n_\ms{LWE}^{(2)}}$, $\#\mc{S}_\ms{LWE} =m_\ms{LWE}^{(2)}$ and $\mb{x}_\ms{LWE} \in \{0, 1\}^{m_\ms{LWE}^{(2)}}$. Here, $n_\ms{LWE}^{(2)} = m_\mb{M} = 3n$ and $m_\ms{LWE}^{(2)} \approx n \log q^\prime + m_\mb{M}\log(2B_\ms{LWE}) + m_\mb{M}\log(q^\prime) \approx 4n \log q^\prime$;
\end{itemize}
From the matrices $\mb{A}_\ms{SIS}, \mb{A}_\ms{sig}, \mb{A}_\ms{LWE}^{(1)}, \mb{A}_\ms{enc}, \mb{A}_\ms{LWE}^{(2)}$; the column vectors $\mb{y}_\ms{SIS}, \mb{y}_\ms{sig}, \mb{y}_\ms{LWE}^{(1)}, \mb{y}_\ms{enc}, \mb{y}_\ms{LWE}^{(2)}$; the sets $\mc{S}_\ms{SIS}$, $\mc{S}_\ms{sig}$, $\mc{S}_\ms{LWE}^{(1)}$, $\mc{S}_\ms{enc}$, $\mc{S}_\ms{LWE}^{(2)}$ and the witness $\mb{x}_\ms{SIS}, \mb{x}_\ms{sig}, \mb{x}_\ms{LWE}^{(1)}, \mb{x}_\ms{enc}, \mb{x}_\ms{LWE}^{(2)}$; we can form a matrix $\mb{A}_\ms{Sign} \in \mathbb{Z}_q^{n_\ms{Sign} \times m_\ms{Sign}}$, a vector $\mb{y}_\ms{Sign} \in \mathbb{Z}_q^{n_\ms{Sign}}$, a set $\mc{S}_\ms{Sign}$ and a witness $\mb{x}_\ms{Sign} \in \{0, 1\}^{m_\ms{Sign}}$ such that $\mb{A}_\ms{Sign} \cdot \mb{x}_\ms{Sign} = \mb{y}_\ms{Sign} \bmod q$ and the set $\mc{S}_\ms{Sign} = \{(i, j, k): i, j, k \in [m_\ms{Sign}]\}$ enforces the quadratic constraints over the coordinates of $\mb{x}_\ms{Sign}$. In particular, we have \[
n_\ms{Sign} = n_\ms{SIS} + n_\ms{sig} + n_\ms{LWE}^{(1)} +  n_\ms{enc} + n_\ms{LWE}^{(2)} =\mc{O}(n \log q^\prime) = \mc{O}(\lambda \log \lambda),
\] 
and \[ m_\ms{Sign} = 
 m_\ms{SIS} + m_\ms{sig} + m_\ms{LWE}^{(1)} + m_\ms{enc} + m_\ms{LWE}^{(2)} = \mc{O}(n(\log q^\prime)^2) = \mc{O}(\lambda \log^2 \lambda),
\]
and the set $\mc{S}_\ms{Sign}$ has size $\ell_\ms{Sign} \approx m_\ms{Sign}$. By Proposition~\ref{prop:yaz-protocol}, the bit size of prover's response in the ZK argument proving $\mc{R}_\ms{Sign}$ is of order $\mc{O}(\lambda \log^3 \lambda)$.

Similarly, we can transform statement $\mc{R}_\ms{Claim}$ (Definition~\ref{def:ts-relation-claim}) by transforming each of the statements defining $\mc{R}_\ms{Claim}$ to a case of \eqref{eq:quadratic-relation}. The final statement has the form $((\mb{A}_\ms{Claim}, \mb{y}_\ms{Claim}, \mc{S}_\ms{Claim}), \mb{x}_\ms{Claim})$, where $\mb{A}_\ms{Claim} \in \mathbb{Z}_q^{n_\ms{Claim} \times m_\ms{Claim}}$, $\mb{y}_\ms{Claim} \in \mathbb{Z}_q^{n_\ms{Claim}}$, $\mb{x}_\ms{Claim} \in \{0, 1\}^{m_\ms{Claim}}$ and the set $\mc{S}_\ms{Claim}$ enforces the binary constraints over the coordinates of $\mb{x}_\ms{Claim}$. In particular, we have \[
n_\ms{Claim} = n_\ms{SIS} + n_\ms{LWE}^{(2)} = n + 3n = 4n = \mc{O}(\lambda),
\] 
and \[
m_\ms{Claim} = m_\ms{SIS} + m_\ms{LWE}^{(2)} \approx n + 4n \log q^\prime = \mc{O}(n\log q^\prime) = \mc{O}(\lambda \log \lambda),
\] 
and $\mc{S}_\ms{Claim}$ has size $\ell_\ms{Claim} = 4n = \mc{O}(\lambda)$. By Proposition~\ref{prop:yaz-protocol}, the bit size of prover's response in the ZK argument proving $\mc{R}_\ms{Claim}$ is of order $\mc{O}(\lambda \log^2 \lambda)$.

From $(n_\ms{Sign}, m_\ms{Sign}, \ell_\ms{Sign})$ and $(n_\ms{Claim}, m_\ms{Claim}, \ell_\ms{Claim})$, one sets up the common reference string $\ms{crs}$ in the ZK argument system by choosing an auxiliary string commitment schemes and public parameters of BDLOP commitment scheme \cite{BDLOP18}. To prove $\mc{R}_\ms{Sign}$ and $\mc{R}_\ms{Claim}$, signers run the ZK protocol with generalized Unruh transformations using the respective hash functions.
\end{document}